\documentclass{article}
\usepackage{geometry}
\usepackage[utf8]{inputenc}
\usepackage[T1]{fontenc}
\usepackage{lmodern}
\usepackage[dvipsnames]{xcolor}

\usepackage{nicefrac}
\usepackage{graphicx}%
\usepackage{multirow}
\usepackage{amsmath,amssymb,mathrsfs}
\usepackage{tikz}
\usetikzlibrary{positioning,calc}

\usepackage{hyperref}

\newcommand{\ignore}[1]{}

\newcommand{\R}{\mathbb{R}}

\newcommand{\C}{\mathbb{C}}

\DeclareMathOperator{\SO}{SO}
\DeclareMathOperator{\OO}{O}
\DeclareMathOperator{\Sp}{Sp}
\newcommand{\dd}{\text{d}}
\newcommand{\zz}{{\overline{z}}}
\newcommand{\sdd}{\Delta_\Gamma}

\newcommand{\bb}{\boldsymbol}

\newcommand{\MSbar}{\overline{\mbox{MS}}}

\title{Five loop renormalization of $\phi^3$ theory with applications to the Lee--Yang edge singularity and percolation theory}

\author{M.~Borinsky,$^a$ J.~A.~Gracey,$^b$ M.~V.~Kompaniets$^c$ and O.~Schnetz$^d$}

\date{%
    $^a$\small{\textit{Nikhef Theory Group, Science Park 105, 1098 XG Amsterdam, The Netherlands}}\\%
    $^b$\small{\textit{Theoretical Physics Division, Department of Mathematical Sciences, \\University of Liverpool, P.O. Box 147, Liverpool, L69 3BX, United Kingdom}}\\%
    $^c$\small{\textit{St. Petersburg State University, 7/9 Universitetskaya nab., \\St. Petersburg 199034, Russia}}\\%
    $^d$\small{\textit{Department Mathematik, Friedrich Alexander Universit\"{a}t Erlangen-N\"{u}rnberg, \\Cauerstra\ss{}e 11, 91058 Erlangen, Germany}}%
}

\hypersetup{
  pdfauthor={M. Borinsky, J. A. Gracey, M. V. Kompaniets, O. Schnetz},
  pdftitle={Five loop renormalization of phi3 theory with applications to the Lee--Yang edge singularity and percolation theory},
  pdfsubject={},
  pdfkeywords={},
  linkcolor  = RedViolet!85!black,
  citecolor  = BlueGreen!85!black,
  urlcolor   = YellowOrange!85!black,
  colorlinks = true,
}

\begin{document}

\vspace*{-2\baselineskip}%
\hspace*{\fill} \mbox{\footnotesize{\textsc{Nikhef-2021-006, LTH-1254}}}

\begingroup
\let\newpage\relax
\maketitle
\endgroup

\abstract{ %
We apply the method of graphical functions that was recently extended to
six dimensions for scalar theories, to $\phi^3$ theory and compute the
$\beta$ function, the wave function anomalous dimension as well as the mass
anomalous dimension in the $\MSbar$ scheme to five loops. From the results
we derive the corresponding renormalization group functions for the Lee--Yang
edge singularity problem and percolation theory. After determining the
$\varepsilon$ expansions of the respective critical exponents to $\mathcal{O}(\varepsilon^5)$
we apply recent resummation technology to obtain improved exponent estimates
in $3$, $4$ and $5$ dimensions. These compare favourably with estimates from 
fixed dimension numerical techniques and refine the four loop results. To assist with this
comparison we collated a substantial amount
of data from numerical techniques which are
included in tables for each exponent.

}

\section{Introduction}
One of the core quantum field theories with applications to various areas of
physics is that of a scalar field with a cubic self-interaction. It has many
interesting properties. For instance, it is known to be asymptotically free in six dimensions
\cite{Macfarlane:1974vp,Ma:1975vn} which is also its critical dimension.
As such it offered a much simpler forum to study this property
rather than in the more complicated non-abelian gauge field theory underlying
the strong interactions which also has this property. Another major area where
scalar $\phi^3$ theory has important consequences is that of condensed matter
physics. For example, the non-unitary version of the model,
\cite{Fisher_1978}, describes the phase transitions of the \emph{Lee--Yang edge
singularity problem}. In particular the critical exponents computed in the cubic
theory produce estimates which are not out of line with those of other methods,
\cite{Bonfim_1980,Bonfim_1981}. Having an accurate
estimate for the exponent $\sigma$ for the Lee--Yang
edge singularity is important in lattice gauge
theory studies of Quantum Chromodynamics (QCD),
\cite{Connelly:2020pno,Connelly:2020gwa}.
Specifically it governs the analytic behaviour of the
partition function in the smooth chiral 
symmetry crossover when there is a non-zero
chemical potential. The latter is included in
QCD as an purely imaginary parameter that
leads to loss of unitarity. Though for the
region of application in 
\cite{Connelly:2020pno,Connelly:2020gwa}
physical meaningful results can still be extracted.
In addition endowing the cubic model with specific
symmetries means it can also describe phase transitions in \emph{percolation problems}.
This follows from taking the replica limit of the $(N+1)$-state Potts model
\cite{potts_1952}. The critical dynamics of the phase transition in percolation
has been widely studied using Monte Carlo or series methods, 
on discrete or spin
systems. References for this comprehensive
body of work will be provided in later sections. 
Indeed such analyses proceed over a range of integer dimensions from
$2$ to $6$ inclusive where the latter would correspond to the mean field
approximation given its relation to the critical dimension. The relation of the
discrete percolation theory models to that of a continuum field theory resides
in the fact that at the phase transition both scalar $\phi^3$ theory and the
spin models lie in the same universality class. What differs of course in both
approaches are the techniques used to estimate the physically measurable
critical exponents. On the continuum quantum field theory side these are
\emph{renormalization group invariants} that are determined from high perturbative
loop order renormalization group functions. While these functions are scheme
dependent, the critical exponents at the $d$ dimensional Wilson--Fisher fixed
point, \cite{Wilson:1971dc}, are scheme independent. In more recent years other
continuum field theory techniques have been developed. Two of the main ones are
the functional renormalization group and the modern manifestation of the
conformal bootstrap and applied to the Lee--Yang and percolation problem for
example in \cite{Gliozzi:2014jsa,An:2016lni,Zambelli_2017} and
\cite{LeClair:2018edq,hikami_2018,Pismensky_2015} respectively.

In terms of basic scalar $\phi^3$ theory the multiloop renormalization of the
model has proceeded in stages over the last half century or so. The one and two
loop renormalization group functions were determined in
\cite{Macfarlane:1974vp}. This was extended to the $\OO(N)$ group in
\cite{Ma:1975vn} where the leading order value of $N$ for the existence of the conformal window was
determined.
The extension to three loops for both the
Lee--Yang edge singularity and percolation problems was carried out in
\cite{Bonfim_1980,Bonfim_1981}. From the point of view of hindsight that
computation was well ahead of its time given the difficulty of several of the
three loop vertex graphs that needed to be evaluated. Moreover, estimates for
the exponents in the percolation and Lee--Yang problems were extracted in
dimensions $d$ in the range $2$~$<$~$d$~$<$~$6$ that were competitive with
other results available then. To achieve a high
degree of accuracy the analysis
benefited
from improved resummation techniques such as Pad\'{e} approximants
and Borel transformations, where in the latter conformal mappings were applied
and the  behaviour of the $\varepsilon$ expansion was incorporated.
Thereafter progress in systematically renormalizing theories to higher loop order was
hindered in general by a lack of technology to push to four loops. 
Indeed three loop calculations were only 
viable due to the integration by parts (IBP) method
introduced in \cite{Chetyrkin:1981qh}. However,
with the
development of Laporta's integration by parts algorithm, \cite{Laporta:2001dd},
and its implementation within a variety of publicly available packages, the
four loop renormalization of $\phi^3$ was carried out in \cite{Gracey:2015tta}.
That article covered a range of applications to various problems that had
emerged in the interim. For instance, in recent
years it has been shown that there
is a connection of $\phi^3$ theory with dualities in higher spin AdS/CFTs,
\cite{Klebanov:2002ja,Giombi:2014iua}. This generated an interest in
understanding the conformal window of $\phi^3$ theories for various symmetry
groups such as $\OO(N)$ and $\Sp(N)$, \cite{Fei2014On,Fei2015On,Fei2015SPn}. One
highlight of the four loop result of \cite{Gracey:2015tta} was the improvement in
estimates for the Lee--Yang and percolation theory exponents. What this
analysis benefited immensely from was the progress in classifying two
dimensional conformal field theories in the years after
\cite{Bonfim_1980,Bonfim_1981}. Specifically the values of the exponents of
each problem in two dimensions were found {\em exactly}. For instance, the exponent $\sigma$ in
one and two dimensions was determined exactly in
\cite{Cardy:1985yy}. For percolation theory
the two dimensional values 
can be derived from
the unitary minimal conformal field theories
of \cite{Friedan:1983xq} when $m$~$=$~$2$ and
the central charge is $c$~$=$~$0$. Therefore it
was possible to use that data, together with
hyperscaling relations, to construct constrained
Pad\'{e} approximants motivated by the
application of this idea given in \cite{Butera:2012tq}. The
upshot was that exponent estimates based on four loop results for three
dimensions were within one standard deviation of Monte Carlo and series results. This
is all the more remarkable when one recalls this is the resummation of a series
that is more reliable near six dimensions down to three dimensions. We note that for brevity
we will refer to results from non-continuum field theories as Monte Carlo but 
this will cover those from series methods too. We 
note that in this respect we have 
compiled exponent estimates from as many sources
as we could. These will also include strong
coupling methods, functional renormalization group
techniques and several specialized approaches. An
excellent live source for percolation exponents
is \cite{percolationtable}. 

While the Laporta approach has revolutionized our ability to extend results
for many theories to loop orders that what would
have been impossible to conceive of a decade ago,
one is always interested in going beyond even
those orders. In the short
term any such developments have to proceed in simpler theories. Indeed this
has been the case for scalar $\phi^4$ theory which was renormalized at five
loops in the $90$'s in \cite{Chetyrkin:1981jq,Kleinert:1991rg} but not extended
to six loops until around a quarter of a century later,
\cite{batkovich2016six,KompanietsPanzer2017}.
Moreover the latter article \cite{KompanietsPanzer2017} contained a most
comprehensive analysis of the resummation of exponents using the asymptotic
properties of the series. Such an analysis was much-needed after the huge jump
in precision. Indeed it revisited the assumptions made in earlier approaches in
the literature. Subsequent to 
\cite{batkovich2016six,KompanietsPanzer2017}
a novel method was applied to scalar $\phi^4$
theory in \cite{Schnetz_2018} which is termed graphical functions. This
extended the renormalization group functions to the staggeringly high
{\em seven}
 loop order\footnote{
The eight loop evaluation of the field anomalous dimension has now been determined \cite{Schnetz2018hyperlog}.
}.
One highlight was the appearance of a new period in the
$\beta$ function which is conjectured to not be expressible in terms of a
multiple zeta value (MZV), \cite{PanzerSchnetz:2017coact,Schnetz_2018}. %
En route expectations
concerning the non-MZV 
content predicted in \cite{Broadhurst:1995km} at
high loop order were confirmed.

One lesson from \cite{Schnetz_2018} was the potential usefulness of the
graphical function technique to extend the renormalization group functions.
An obstruction that limited the technique to four dimensional problems 
was overcome by the first and fourth author, who
extended the method to arbitrary even dimensions.
The details of this extension will be given
elsewhere \cite{Borinsky2020prep}. 
This availability of graphical functions in arbitrary even dimensions immediately opened up
new possibilities for the computation of the renormalization group functions in $\phi^3$ theory in six dimensions.
In this article we make use of this new tool and provide the renormalization group
functions of $\phi^3$ theory up to five loops.
We emphasise that we will make no use at
all of integration by parts in our computation.
Consequently we will find the next term in the
$\varepsilon$ expansion
of the critical exponents for the Lee--Yang singularity and percolation problems. This will
form the first part of the article.
The second part
will be devoted to a comprehensive resummation
analysis of the respective critical exponents
from the $\varepsilon$
expansion in the region between $2$ and $6$ dimensions. 
This resummation analysis will be based on technology from
\cite{KompanietsPanzer2017} 
but combined with
constraints from two dimensional theories as in \cite{Gracey:2015tta}.
Thanks to the additional perturbative order
and the more advanced resummation technology we are able to improve on the estimates obtained
in \cite{Gracey:2015tta}. In broad terms 
our estimates of the critical exponents are consistent with Monte Carlo and series
results which have
been similarly refined in recent years. In order
to make this comparison of our five loop
estimates with numerical data we have carried out
an analysis on all exponent estimates in the 
literature that had error bars and produced a
global average.

The paper is organized as follows. In Section~2 we introduce the scalar cubic
theory for the cases where there is one scalar field and the Lagrangian which is 
relevant for the percolation problem. The core machinery of the graphical
function formalism used to compute the five loop renormalization group
functions is discussed in depth in Section~3. The outcome of this mammoth
task is provided via the explicit five loop expressions for the
$\beta$ function, field anomalous dimension and the mass anomalous
dimension in Section~4 for both the Lee--Yang and percolation problems. As a
corollary the $\varepsilon$ expansion of the three corresponding critical
exponents are determined to $\mathcal{O}(\varepsilon^5)$ too.
In the subsequent section we introduce and discuss aspects of the different resummation
methods we use to extract exponent estimates. 
Then Section~6 will provide the full 
consequences of that analysis for both the
Lee--Yang problem and percolation theory.
A central element of this section is the
compilation of tables for each exponent for
both problems. Each table provides comprehensive
data of fixed dimension exponent estimates
as well as the outcome of each of our 
resummations. An overall average is provided
for each dimension for both Monte Carlo data and
our five loop results in order to have a common
comparison point. 
Concluding remarks are provided in Section~7.
Throughout the article we will in general follow
the notation of~\cite{Vasilev2004}.

\section{Background}
\label{sec:background}

We begin by outlining the essential background to the problem of extracting
critical exponents for both the Lee--Yang and percolation problems. First the
action for the basic cubic theory is
\begin{equation}
S(\phi_0) ~=~ -~ \int \dd^d x \left[ \frac{1}{2}(\partial \phi_0)^2
+ \frac{1}{2}m_0^2\phi_0^2 + \frac{g_0}{3!} \phi_0^3 \right]~,
\end{equation}
where $\phi_0$, $m_0$ and $g_0$ are the bare field, mass and coupling constant.
It is this version of the theory that was shown to be asymptotically free in
 six dimensions, \cite{Macfarlane:1974vp}. The connection to the action
underlying the Lee--Yang edge singularity problem is given by
the coupling constant mapping $g_0$~$\to$~$ig_0$, %
which yields a \emph{non-unitary theory}, \cite{Fisher_1978}.
The analogous action for
percolation theory requires $N$ fields $\phi_i$ prior to taking the replica
limit defined formally by $N$~$\to$~$0$. First we note that the most general
renormalizable $\phi^3$ theory type action in six dimensions is
\begin{equation}
S(\phi_0) ~=~ -~ \int \dd^d x \left[\frac{1}{2}(\partial \phi_{0i})^2
+ \frac{1}{2} m_0^2 \phi_{0i}^2
+ \frac{g_0}{3!} d_{ijk} \phi_{0i} \phi_{0j} \phi_{0k} \right]~,
\end{equation} 
where the indices run from $1$ to $N$. Specific values for the fully symmetric tensor $d_{ijk}$
correspond to different problems and eventually lead to different prefactors  
which multiply the individual scalar Feynman integrals in the respective perturbative 
expansions. 
For percolation theory the $d_{ijk}$ tensor is related to the vertices of an
$N$-dimensional tetrahedron, \cite{Zia:1975ha}, and corresponds to the
$(N+1)$-state Potts model, \cite{Fortuin:1971dw}. In particular $d_{ijk}$ is given by
\begin{equation}
\label{eq:dijk_def}
    d_{ijk} = \sum_{\alpha=1}^{N+1} e_\alpha^i e_\alpha^j e_\alpha^k~,
\end{equation}
where the $N$-dimensional vectors $e_1^i,\ldots,e_{N+1}^i$ satisfy the algebra, \cite{Zia:1975ha},
\begin{equation}
\label{eq:percolationrelation}
    \sum_{\alpha=1}^{N+1} e_\alpha^i = 0~, \quad
\sum_{\alpha=1}^{N+1}e_\alpha^i e_\alpha^j = (N+1) \delta^{ij}~,\quad
\sum_{i=1}^N e_\alpha^i e_\beta^i = (N+1)\delta_{\alpha\beta} - 1 ~.
\end{equation}
In previous calculations \cite{Bonfim_1981,Gracey:2015tta} these algebraic
rules were used to compute the $N$ dependence of each individual graph. For instance in \cite{Bonfim_1981} the renormalization group
functions were written in terms of invariants which corresponded with the primitively divergent
Feynman integrals.  Therefore the algebra was used to determine the $N$ dependence of the
invariants as well as their subsequent value in the replica limit. Here we use a presumably new
diagrammatic approach which is based on a 
similar one given in \cite{Bonfim_1981} to calculate the necessary prefactors for each
individual graph. The details of this approach
will be given in Section~\ref{sec:combinatorial_factors}.

For both the Lee--Yang and the percolation problem we renormalize in the modified minimal subtraction ($\MSbar$)
scheme in the dimensionally regularized theory in $d$~$=$~$6$~$-$~$\varepsilon$
which retains multiplicative renormalizability\footnote{The three loop renormalization group calculations~\cite{Bonfim_1980,Bonfim_1981} were also performed in $d=6-\varepsilon$, while the most recent four loop computations \cite{Gracey:2015tta} used $d=6-2\varepsilon$, which is more common in high energy physics.%
}. This means that the number of
independent renormalization constants equates to the number of terms in each
action. Thus the renormalized action in terms of renormalized entities for the general
theory is
\begin{equation}\label{Zfactors}
    S^R(\phi) ~=~ -~ \int \dd^d x \left[ \frac{1}{2} Z_1 (\partial \phi_i)^2
+ \frac{1}{2} Z_2 m^2\phi_i^2
+ \frac{g \mu^{\varepsilon/2}}{3!} Z_3 d_{ijk} \phi_i \phi_j \phi_k\right] ~,
\end{equation}
where
\begin{align} \begin{aligned}\label{renfkt} \phi_{0i} &= \phi_i Z_\phi~,&&& m_0^2&=m^2 Z_{m^2}~, &&& g_0 &= g\mu^{\varepsilon/2} Z_g,\\
 Z_1 &= Z_\phi^2~, &&& Z_2 &= Z_{m^2} Z_\phi^2~, &&& Z_3 &= Z_g Z_\phi^3 ~, \end{aligned} \end{align}
and we have defined the field renormalization constant in the way which is more common in
statistical physics. We highlight this explicitly in
contradistinction to the high energy physics convention where it is usually
defined as ``$\phi_{0i}$~$=$~$\phi_i Z_\phi^{1/2}\,$''.
As we use the $\MSbar$ scheme
we recall that the coupling constant and $\varepsilon$ dependence of the
renormalization constants $Z_n$ is given by
\begin{equation}
    Z_n = Z_n(g,\varepsilon) = 1+ \sum_{k=1}^{\infty}
\frac{Z_{n,k}(g)}{\varepsilon^k} ~.
\end{equation}
These are determined perturbatively by requiring the
finiteness of the renormalized 1-PI Green's
functions $\Gamma^R_n(g,m,\mu) = \left(Z_\phi\right)^n\Gamma_n(g_0,m_0)$. The 
conventional method to ensure finiteness is to use the 
Bogoliubov--Parasiuk $R$-operation \cite{bogoliubow1957multiplikation,BogShirk}, 
as well as the $R^*$-operation \cite{Chetyrkin:RRR,ChetyrkinSmirnov:Rcorrected,ChetyrkinTkachov:InfraredR,ChetyrkinGorishnyLarinTkachov:Analytical5loop}.
These allow for the use of infrared
regularization or infrared
rearrangement for Feynman diagrams on an
individual basis. This significantly simplifies the
calculations. In this article however we will use the much more powerful technique of graphical
functions to calculate the divergences of all the diagrams. This will be
discussed in the next section. One major advantage of that approach is that it
enables us to compute all diagrams in a straightforward way {\em without} any
infrared rearrangement and $R/R^*$-operations. The only simplification we use
is to consider purely massless graphs throughout. To extract the mass
renormalization constant we evaluate a $2$-point Green's function with the
mass operator inserted at zero momentum.

Once we have extracted the renormalization constants to five loops the next
stage is to produce the corresponding renormalization group functions
$\beta(g)$, $\gamma_\phi(g)$ and $\gamma_{m^2}(g)$ which are defined by
\begin{gather} \begin{gathered} \beta(g) := \mu\frac{\partial g}{\partial \mu}\bigg|_{g_0} = \frac{-\varepsilon g/2}{1+g{\partial_g} \ln{(Z_g)}}~, \quad \gamma_\phi(g) := \mu\frac{\partial}{\partial \mu} \ln(Z_\phi)\bigg|_{g_0} = \beta(g)\frac{\partial}{\partial g}\ln(Z_\phi)~, \\
\gamma_{m^2}(g) := \mu\frac{\partial}{\partial \mu} \ln(Z_{m^2})\bigg|_{g_0} = \beta(g)\frac{\partial}{\partial g}\ln(Z_{m^2}) ~. \end{gathered} \end{gather}
These ensure that after renormalization all finite renormalized $n$-point 1-PI Green's functions
$\Gamma_n^R$ will satisfy the renormalization group equation
\begin{equation}
    \left[\mu\partial_\mu +\beta(g)\partial_g-\gamma_{m^2}(g)m^2\partial_{m^2}-n\gamma_\phi(g)\right]\Gamma_n^R = 0~,
    \label{rg}
\end{equation}
where $\partial_g$~$=$~$\frac{\partial }{\partial g}$ for instance.
Once the
renormalization group functions have been established, they lay the foundation
for our application to critical exponents. In general if there is a non-trivial
infrared fixed point of the $\beta$ function at $g^\ast$ with
$\beta(g^\ast)$~$=$~$0$ then in the limit $g$~$\to$~$g^\ast$ eq.~\eqref{rg}
transforms to an equation describing critical scaling with exponents
$\gamma_\phi^\ast$~$=$~$\gamma_\phi(g^\ast)$ and
$\gamma_{m^2}^\ast$~$=$~$\gamma_{m^2}(g^\ast)$. In addition the correction to
scaling exponent $\beta^{\prime \, \ast}$~$=$~$\partial_g\beta(g)|_{g=g^*}$ corresponding to the $\beta$ function slope at criticality
will be of interest. In terms of the notation used in statistical physics the
connection between these critical point renormalization group functions and the
critical exponents is
\begin{gather} \begin{gathered} \eta ~=~ 2\gamma_\phi^*~, \quad 1/\nu ~=~ 2+\gamma_{m^2}^* ~=~ \eta_O-\eta+2~, \quad \omega ~=~ \beta^{\prime \, \ast}~, \end{gathered} \end{gather}
where $\eta_O$ is the anomalous dimension derived from $Z_2$ evaluated at
$g^\ast$ and $\nu$ corresponds to the correlation length exponent. Knowledge of
the two basic exponents $\eta$ and $\nu$ means that other critical exponents
can be accessed via hyperscaling relations. These are given by
\begin{gather} \begin{gathered} \alpha ~=~ 2 ~-~ d \nu~,\quad \beta ~=~ \frac{1}{2} (d-2+\eta) \nu~,\quad \gamma ~=~ (2-\eta)\nu~, \quad \delta ~=~ \frac{d+2-\eta}{d-2+\eta}~, \\
\sigma ~=~ \frac{2}{(d+2-\eta)\nu}~, \quad \tau ~=~ 1 ~+~ \frac{2d}{d+2-\eta}~, \quad \Omega ~=~ \frac{2\omega}{d+2-\eta} ~. \end{gathered} \label{scalerelns} \end{gather}
These will be our main focus for the percolation problem. For
the Lee--Yang problem we will concentrate on
$\eta$, $\nu$, $\sigma$ and $\omega$
specifically.

\section{Computational technique} 
\label{sec:computational}

\subsection{Graphical Functions}
For the evaluation of the required Feynman integrals, we made heavy use of the graphical function technique that has been introduced in $d=4$ by the fourth author in \cite{Schnetz:2013hqa},
extended to $d=4-\varepsilon$ in \cite{Schnetz_2018} and recently generalized to all even dimensions $\geq4$ by the first and the fourth author \cite{Borinsky2020prep}.

Recall that the massless scalar propagator in $d$-dimensional Euclidean space time is the Green's function for the respective massless Klein--Gordon equation,
\begin{align} \label{eq:propagator} \Box_{\bb x} \frac{1}{|\bb x -\bb y|^{d-2}} =-\frac{4}{\Gamma(d/2-1)}\delta^{(d)}(\bb x -\bb y), \end{align}
where $\Gamma(x)=\int_0^\infty t^{x-1}\exp(-t)\dd t$ is the gamma function. 
Up to a rescaling and a reparametrization which will be specified later, a graphical function is an Euclidean massless position space three-point correlation function which can be written as an integral over a product of such propagators,
\begin{align} \label{eq:threepointfunction} G_\Gamma(\bb x_a,\bb x_b,\bb x_c)= \left( \prod_{v \in V_\Gamma^\text{int}}\int_{\R^d} \frac{\dd \bb x_v}{\pi^{d/2}} \right) \prod_{\{v,w\} \in E_\Gamma} \frac{1}{|\bb x_v-\bb x_w|^{d-2}}. \end{align}
It is determined by a graph $\Gamma$ with edges $E_\Gamma$, internal vertices in $V_\Gamma^\text{int}$ and external vertices $ \{a,b,c\} = V_\Gamma^\text{ext}$ such that $V_\Gamma^\text{int} \cap V_\Gamma^\text{ext} = \emptyset$.
Note that in our position space setting, external vertices can have any number of incident edges.
Such a three-point function $G_\Gamma$ has translation, rotation and scaling symmetries, such that for all $\bb x_a, \bb x_b, \bb x_c \in \R^d$,
\begin{align} G_\Gamma(\bb x_a,\bb x_b,\bb x_c) &=G_\Gamma(\bb x_a + \bb v, \bb x_b + \bb v, \bb x_c + \bb v) & &\text{for all } \bb v \in \R^d \notag \\
 \label{eqn:symmetry} &= G_\Gamma(\Lambda \bb x_a, \Lambda \bb x_b, \Lambda \bb x_c) & & \text{for all }\Lambda \in \SO(d) \\
 &= \xi^{-\sdd}G_\Gamma(\xi \bb x_a, \xi \bb x_b, \xi \bb x_c) & & \text{for all }\xi \in \R_{>0}, \notag \end{align}
with the \emph{superficial degree of divergence} given by
$$
\sdd=d |V_\Gamma^\text{int}| - (d-2) |E_\Gamma|.
$$
It follows that two degrees of freedom together with $\sdd$ are sufficient to parameterize the three-point function $G_\Gamma$. A convenient parameterization is given by a single complex variable $z \in \C$,
\begin{align} \label{eqn:z_parametrization_implicit} z \zz &= \frac{\bb x_{ac}^2}{\bb x_{ab}^2}, & (1-z)(1-\zz) &= \frac{\bb x_{bc}^2}{\bb x_{ab}^2}, \end{align}
where $\bb x_{ij} = \bb x_j - \bb x_i$. Using this parameterization, we can write $G_\Gamma$ as
\begin{align} \label{eq:defgf} G_\Gamma(\bb x_a, \bb x_b, \bb x_c ) = |\bb x_{ab}|^{\sdd} f_\Gamma(z), \end{align}
where $f_\Gamma: \C \rightarrow \R$, the \emph{graphical function}, is independent of the overall scale.

An important feature is that $f_\Gamma(z)$ is a \emph{single-valued} real analytic function on $\C\backslash\{0,1\}$, \cite{Golz:2015rea}. Furthermore, a graphical function admits expansions of $\log$-Laurent type at the singular points $0$, $1$ and $\infty$, \cite{Schnetz:2013hqa,Borinsky2020prep},
\begin{gather}\label{aexp} f_\Gamma(z)=\sum_{\ell=0}^{L_a}\sum_{m,\overline{m}=M_a}^\infty c^a_{\ell,m,\overline{m}}[\log(z-a)(\zz-a)]^\ell(z-a)^m(\zz-a)^{\overline{m}},\quad|z-a|<1,\quad a\in \{0,1\} \\
\label{infexp} f_\Gamma(z)=\sum_{\ell=0}^{L_\infty}\sum_{m,\overline{m}=-\infty}^{M_\infty} c^\infty_{\ell,m,\overline{m}}[\log(z\zz)]^\ell z^m \zz^{\overline{m}},\quad|z|>1. \end{gather}
Due to the existence of the expansion at infinity in eq.~\eqref{infexp} the graphical function naturally lives on the Riemann sphere $\C\cup\{\infty\}$.

These fundamental structures of graphical functions are vital for the efficiency of the graphical function method.
After going from the $\bb x_a,\bb x_b,\bb x_c \in \R^d$ coordinates to $z,\zz$ via the parameterization in eq.~\eqref{eqn:z_parametrization_implicit}, it is convenient to rename the external vertices $a,b$ and $c$ to $0, 1$ and $z$. The reason for this is that eq.~\eqref{eqn:z_parametrization_implicit} effectively identifies the plane that is spanned in $d$-dimensional space by the points $\bb x_a,\bb x_b,\bb x_c$ with the complex plane $\C$ such that $0$ is mapped to $\bb x_a$, $1$ is mapped to $\bb x_b$ and $z$ to $\bb x_c$.

Graphical functions fulfill a large number of \emph{combinatorial} identities. We can depict a general graph with the three labeled external vertices as $$\Gamma = 
\begin{tikzpicture}[x=2ex,y=2ex,baseline={([yshift=-.6ex]current bounding box.center)}] \coordinate (v0); \coordinate [above right=1.41 and 2 of v0] (v1); \coordinate [below right=1.41 and 2 of v0] (v2); \node [left =.1 of v0] (v0n) {$z$}; \coordinate [right =1 of v0] (v0u); \node [right=.07 of v1] (v1n) {$1$}; \coordinate [below left=.71 and .71 of v1] (v1u); \node [right=.07 of v2] (v2n) {$0$}; \coordinate [above left=.71 and .71 of v2] (v2u); \draw[fill=gray] (v0) .. controls (v0u) and (v1u) .. (v1) .. controls (v1u) and (v2u) .. (v2) .. controls (v2u) and (v0u) .. (v0); \filldraw (v0) circle (1pt); \filldraw (v1) circle (1pt); \filldraw (v2) circle (1pt); \end{tikzpicture}~,%
$$ and identify the graph with its associated graphical function (as long as no confusion is possible), 
$$f_\Gamma(z) = 
\left[
\begin{tikzpicture}[x=1.5ex,y=1.5ex,baseline={([yshift=-.6ex]current bounding box.center)}] \coordinate (v0); \coordinate [above right=1.41 and 2 of v0] (v1); \coordinate [below right=1.41 and 2 of v0] (v2); \node [left =.1 of v0] (v0n) {$z$}; \coordinate [right =1 of v0] (v0u); \node [right=.07 of v1] (v1n) {$1$}; \coordinate [below left=.71 and .71 of v1] (v1u); \node [right=.07 of v2] (v2n) {$0$}; \coordinate [above left=.71 and .71 of v2] (v2u); \draw[fill=gray] (v0) .. controls (v0u) and (v1u) .. (v1) .. controls (v1u) and (v2u) .. (v2) .. controls (v2u) and (v0u) .. (v0); \filldraw (v0) circle (1pt); \filldraw (v1) circle (1pt); \filldraw (v2) circle (1pt); \end{tikzpicture}%
\right].
$$ 
In this notation we have the following identities which can be used to add and remove edges between external vertices:
\begin{align} \label{eq:extidentity} \left[ \begin{tikzpicture}[x=1.5ex,y=1.5ex,baseline={([yshift=-.6ex]current bounding box.center)}] \coordinate (v0); \coordinate [above right=1.41 and 2 of v0] (v1); \coordinate [below right=1.41 and 2 of v0] (v2); \node [left =.1 of v0] (v0n) { $z$}; \coordinate [right =1 of v0] (v0u); \node [right=.07 of v1] (v1n) { $1$}; \coordinate [below left=.71 and .71 of v1] (v1u); \node [right=.07 of v2] (v2n) { $0$}; \coordinate [above left=.71 and .71 of v2] (v2u); \draw[fill=gray] (v0) .. controls (v0u) and (v1u) .. (v1) .. controls (v1u) and (v2u) .. (v2) .. controls (v2u) and (v0u) .. (v0); \filldraw (v0) circle (1pt); \filldraw (v1) circle (1pt); \filldraw (v2) circle (1pt); \end{tikzpicture} \right] &= \left[ \begin{tikzpicture}[x=1.5ex,y=1.5ex,baseline={([yshift=-.6ex]current bounding box.center)}] \coordinate (v0); \coordinate [above right=1.41 and 2 of v0] (v1); \coordinate [below right=1.41 and 2 of v0] (v2); \node [left =.1 of v0] (v0n) {$z$}; \coordinate [right =1 of v0] (v0u); \node [right=.07 of v1] (v1n) { $1$}; \coordinate [below left=.71 and .71 of v1] (v1u); \node [right=.07 of v2] (v2n) { $0$}; \coordinate [above left=.71 and .71 of v2] (v2u); \draw[fill=gray] (v0) .. controls (v0u) and (v1u) .. (v1) .. controls (v1u) and (v2u) .. (v2) .. controls (v2u) and (v0u) .. (v0); \draw (v1) to [bend left=90] (v2); \filldraw (v0) circle (1pt); \filldraw (v1) circle (1pt); \filldraw (v2) circle (1pt); \end{tikzpicture} \right] = (z\zz)^{d/2-1} \left[ \begin{tikzpicture}[x=1.5ex,y=1.5ex,baseline={([yshift=-.6ex]current bounding box.center)}] \coordinate (v0); \coordinate [above right=1.41 and 2 of v0] (v1); \coordinate [below right=1.41 and 2 of v0] (v2); \node [left =.1 of v0] (v0n) { $z$}; \coordinate [right =1 of v0] (v0u); \node [right=.07 of v1] (v1n) { $1$}; \coordinate [below left=.71 and .71 of v1] (v1u); \node [right=.07 of v2] (v2n) { $0$}; \coordinate [above left=.71 and .71 of v2] (v2u); \draw[fill=gray] (v0) .. controls (v0u) and (v1u) .. (v1) .. controls (v1u) and (v2u) .. (v2) .. controls (v2u) and (v0u) .. (v0); \draw (v2) to [bend left=90] (v0); \filldraw (v0) circle (1pt); \filldraw (v1) circle (1pt); \filldraw (v2) circle (1pt); \end{tikzpicture} \right] = \left[(z-1)(\zz-1) \right]^{d/2-1} \left[ \begin{tikzpicture}[x=1.5ex,y=1.5ex,baseline={([yshift=-.6ex]current bounding box.center)}] \coordinate (v0); \coordinate [above right=1.41 and 2 of v0] (v1); \coordinate [below right=1.41 and 2 of v0] (v2); \node [left =.1 of v0] (v0n) { $z$}; \coordinate [right =1 of v0] (v0u); \node [right=.07 of v1] (v1n) { $1$}; \coordinate [below left=.71 and .71 of v1] (v1u); \node [right=.07 of v2] (v2n) { $0$}; \coordinate [above left=.71 and .71 of v2] (v2u); \draw[fill=gray] (v0) .. controls (v0u) and (v1u) .. (v1) .. controls (v1u) and (v2u) .. (v2) .. controls (v2u) and (v0u) .. (v0); \draw (v0) to [bend left=90] (v1); \filldraw (v0) circle (1pt); \filldraw (v1) circle (1pt); \filldraw (v2) circle (1pt); \end{tikzpicture} \right]. \end{align}
A permutation of the external vertices $0,1$ and $z$ corresponds to a specific M\"obius transformation of the $z$ variables together with an overall conformal rescaling,
\begin{align} \label{eq:permidentity} \left[ \begin{tikzpicture}[x=1.5ex,y=1.5ex,baseline={([yshift=-.6ex]current bounding box.center)}] \coordinate (v0); \coordinate [above right=1.41 and 2 of v0] (v1); \coordinate [below right=1.41 and 2 of v0] (v2); \node [left =.1 of v0] (v0n) { $z$}; \coordinate [right =1 of v0] (v0u); \node [right=.07 of v1] (v1n) { $0$}; \coordinate [below left=.71 and .71 of v1] (v1u); \node [right=.07 of v2] (v2n) { $1$}; \coordinate [above left=.71 and .71 of v2] (v2u); \draw[fill=gray] (v0) .. controls (v0u) and (v1u) .. (v1) .. controls (v1u) and (v2u) .. (v2) .. controls (v2u) and (v0u) .. (v0); \filldraw (v0) circle (1pt); \filldraw (v1) circle (1pt); \filldraw (v2) circle (1pt); \end{tikzpicture} \right] &= \left[ \begin{tikzpicture}[x=1.5ex,y=1.5ex,baseline={([yshift=-.6ex]current bounding box.center)}] \coordinate (v0); \coordinate [above right=1.41 and 2 of v0] (v1); \coordinate [below right=1.41 and 2 of v0] (v2); \node [left =.1 of v0] (v0n) { $z$}; \coordinate [right =1 of v0] (v0u); \node [right=.07 of v1] (v1n) { $1$}; \coordinate [below left=.71 and .71 of v1] (v1u); \node [right=.07 of v2] (v2n) { $0$}; \coordinate [above left=.71 and .71 of v2] (v2u); \draw[fill=gray] (v0) .. controls (v0u) and (v1u) .. (v1) .. controls (v1u) and (v2u) .. (v2) .. controls (v2u) and (v0u) .. (v0); \filldraw (v0) circle (1pt); \filldraw (v1) circle (1pt); \filldraw (v2) circle (1pt); \end{tikzpicture} \right] = (z\zz)^{\sdd/2} \left[ \begin{tikzpicture}[x=1.5ex,y=1.5ex,baseline={([yshift=-.6ex]current bounding box.center)}] \coordinate (v0); \coordinate [above right=1.41 and 2 of v0] (v1); \coordinate [below right=1.41 and 2 of v0] (v2); \node [left =.1 of v0] (v0n) { $1$}; \coordinate [right =1 of v0] (v0u); \node [right=.07 of v1] (v1n) { $0$}; \coordinate [below left=.71 and .71 of v1] (v1u); \node [right=.07 of v2] (v2n) { $\frac{1}{z}$}; \coordinate [above left=.71 and .71 of v2] (v2u); \draw[fill=gray] (v0) .. controls (v0u) and (v1u) .. (v1) .. controls (v1u) and (v2u) .. (v2) .. controls (v2u) and (v0u) .. (v0); \filldraw (v0) circle (1pt); \filldraw (v1) circle (1pt); \filldraw (v2) circle (1pt); \end{tikzpicture} \right].                       \end{align}
These two identities generate the full permutation group of the external vertices. 
The factorization rules in eq.~\eqref{eq:extidentity} and the permutation identities in eq.~\eqref{eq:permidentity} follow immediately from eq.~\eqref{eq:defgf}, the definition of a position space three-point function in eq.~\eqref{eq:threepointfunction} and the parametrization in eq.~\eqref{eqn:z_parametrization_implicit}.

Up to this point, all statements on graphical functions are valid in even dimensions $\geq4$. To handle QFTs with subdivergences we need to use a regulator.  It is convenient to use dimensional regularization because the concept of graphical functions is based on the complex plane which is independent of the ambient space.
Hence, it is stable under the deformation of the dimension to real numbers. In fact, all the previous identities work for general dimensions. We use the notation
$$
d=2n+4-\varepsilon,\quad n\in\{0,1,2,3,\ldots\},\quad\varepsilon\in\R~.
$$
The general idea is to consider graphical functions in $d$ dimensions as Laurent series in $\varepsilon$. The coefficient of every power in $\varepsilon$ conjecturally reflects the
structure of graphical functions in fixed even dimensions. In particular, every coefficient is a single-valued real analytic function on $\C\backslash\{0,1\}$ and admits
$\log$-Laurent expansions (eqs.~\eqref{aexp} and \eqref{infexp}) at $0$, $1$, and $\infty$.

The most important identity for the graphical function technique follows from the definition of the propagator, eq.~\eqref{eq:propagator}, combined with eqs.~\eqref{eq:threepointfunction}
and \eqref{eqn:z_parametrization_implicit}. The intuition behind this identity is that, according to eq.~\eqref{eq:propagator}, the box operator can be used to `amputate' single external
edges of a position space Feynman diagram:
Two position space three-point functions $G_{\tilde \Gamma}$ and $G_{\Gamma}$ 
whose underlying Feynman graphs $\widetilde \Gamma$ and $\Gamma$ only differ by an appended edge along the external vertex $c$,
\begin{align*} \widetilde{\Gamma} = \begin{tikzpicture}[x=1.5ex,y=1.5ex,baseline={([yshift=-.6ex]current bounding box.center)}] \coordinate (v0); \coordinate [above right=1.41 and 2 of v0] (v1); \coordinate [below right=1.41 and 2 of v0] (v2); \coordinate [left =3 of v0] (v00); \node [left =.1 of v00] (v0n) { $c$}; \coordinate [right =1 of v0] (v0u); \node [right=.07 of v1] (v1n) { $b$}; \coordinate [below left=.71 and .71 of v1] (v1u); \node [right=.07 of v2] (v2n) { $a$}; \coordinate [above left=.71 and .71 of v2] (v2u); \draw[fill=gray] (v0) .. controls (v0u) and (v1u) .. (v1) .. controls (v1u) and (v2u) .. (v2) .. controls (v2u) and (v0u) .. (v0); \draw (v00) -- (v0); \filldraw (v00) circle (1pt); \filldraw (v0) circle (1pt); \filldraw (v1) circle (1pt); \filldraw (v2) circle (1pt); \end{tikzpicture}~, \qquad \Gamma = \begin{tikzpicture}[x=1.5ex,y=1.5ex,baseline={([yshift=-.6ex]current bounding box.center)}] \coordinate (v0); \coordinate [above right=1.41 and 2 of v0] (v1); \coordinate [below right=1.41 and 2 of v0] (v2); \node [left =.1 of v0] (v0n) { $c$}; \coordinate [right =1 of v0] (v0u); \node [right=.07 of v1] (v1n) { $b$}; \coordinate [below left=.71 and .71 of v1] (v1u); \node [right=.07 of v2] (v2n) { $a$}; \coordinate [above left=.71 and .71 of v2] (v2u); \draw[fill=gray] (v0) .. controls (v0u) and (v1u) .. (v1) .. controls (v1u) and (v2u) .. (v2) .. controls (v2u) and (v0u) .. (v0); \filldraw (v0) circle (1pt); \filldraw (v1) circle (1pt); \filldraw (v2) circle (1pt); \end{tikzpicture}, \end{align*}
fulfill the partial differential equation
\begin{align} \Box_{\bb x_c} G_{\tilde \Gamma} (\bb x_a, \bb x_b, \bb x_c) = -\frac{4}{\Gamma(d/2-1)} G_{\Gamma} (\bb x_a, \bb x_b, \bb x_c)~. \end{align}
In fact, such combinatorial differential equations hold for arbitrary $n$-point functions. In our case of the three-point function we can translate the Laplacian $\Box_{\bb x_c}$ via eq.~\eqref{eqn:z_parametrization_implicit} into an operator on the space of complex functions in $z$ and $\zz$ to get an effective Laplacian, which operates on graphical functions,
\begin{gather} \begin{gathered} \label{eq:effLap} \left(\frac{1}{(z-\zz)^{n+1}}\Delta_n(z-\zz)^{n+1}+\frac{\varepsilon/2}{z-\zz}(\partial_{z} - \partial_{\zz})\right) \left[ \begin{tikzpicture}[x=1.5ex,y=1.5ex,baseline={([yshift=-.6ex]current bounding box.center)}] \coordinate (v0); \coordinate [above right=1.41 and 2 of v0] (v1); \coordinate [below right=1.41 and 2 of v0] (v2); \coordinate [left =3 of v0] (v00); \node [left =.1 of v00] (v0n) { $z$}; \coordinate [right =1 of v0] (v0u); \node [right=.07 of v1] (v1n) { $1$}; \coordinate [below left=.71 and .71 of v1] (v1u); \node [right=.07 of v2] (v2n) { $0$}; \coordinate [above left=.71 and .71 of v2] (v2u); \draw[fill=gray] (v0) .. controls (v0u) and (v1u) .. (v1) .. controls (v1u) and (v2u) .. (v2) .. controls (v2u) and (v0u) .. (v0); \draw (v00) -- (v0); \filldraw (v00) circle (1pt); \filldraw (v0) circle (1pt); \filldraw (v1) circle (1pt); \filldraw (v2) circle (1pt); \end{tikzpicture} \right] =-\frac{1}{\Gamma(n+1-\varepsilon/2)} \left[ \begin{tikzpicture}[x=1.5ex,y=1.5ex,baseline={([yshift=-.6ex]current bounding box.center)}] \coordinate (v0); \coordinate [above right=1.41 and 2 of v0] (v1); \coordinate [below right=1.41 and 2 of v0] (v2); \node [left =.1 of v0] (v0n) { $z$}; \coordinate [right =1 of v0] (v0u); \node [right=.07 of v1] (v1n) { $1$}; \coordinate [below left=.71 and .71 of v1] (v1u); \node [right=.07 of v2] (v2n) { $0$}; \coordinate [above left=.71 and .71 of v2] (v2u); \draw[fill=gray] (v0) .. controls (v0u) and (v1u) .. (v1) .. controls (v1u) and (v2u) .. (v2) .. controls (v2u) and (v0u) .. (v0); \filldraw (v0) circle (1pt); \filldraw (v1) circle (1pt); \filldraw (v2) circle (1pt); \end{tikzpicture} \right] \\
\text{with } \Delta_n = \partial_{z}\partial_{\zz}+\frac{n(n+1)}{(z-\zz)^2}. \end{gathered} \end{gather}
To obtain a graphical function of higher complexity from a graphical function of lower complexity, we would like to solve this partial differential equation for the graphical function with the appended edge. To do this, we need to invert the effective Laplacian. This inversion can be roughly separated into two related problems:
First, we have to find a general solution of the differential equation. Second, we need to specify the solution that gives the wanted graphical function.

The first problem is reasonably easy in four dimensions, i.e.\ $n=0$. As long as $\varepsilon=0$ the partial differential operator $\Delta_0$ factorizes into $\partial_z$ and $\partial_\zz$. Even though, naively integrating with respect to $z$ and $\zz$ results in undetermined integration constants which may be arbitrary functions of $\zz$ and $z$ respectively, these integration constants
are restricted if we require the result to be single-valued. Single-valued integration needs to be performed on a suitable class of functions. Because of the denominator $z-\zz$ the class of single-valued multiple polylogarithms as studied in \cite{BrSVMP,brown2004single} is not enough. However, there exists a generalization by the fourth author to \emph{generalized single-valued hyperlogarithms} (GSVHs) which exactly accommodates the situation \cite{Schnetz2020gsvh}.
This theory comes with general and conveniently fast algorithms for single-valued integration in $z$ and $\zz$.
For $\varepsilon>0$ the generalization is straightforward: We are not interested in a full result but rather in a Laurent series in $\varepsilon$. This allows us
to treat the $\varepsilon$ term in eq.~\eqref{eq:effLap} as a perturbation and to solve the differential equation by iteratively inverting $\Delta_0$.

The situation is much more complicated for $n\geq1$ (i.e.\ $d \geq 6$). Note that the effective Laplacian $\Delta_n$ is a partial differential operator in $z$ and $\zz$ which in general does not admit a simple solution by integration. However, somewhat surprisingly, a general solution of eq.~\eqref{eq:effLap} for all $n=0,1,2,3,\ldots$ in the case $\varepsilon=0$ was found by the first and the fourth author \cite{Borinsky2020prep}.  The function space of GSVHs is perfectly suitable for general $n$. The case $\varepsilon>0$ is again treated as a perturbation.

The second problem---specifying the exact solution---is solved for $\varepsilon=0$ by a theorem \cite{Schnetz:2013hqa,Borinsky2020prep}: the structures of graphical functions---single-valuedness and $\log$-Laurent expansions---are so
restrictive that they fully specify the solution. That means there is always only one special solution among the family of general solutions that behaves like a graphical function.
Finding this special solution is solved by an algorithm given in \cite{Borinsky2020prep}.
For $\varepsilon>0$, handling subdivergences in this context is a bit subtle but always possible. Altogether we obtain a general and surprisingly efficient algorithm to append an edge to the external
label $z$ of a graphical function.

Using the three basic operations---adding edges between external vertices, permuting external vertices, and appending an edge---allows one to construct
a wide class of graphical functions from the empty graphical function (the graphical function with neither edges nor internal vertices), which is the constant $1$. See Fig.~\ref{fig:GFreduction} for a non-trivial example of a two-point function in $\phi^4$ theory that can be constructed from these basic operations.

In practice, there exist a few more elementary operations (products, factors, completion) that provide combinatorial relations between different graphical functions. A graphical function that can be expressed as a sequence these elementary operation applied to the trivial graphical function is called \emph{constructible}
and can be calculated to any reasonable order in $\varepsilon$ (limited by time and memory demands). We want to emphasize that this concept of constructible graphical functions can easily be applied to any massless QFT in even dimensions.

Still, starting at some loop order, there exist graphical functions which cannot be constructed. Some of these graphical functions may be amendable to some reductions by elementary operations, but eventually a non-empty irreducible graphical function will be reached that has to be calculated by other means.  Beyond constructability there exists a toolbox of additional identities which has some resemblance to standard momentum space techniques (e.g.~momentum space IBPs). Explicitly we have special identities, approximations in $\varepsilon$ and position space IBPs that can be used to further simplify the calculations.  However this toolbox is still in its infancy and small additions may lead to dramatic increases in the applicability of the overall graphical function technique.

Eventually, some (typically small) set of (typically small) graphical functions remains that has to be integrated by different means.
A brute force approach is to write the respective graphical functions as parametric integrals \cite{Golz:2015rea} which in many cases can be integrated using an algorithm by F.~Brown \cite{Brown:2008um} which is implemented as {\tt HyperInt} by E.~Panzer \cite{Panzer:2014caa}. In four dimensions this parametric integration often works quite well, even though it is always very slow compared to operations on graphical functions. In six or higher dimensions the use of parametric integration is limited by the existence on squares and higher powers in the denominators of the parametric integrals as this causes problems in the implementation of {\tt HyperInt} and results in large time and memory consumption. 
For the present fifth order computation in $\phi^3$ theory the use of {\tt HyperInt} was not necessary as all contributing graphical functions can be reduced to the trivial one via identities.

For six loops approximately $80\%$ of the Feynman integrals are calculated. Because of the vast number and the higher
complexity of graphs at six loops it is necessary to make
better use of some techniques in the theory of graphical functions. Most prominently, making full use of a generalized $\Delta-Y$ identity and position space integration by parts (IBP)
requires the (notoriously tedious) solution of large linear systems. In the primitive case these identities (with some others) were implemented by the first author. This solved all
primitive six loop integrals in $\phi^3$ theory \cite{Borinsky2020prep2}. Even at $7$ loops $92\%$ of the primitive integrals could be calculated \cite{Borinsky2020prep2}.
The implementation for subdivergent graphs is straight forward but still lacking. We expect that the implementation of position space IBP will solve $\phi^3$ theory up to six loops in perturbation theory.
It might be necessary to use some minor additions from other techniques such as $R/R^*$ \cite{ChetyrkinTkachov:InfraredR,ChetyrkinSmirnov:Rcorrected,Chetyrkin:RRR} or its
Hopf algebraic version \cite{Beekveldt:2020kzk} that make it possible to exclusively work with finite expressions. A hypothetical extension to $7$ loops might already require
the addition of numerical methods such as tropical parametric integration \cite{borinsky2020tropical} to evaluate a small number of primitive graphs that cannot be evaluated via GSVHs.
Such graphs are known to appear in $\phi^4$ theory at $8$ loops \cite{Brown:2010bw}.

\subsection{Periods and renormalization group functions}
\begin{figure}
\centering
\begin{tikzpicture}[x=2.3ex,y=2.3ex] \begin{scope}[local bounding box=fullcateye] \coordinate (vm); \coordinate [above=1 of vm] (vo); \coordinate [left =2 of vm] (vl); \coordinate [right=2 of vm] (vr); \coordinate [left =1 of vl] (vll); \coordinate [right=1 of vr] (vrr); \draw (vm) to [bend left =90] (vo); \draw (vm) to [bend right=90] (vo); \draw (vl) to [bend left =20] (vo); \draw (vr) to [bend right=20] (vo); \draw (vl) to [bend right=50] (vr); \draw (vl) -- (vm); \draw (vr) -- (vm); \draw (vll) -- (vl); \draw (vrr) -- (vr); \node [left =.2 of vll] {$0$}; \node [right=.2 of vrr] {$1$}; \filldraw (vm) circle (1pt); \filldraw (vo) circle (1pt); \filldraw (vl) circle (1pt); \filldraw (vr) circle (1pt); \filldraw (vll) circle (1pt); \filldraw (vrr) circle (1pt); \end{scope} \begin{scope}[xshift=130,local bounding box=fullcateyez] \coordinate (vm); \coordinate [above=1 of vm] (vo); \coordinate [left =2 of vm] (vl); \coordinate [right=2 of vm] (vr); \coordinate [left =1 of vl] (vll); \coordinate [right=1 of vr] (vrr); \draw (vm) to [bend left =90] (vo); \draw (vm) to [bend right=90] (vo); \draw (vl) to [bend left =20] (vo); \draw (vr) to [bend right=20] (vo); \draw (vl) to [bend right=50] (vr); \draw (vl) -- (vm); \draw (vr) -- (vm); \draw (vll) -- (vl); \draw (vrr) -- (vr); \node [left =.2 of vll] {$0$}; \node [right=.2 of vrr] {$1$}; \node [above=.2 of vo ] {$z$}; \filldraw (vm) circle (1pt); \filldraw (vo) circle (1pt); \filldraw (vl) circle (1pt); \filldraw (vr) circle (1pt); \filldraw (vll) circle (1pt); \filldraw (vrr) circle (1pt); \end{scope} \draw[-latex,thick] (fullcateye-|fullcateyez.west) -- (fullcateye) node[midway,above]{$\int \dd z \dd \zz$}; \draw[-latex,thick,dashed] ([xshift=30]fullcateye-|fullcateyez.east) -- (fullcateye-|fullcateyez.east) node[midway,above]{}; \begin{scope}[yshift=-40,local bounding box=fullcateyez2] \coordinate (vm); \coordinate [above=1 of vm] (vo); \coordinate [below=1 of vm] (vu); \coordinate [left =2 of vm] (vl); \coordinate [right=2 of vm] (vr); \coordinate [left =1 of vl] (vll); \coordinate [right=1 of vr] (vrr); \draw (vm) to [bend left =90] (vo); \draw (vm) to [bend right=90] (vo); \draw (vl) to [bend left =20] (vo); \draw (vr) to [bend right=20] (vo); \draw (vl) to [bend right=50] (vr); \draw (vl) -- (vm); \draw (vr) -- (vm); \draw (vll) -- (vl); \draw (vrr) -- (vr); \node [left =.2 of vll] {$z$}; \node [right=.2 of vrr] {$1$}; \node [above=.2 of vo ] {$0$}; \node [below=.2 of vu ] {\phantom{$0$}}; \filldraw (vm) circle (1pt); \filldraw (vo) circle (1pt); \filldraw (vl) circle (1pt); \filldraw (vr) circle (1pt); \filldraw (vll) circle (1pt); \filldraw (vrr) circle (1pt); \end{scope} \draw[-latex,thick] (fullcateyez2) -- ([xshift=-30]fullcateyez2-|fullcateyez2.west) node[midway,above]{$0 \leftrightarrow z$}; \begin{scope}[yshift=-40,xshift=130,local bounding box=cateyeredl] \coordinate (vm); \coordinate [above=1 of vm] (vo); \coordinate [below=1 of vm] (vu); \coordinate [left =2 of vm] (vl); \coordinate [right=2 of vm] (vr); \coordinate [left =1 of vl] (vll); \coordinate [right=1 of vr] (vrr); \draw (vm) to [bend left =90] (vo); \draw (vm) to [bend right=90] (vo); \draw (vl) to [bend left =20] (vo); \draw (vr) to [bend right=20] (vo); \draw (vl) to [bend right=50] (vr); \draw (vl) -- (vm); \draw (vr) -- (vm); \draw (vrr) -- (vr); \node [left =.2 of vl] {$z$}; \node [right=.2 of vrr] {$1$}; \node [above=.2 of vo ] {$0$}; \node [below=.2 of vu ] {\phantom{$0$}}; \node [right=.2 of vrr ] {\phantom{$0$}}; \node [left=.2 of vll ] {\phantom{$0$}}; \filldraw (vm) circle (1pt); \filldraw (vo) circle (1pt); \filldraw (vl) circle (1pt); \filldraw (vr) circle (1pt); \filldraw (vrr) circle (1pt); \end{scope} \draw[-latex,thick] (fullcateyez2-|cateyeredl.west) -- (fullcateyez2) node[midway,above]{$\Box_{z}^{-1}$}; \draw[-latex,thick,dashed] ([xshift=30]fullcateyez2-|cateyeredl.east) -- (fullcateyez2-|cateyeredl.east) node[midway,above]{}; \begin{scope}[yshift=-80,xshift=0,local bounding box=cateyeredl2] \coordinate (vm); \coordinate [above=1 of vm] (vo); \coordinate [below=1 of vm] (vu); \coordinate [left =2 of vm] (vl); \coordinate [right=2 of vm] (vr); \coordinate [left =1 of vl] (vll); \coordinate [right=1 of vr] (vrr); \draw (vm) to [bend left =90] (vo); \draw (vm) to [bend right=90] (vo); \draw (vr) to [bend right=20] (vo); \draw (vl) to [bend right=50] (vr); \draw (vl) -- (vm); \draw (vr) -- (vm); \draw (vrr) -- (vr); \node [left =.2 of vl] {$z$}; \node [right=.2 of vrr] {$1$}; \node [above=.2 of vo ] {$0$}; \node [below=.2 of vu ] {\phantom{$0$}}; \node [right=.2 of vrr ] {\phantom{$0$}}; \node [left=.2 of vll ] {\phantom{$0$}}; \filldraw (vm) circle (1pt); \filldraw (vo) circle (1pt); \filldraw (vl) circle (1pt); \filldraw (vr) circle (1pt); \filldraw (vrr) circle (1pt); \end{scope} \draw[-latex,thick] (cateyeredl2) -- ([xshift=-30]cateyeredl2-|cateyeredl2.west) node[midway,above]{\text{ext.~edges}}; \begin{scope}[yshift=-80,xshift=130,local bounding box=cateyeredl3] \coordinate (vm); \coordinate [above=1 of vm] (vo); \coordinate [below=1 of vm] (vu); \coordinate [left =2 of vm] (vl); \coordinate [right=2 of vm] (vr); \coordinate [left =1 of vl] (vll); \coordinate [right=1 of vr] (vrr); \draw (vm) to [bend left =90] (vo); \draw (vm) to [bend right=90] (vo); \draw (vr) to [bend right=20] (vo); \draw (vl) to [bend right=50] (vr); \draw (vl) -- (vm); \draw (vr) -- (vm); \draw (vrr) -- (vr); \node [left =.2 of vl] {$1$}; \node [right=.2 of vrr] {$z$}; \node [right=.2 of vrr ] {\phantom{$0$}}; \node [left=.2 of vll ] {\phantom{$0$}}; \node [above=.2 of vo ] {$0$}; \node [below=.2 of vu ] {\phantom{$0$}}; \filldraw (vm) circle (1pt); \filldraw (vo) circle (1pt); \filldraw (vl) circle (1pt); \filldraw (vr) circle (1pt); \filldraw (vrr) circle (1pt); \end{scope} \draw[-latex,thick] (cateyeredl2-|cateyeredl3.west) -- (cateyeredl2) node[midway,above]{$1 \leftrightarrow z$}; \draw[-latex,thick,dashed] ([xshift=30]cateyeredl3-|cateyeredl3.east) -- (cateyeredl3) node[midway,above]{}; \begin{scope}[yshift=-120,local bounding box=cateyeredlr] \coordinate (vm); \coordinate [above=1 of vm] (vo); \coordinate [below=1 of vm] (vu); \coordinate [left =2 of vm] (vl); \coordinate [right=2 of vm] (vr); \coordinate [left =1 of vl] (vll); \coordinate [right=1 of vr] (vrr); \draw (vm) to [bend left =90] (vo); \draw (vm) to [bend right=90] (vo); \draw (vr) to [bend right=20] (vo); \draw (vl) to [bend right=50] (vr); \draw (vl) -- (vm); \draw (vr) -- (vm); \node [left =.2 of vl] {$1$}; \node [right=.2 of vr] {$z$}; \node [above=.2 of vo ] {$0$}; \node [below=.2 of vu ] {\phantom{$0$}}; \node [right=.2 of vrr ] {\phantom{$0$}}; \node [left=.2 of vll ] {\phantom{$0$}}; \filldraw (vm) circle (1pt); \filldraw (vo) circle (1pt); \filldraw (vl) circle (1pt); \filldraw (vr) circle (1pt); \end{scope} \draw[-latex,thick] (cateyeredlr) -- ([xshift=-30]cateyeredlr-|cateyeredlr.west) node[midway,above]{$\Box_{z}^{-1}$}; \begin{scope}[yshift=-120,xshift=130,local bounding box=cateyeredlr2] \coordinate (vm); \coordinate [above=1 of vm] (vo); \coordinate [below=1 of vm] (vu); \coordinate [left =2 of vm] (vl); \coordinate [right=2 of vm] (vr); \coordinate [left =1 of vl] (vll); \coordinate [right=1 of vr] (vrr); \draw (vm) to [bend left =90] (vo); \draw (vm) to [bend right=90] (vo); \draw (vl) -- (vm); \draw (vr) -- (vm); \node [left =.2 of vl] {$1$}; \node [right=.2 of vr] {$z$}; \node [above=.2 of vo ] {$0$}; \node [below=.2 of vu ] {\phantom{$0$}}; \node [right=.2 of vrr ] {\phantom{$0$}}; \node [left=.2 of vll ] {\phantom{$0$}}; \filldraw (vm) circle (1pt); \filldraw (vo) circle (1pt); \filldraw (vl) circle (1pt); \filldraw (vr) circle (1pt); \end{scope} \draw[-latex,thick] (cateyeredlr-|cateyeredlr2.west) -- (cateyeredlr) node[midway,above]{\text{ext.~edges}}; \draw[-latex,thick,dashed] ([xshift=30]cateyeredlr2-|cateyeredlr2.east) -- (cateyeredlr2) node[midway,above]{}; \begin{scope}[yshift=-160,xshift=0,local bounding box=cateyeredlr3] \coordinate (vm); \coordinate [above=1 of vm] (vo); \coordinate [below=1 of vm] (vu); \coordinate [left =2 of vm] (vl); \coordinate [right=2 of vm] (vr); \coordinate [left =1 of vl] (vll); \coordinate [right=1 of vr] (vrr); \draw (vm) to [bend left =90] (vo); \draw (vm) to [bend right=90] (vo); \draw (vl) -- (vm); \node [left =.2 of vl] {$1$}; \node [below right=.2 of vm] {$z$}; \node [above=.2 of vo ] {$0$}; \node [below=.2 of vu ] {\phantom{$0$}}; \node [right=.2 of vrr ] {\phantom{$0$}}; \node [left=.2 of vll ] {\phantom{$0$}}; \filldraw (vm) circle (1pt); \filldraw (vo) circle (1pt); \filldraw (vl) circle (1pt); \end{scope} \draw[-latex,thick] (cateyeredlr3) -- ([xshift=-30]cateyeredlr3-|cateyeredlr3.west) node[midway,above]{$\Box_{z}^{-1}$}; \begin{scope}[yshift=-160,xshift=130,local bounding box=one] \coordinate (vm); \coordinate [above=1 of vm] (vo); \coordinate [below=1 of vm] (vu); \coordinate [left =2 of vm] (vl); \coordinate [right=2 of vm] (vr); \coordinate [left =1 of vl] (vll); \coordinate [right=1 of vr] (vrr); \node [left =.2 of vl] {$1$}; \node [below right=.2 of vm] {$z$}; \node [above=.2 of vo ] {$0$}; \node [below=.2 of vu ] {\phantom{$0$}}; \node [right=.2 of vrr ] {\phantom{$0$}}; \node [left=.2 of vll ] {\phantom{$0$}}; \filldraw (vm) circle (1pt); \filldraw (vo) circle (1pt); \filldraw (vl) circle (1pt); \end{scope} \draw[-latex,thick] (one) -- ([xshift=-42]one-|one.west) node[midway,above]{\text{ext.~edges}}; \end{tikzpicture}

\caption{Example of the graphical reduction of a two-point function in $\phi^4$ theory.
The explicit expression can be obtained by following the arrows which start from the trivial graphical function which is equal to $1$.}
\label{fig:GFreduction}
\end{figure}
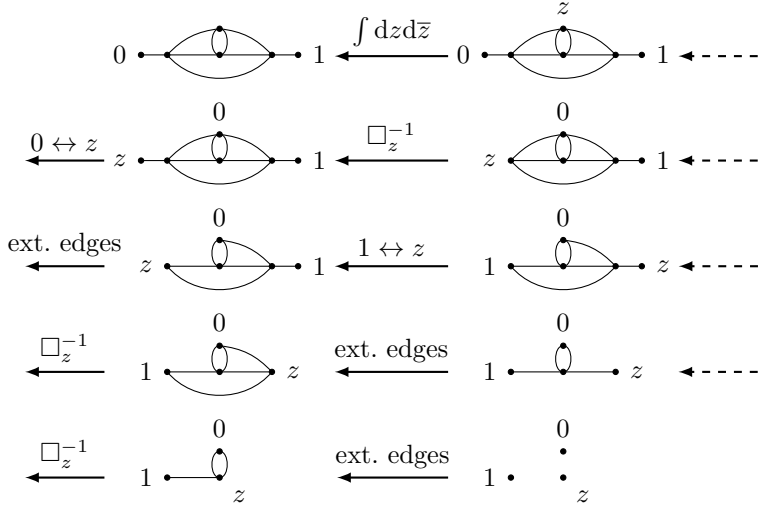

To determine renormalization functions it is sufficient to calculate ($\varepsilon$-dependent) periods, i.e.\ the Feynman integrals of graphs with two external vertices.  If a graph has two external vertices $\bb x_a,\bb x_b$ then translation and scale-invariance (eq.~\eqref{eqn:symmetry}) forces the respective position space Feynman integral to evaluate to $|\bb x_{ab}|^{\sdd} P_\Gamma(\varepsilon)$
where the \emph{period}, $P_\Gamma(\varepsilon)$, is only a function of $\varepsilon$.

The calculation of periods is much simpler than the calculation of graphical functions: 
one can \emph{complete} a period by adding an edge between $\bb x_a$ and $\bb x_b$ with a specific weight so that one is free to choose an arbitrary set of three external vertices $0$, $1$, $z$ to obtain a graphical function. If this graphical function can be calculated it is always possible to efficiently integrate over the third vertex $z$ to obtain the period. This freedom is of great benefit for calculating periods. Often there exists a particularly convenient choice which facilitates the calculation. In a certain sense graphical functions were originally invented to use them for the calculation of periods with exactly this concept. Fig.~\ref{fig:GFreduction} shows an example of a full reduction chain of a period in $\phi^4$ theory.

The graphical function technique is powerful enough to take a naive approach to renormalization. We calculate each amplitude to the necessary order in $\varepsilon$. Instead of calculating the three-point function directly in terms of graphical functions, we set one external vertex to infinity and calculate the simpler period of the resulting truncated three-point function as an effective two-point function.  We add all amplitudes to get regularized (effective) two-point functions from which we read off the $Z$ factors by the condition that renormalization renders the (effective) two-point functions finite, see eq.~\eqref{Zfactors}. From the $Z$ factors we read off the renormalization functions, see eq.~\eqref{renfkt}.

A detailed explanation of the graphical function method will be in \cite{Schnetz2020phi4loop7}. All algorithms are implemented as Maple procedures in {\tt HyperlogProcedures} by the fourth author \cite{Schnetz2018hyperlog}.  The calculation of the fifth order result in six dimensional $\phi^3$ theory is fully automated in {\tt HyperlogProcedures} and takes about two days on a single core consuming 38 GB of memory.  It can easily be parallelized.

\subsection{Percolation theory prefactors}
\label{sec:combinatorial_factors}
Our diagrammatic approach to compute the combinatorial prefactors $\mathscr{C}_\Gamma$ for percolation theory follows from eqs.~\eqref{eq:dijk_def} and \eqref{eq:percolationrelation} via the diagrammatic discussion in \cite[eq.\ (3.21) and the paragraph that follows]{Bonfim_1981}. For a graph $\Gamma$, we write the associated \emph{completed graph} as $\overline{\Gamma}$. A graph is completed by attaching all external legs to an additional vertex \cite{Schnetz:2008mp}. 
Completion utilizes the fact that the tensor structure of two- and three-point graphs is fixed up to a constant. Adapting the notation of \cite{KompanietsPanzer2017}, for the percolation theory problem we find the following relations of the combinatorial factors $\mathscr{C}_\Gamma$, which follow from the algebraic rules above
\begin{equation}
\mathscr{C}_\Gamma(\alpha) =
\mathscr{C}_{\overline{\Gamma}}
\cdot
\begin{cases}
\frac{(N+1)\delta_{\alpha_1,\alpha_2}-1}{(N+1)N} &\text{ if $\Gamma$ is a $2$-point function,}\\
\frac{(N+1)^2\delta_{\alpha_1,\alpha_2}\delta_{\alpha_1,\alpha_3}-(N+1)(\delta_{\alpha_1,\alpha_2}+\delta_{\alpha_1,\alpha_3}+\delta_{\alpha_2,\alpha_3})+2}{(N+1)N(N-1)}
&\text{ if $\Gamma$ is a $3$-point function,}
\end{cases}
\end{equation}
where $\alpha_1, \alpha_2$ and $\alpha_3$ are associated with the two or three external tensor indices of the respective $n$-point functions.
The combinatorial factors of the completed graphs are readily computed by \emph{contraction-deletion}.
\begin{equation}
\begin{array}{rll}
\mathscr{C}_{\overline{\Gamma}} &=\;\; (N+1) \mathscr{C}_{\overline{\Gamma} / e} - \mathscr{C}_{\overline{\Gamma} \setminus e}&\text{ for each edge } e \in \overline{\Gamma},\\
\mathscr{C}_{\overline{\Gamma}_1\,\overline{\Gamma}_2} &=\;\; \mathscr{C}_{\overline{\Gamma}_1} \cdot \mathscr{C}_{\overline{\Gamma}_2}&\text{ for the disjoint union $\overline{\Gamma}_1\,\overline{\Gamma}_2$,}\\
\mathscr{C}_\bullet &=\;\; N+1&\text{ for the single vertex $\bullet$,}
\end{array}
\end{equation}
where $\overline{\Gamma} / e$ and $\overline{\Gamma} \setminus e$ denote the contraction and deletion of the edge $e$ in the graph $\overline{\Gamma}$. 
The contraction of a self-loop is the deletion of the loop in this context. 

The percolation theory problem is described by subsequently taking the $N \rightarrow 0$ limit for each prefactor. 

\section{Five loop renormalization group functions}
\label{sec:results}

With the major task of calculating all the 
integrals and carrying out the renormalization,
we devote this section solely to recording the results of the full five loop
renormalization in the context of the two critical systems that we are
interested in.

\subsection{Lee--Yang edge singularity}

For the Lee--Yang (LY) edge singularity problem the five loop $\MSbar$
renormalization group functions are
\begin{align} \begin{aligned} \beta^{(\mathrm{LY})}(g) = &-\frac{\varepsilon}{2}\,g +\frac{3}{4}\,{g}^{3}-{\frac {125\,{g}^{5}}{144}}+ \left( {\frac{33085}{20736}} +\frac{5\,\zeta_{{3}}}{8} \right) {g}^{7} \\
&+ \left( {\frac {15\,\zeta_4 }{32}}-{\frac{3404365}{746496}}-{\frac {4891\,\zeta_{{3}}}{ 864}}+\frac{5\,\zeta_{{5}}}{3} \right) {g}^{9} \\
& + \left( {\frac {75\,\zeta_6 }{32}}-{\frac {46519\,\zeta_4 }{9216} }+{\frac{102052031}{6718464}}+{\frac {99\,{\zeta_{{3}}}^{2}}{16}}+{ \frac {366647\,\zeta_{{3}}}{6912}}+{\frac {151795\,\zeta_{{5}}}{3456}} -{\frac {5495\,\zeta_{{7}}}{64}} \! \right) {g}^{11} \\
&+\mathcal{O}(g^{13}) \end{aligned} \end{align}
\begin{align} \begin{aligned} \gamma^{(\mathrm{LY})}_\phi(g) =& -\frac{1}{12}\,{g}^{2}+{\frac {13}{432}}\,{g}^{4}+ \left( -{\frac{5195}{62208} }+\frac{1}{24}\,\zeta_{{3}} \right) {g}^{6}+ \left( {\frac{53449}{248832}}+{ \frac {7\,\zeta_{{4}}}{96}}+{\frac {35\,\zeta_{{3}}}{864}} -{\frac {5\,\zeta_{{5}}}{18}} \right) {g}^{8} \\
& +\left( -{\frac {125\, \zeta_{{6}}}{288}}-{\frac {5651\,\zeta_{{4}}}{ 27648}}-{\frac{16492987}{20155392}}-{\frac {25\,{\zeta_{{3}}}^{2}}{144 }}-{\frac {56693\,\zeta_{{3}}}{62208}}+{\frac {4471\,\zeta_{{5}}}{ 10368}}+{\frac {147\,\zeta_{{7}}}{64}} \right) {g}^{10} \\
&+\mathcal{O}(g^{12}) \end{aligned} \end{align}
\begin{align} \begin{aligned} \gamma^{(\mathrm{LY})}_{m^2}(g) = &-\frac{5}{6}\,{g}^{2}+{\frac {97}{108}}\,{g}^{4}+ \left( -{\frac{52225}{31104} }-{\frac {7\,\zeta_{{3}}}{12}} \right) {g}^{6}+ \left( -{\frac {19\, \zeta_{{4}}}{48}}+{\frac{445589}{93312}}+{\frac {821\, \zeta_{{3}}}{144}}-{\frac {35\,\zeta_{{5}}}{18}} \right) {g}^{8} \\
& + \left( -{\frac {25\,\zeta_{{6}}}{9}}+{\frac {66953\,\zeta_{{4}}}{13824}}-{\frac{40331135}{2519424}}-{\frac {229\,{ \zeta_{{3}}}^{2}}{36}}-{\frac {839129\,\zeta_{{3}}}{15552}}-{\frac { 225457\,\zeta_{{5}}}{5184}} \right. \\
& \left. ~~~~~+{\frac {2821\,\zeta_{{7}}}{32}} \right) {g}^{10} +\mathcal{O}(g^{12}) \end{aligned} \end{align}
where $\zeta_z$ is the Riemann zeta function. Each expression 
agrees with the earlier expressions given in
\cite{Macfarlane:1974vp,Bonfim_1980,Bonfim_1981,Gracey:2015tta} up to four loops after the coupling
constant mapping is undone. We note that the 
five loop expression also agrees with the 
recent calculation of \cite{Kompaniets:2021hwg}.
Equipped with these it is straightforward to
derive the respective critical exponents which are
\begin{align} \begin{aligned} \eta^{(\mathrm{LY})}(\varepsilon) =& -\frac{1}{9}\,\varepsilon-{\frac {43}{729}}\,{\varepsilon}^{2}+ \left( -{\frac{8375}{ 236196}}+{\frac {16\,\zeta_{{3}}}{243}} \right) {\varepsilon}^{3}+ \left( -{\frac{3883409}{76527504}}+{\frac {4\,\zeta_4 }{81}}-{\frac {716\,\zeta_{{3}}}{19683}}-{\frac {80\,\zeta_{{5}}}{2187 }} \right) {\varepsilon}^4 \\
& + \left( -{\frac {100\,\zeta_6 }{2187}}-{\frac {179\,\zeta_4 }{6561}}-{\frac{ 1545362585}{24794911296}}+{\frac {136\,{\zeta_{{3}}}^{2}}{2187}}+{ \frac {37643\,\zeta_{{3}}}{59049}}+{\frac {80524\,\zeta_{{5}}}{59049}} \right. \\
& \left. ~~~~~ -{\frac {4172\,\zeta_{{7}}}{2187}} \right) {\varepsilon}^{5} +\mathcal{O}(\varepsilon^6) \\
=& - 0.11111\,\varepsilon- 0.058985\,{\varepsilon}^{2}+ 0.043693\,{\varepsilon}^{3 }- 0.078951\,{\varepsilon}^{4}+ 0.20843\,{\varepsilon}^{5} +\mathcal{O}(\varepsilon^6) \end{aligned} \end{align}
\begin{align} \begin{aligned} \eta_O^{(\mathrm{LY})}(\varepsilon) =& -\frac{2}{3}\,\varepsilon-{\frac {43}{486}}\,{\varepsilon}^{2}+ \left( -{\frac{8375} {157464}}+{\frac {8\,\zeta_{{3}}}{81}} \right) {\varepsilon}^{3}+ \left( {\frac {2\,\zeta_4 }{27}}-{\frac{3883409}{51018336}}-{ \frac {358\,\zeta_{{3}}}{6561}}-{\frac {40\,\zeta_{{5}}}{729}} \right) {\varepsilon}^{4} \\
& + \left( -{\frac {50\,\zeta_6 }{ 729}}-{\frac {179\,\zeta_4 }{4374}}-{\frac{1545362585}{ 16529940864}}+{\frac {68\,{\zeta_{{3}}}^{2}}{729}}+{\frac {37643\, \zeta_{{3}}}{39366}}+{\frac {40262\,\zeta_{{5}}}{19683}} \right. \\
& \left. ~~~~~-{\frac {2086 \,\zeta_{{7}}}{729}} \right) {\varepsilon}^{5} +\mathcal{O}(\varepsilon^6) \\
=& - 0.66667\,\varepsilon- 0.088477\,{\varepsilon}^{2}+ 0.065543\,{\varepsilon}^{3 }- 0.11843\,{\varepsilon}^{4}+ 0.31241\,{\varepsilon}^{5} +\mathcal{O}(\varepsilon^6) \end{aligned} \end{align}
\begin{align} \begin{aligned} 1/\nu^{(\mathrm{LY})}(\varepsilon) =& ~2-{\frac{5}{9}}\varepsilon-{\frac{43}{1458}}{\varepsilon}^{2}+ \left( -{ \frac{8375}{472392}}+{\frac {8\,\zeta_{{3}}}{243}} \right) {\varepsilon}^ {3}+ \\  &+\left( {\frac {2\,\zeta_{{4}}}{81}}-{\frac{3883409}{153055008}}-{ \frac {358\,\zeta_{{3}}}{19683}}-{\frac {40\,\zeta_{{5}}}{2187}} \right) {\varepsilon}^{4}+ \\
 & + \left( -{\frac{1545362585}{49589822592}}-{ \frac {50\,\zeta_{{6}}}{2187}}-{\frac {179\,\zeta_{{4}}}{13122}} +{ \frac {37643\,\zeta_{{3}}}{118098}}+\right.\\
&\left.~~~~~~+{\frac {40262\,\zeta_{{5}}}{59049} }+{\frac {68\,{\zeta_{{3}}}^{2}}{2187}}-{\frac {2086\,\zeta_{{7}}}{ 2187}} \right) {\varepsilon}^{5}+O \left( {\varepsilon}^{6} \right) \\
=&~ 2 -0.55556\, \varepsilon -0.029493\, \varepsilon^2 + 0.021845\, \varepsilon^3 -0.039477 \,\varepsilon^4 +0.10413\, \varepsilon^5 \\
&+\mathcal{O}(\varepsilon^6) \end{aligned} \end{align}

\begin{align} \begin{aligned} \omega^{(\mathrm{LY})}(\varepsilon) =& ~\varepsilon-{\frac {125}{162}}\,{\varepsilon}^{2}+ \left( {\frac{36755}{ 52488}}+{\frac {20\,\zeta_{{3}}}{27}} \right) {\varepsilon}^{3} + \left( \frac{5}{9} \,\zeta_4 -{\frac{31725355}{17006112}}-{\frac {9673\, \zeta_{{3}}}{2187}}+{\frac {160\,\zeta_{{5}}}{81}} \right) {\varepsilon}^ {4} \\
& + \left( {\frac {200\,\zeta_6 }{81}}-{\frac {9673\, \zeta_4 }{2916}}+{\frac{17088604709}{5509980288}}+{ \frac {1384\,{\zeta_{{3}}}^{2}}{243}}+{\frac {12094613\,\zeta_{{3}}}{ 354294}} \right. \\
& \left. ~~~~~ +{\frac {1050770\,\zeta_{{5}}}{19683}}-{\frac {21980\,\zeta_{{ 7}}}{243}} \right) {\varepsilon}^{5} +\mathcal{O}(\varepsilon^6) \\
=& ~\varepsilon- 0.77160\,{\varepsilon}^{2}+ 1.5907\,{\varepsilon}^{3}- 4.5329\,{ \varepsilon}^{4}+ 15.440\,{\varepsilon}^{5} +\mathcal{O}(\varepsilon^6) \end{aligned} \end{align}
\begin{align} \begin{aligned} \sigma^{(\mathrm{LY})}(\varepsilon) =&~ \frac{1}{2}-\frac{1}{12}\,\varepsilon-{\frac {79}{3888}}\,{\varepsilon}^{2}+ \left( -{\frac{ 10445}{1259712}}+{\frac {\zeta_{{3}}}{81}} \right) {\varepsilon}^{3} \\
& + \left( {\frac {\zeta_4 }{108}}-{\frac{4047533}{ 408146688}}-{\frac {161\,\zeta_{{3}}}{26244}}-{\frac {5\,\zeta_{{5}}}{ 729}} \right) {\varepsilon}^{4} \\
& +\left( -{\frac {25\,\zeta_6 }{2916}}-{\frac {161\,\zeta_4 }{34992}}-{\frac {1601178731}{132239526912}}+{\frac {17\,{\zeta_{{3}}}^{2}}{1458}}+{ \frac {112399\,\zeta_{{3}}}{944784}}+{\frac {20101\,\zeta_{{5}}}{78732 }} \right. \\
& \left. ~~~~~-{\frac {1043\,\zeta_{{7}}}{2916}} \right) {\varepsilon}^{5} +\mathcal{O}(\varepsilon^6) \\
=& ~0.50000- 0.083333\,\varepsilon- 0.020319\,{\varepsilon}^{2}+ 0.0065494\,{ \varepsilon}^{3}- 0.014381\,{\varepsilon}^{4}+ 0.038131\,{\varepsilon}^{5} \\
& +\mathcal{O}(\varepsilon^6) \end{aligned} \end{align}
where we have also recorded the numerical value of each coefficient in the
$\varepsilon$ expansion.

\subsection{Percolation theory}

For the percolation problem, denoted by $\mathrm{P}$, the analogous five loop
renormalization group functions after taking the replica limit as described in Section~\ref{sec:combinatorial_factors} are
\begin{align} \begin{aligned} \beta^{(\mathrm{P})}(g) =& -\frac{\varepsilon}{2}\, g-\frac{7}{4}\,{g}^{3}-{\frac {671}{144}}\,{g}^{5} + \left( -{\frac{414031}{ 20736}}-{\frac {93\,\zeta_{{3}}}{8}} \right) {g}^{7} \\
& +\left( {\frac {651\,\zeta_4 }{32}}-{\frac{84156383}{746496}}-{\frac {121109\,\zeta_{{3}}}{864}}-{ \frac {595\,\zeta_{{5}}}{6}} \right) {g}^{9} \\
& +\left( -{\frac{74773124579}{107495424}}+{\frac {930967\,\zeta_4 }{3072}}+{\frac {20825\,\zeta_6 }{64}}-{ \frac {7411\,{\zeta_{{3}}}^{2}}{32}}-{\frac {56477573\,\zeta_{{3}}}{ 27648}} \right. \\
& \left. ~~~~~-{\frac {11846549\,\zeta_{{5}}}{3456}}+{\frac {97293\,\zeta_{{7 }}}{64}} \right) {g}^{11}+ \mathcal{O}(g^{13}) \end{aligned} \end{align}
\begin{align} \begin{aligned} \gamma^{(\mathrm{P})}_\phi(g) =& ~\frac{1}{12}\,{g}^{2}+{\frac {37}{432}}\,{g}^{4} + \left( {\frac{29297}{62208}}-{ \frac {5\,\zeta_{{3}}}{24}} \right) {g}^{6} +\left( {\frac{225455}{82944}}+{\frac {33\, \zeta_4 }{32}}+{\frac {233\,\zeta_{{3}}}{864}}-{\frac { 55\,\zeta_{{5}}}{18}} \right) {g}^{8} \\
& +\left( {\frac{5907303973}{322486272}}+{\frac {169325\,\zeta_4 }{27648}}+{\frac {10675\,\zeta_6 }{576}}+{ \frac {719\,{\zeta_{{3}}}^{2}}{288}}+{\frac {3841369\,\zeta_{{3}}}{ 248832}}+{\frac {5443\,\zeta_{{5}}}{10368}} \right. \\
& \left. ~~~~~-{\frac {3969\,\zeta_{{7}} }{64}} \right) {g}^{10}+ \mathcal{O}(g^{12}) \end{aligned} \end{align}
\begin{align} \begin{aligned} \gamma^{(\mathrm{P})}_{m^2}(g) =& ~\frac{5}{6}\,{g}^{2}+{\frac {193}{108}}\,{g}^{4}+ \left( {\frac {41\,\zeta_{{3 }}}{12}}+{\frac{237751}{31104}} \right) {g}^{6} + \left( {\frac {3299\, \zeta_{{3}}}{72}}+{\frac {190\,\zeta_{{5}}}{9}}-{\frac {33\,\zeta_{{4}}}{8}}+{\frac{4114259}{93312}} \right) {g}^{8} \\
& + \left( {\frac {11167\,{\zeta_{{3}}}^{2}}{144}}+{\frac {91152569\, \zeta_{{3}}}{124416}}+{\frac {5839991\,\zeta_{{5}}}{5184}}-{\frac { 26117\,\zeta_{{7}}}{32}}-{\frac {1100099\,\zeta_{{4}}}{ 13824}}-{\frac {11725\,\zeta_{{6}}}{288}} \right. \\
& \left. ~~~~~+{\frac{ 44535597533}{161243136}} \right) {g}^{10}+\mathcal{O}(g^{12}) ~. \end{aligned} \end{align}
Again the four loop expressions are in agreement
with previous $\MSbar$ computations, \cite{Macfarlane:1974vp,Bonfim_1980,Bonfim_1981,Gracey:2015tta}.
Consequently the critical exponents are
\begin{align} \begin{aligned} \eta^{(\mathrm{P})}(\varepsilon) =& -\frac{1}{21}\varepsilon-{\frac {206}{9261}}\,{\varepsilon}^{2}+ \left( -{\frac{93619 }{8168202}}+{\frac {256\,\zeta_{{3}}}{7203}} \right) {\varepsilon}^{3} \\
& + \left( -{\frac{103309103}{14408708328}}+{\frac {64\,\zeta_4 }{2401}}+{\frac {189376\,\zeta_{{3}}}{9529569}}-{\frac {320\, \zeta_{{5}}}{3087}} \right) {\varepsilon}^{4} \\
& +\left( -{\frac{43137745921 }{3630994498656}}+{\frac {47344\,\zeta_4 }{3176523}}-{ \frac {400\,\zeta_6 }{3087}}-{\frac {187744\,{\zeta_{{3 }}}^{2}}{7411887}}+{\frac {77003747\,\zeta_{{3}}}{600362847}} \right. \\
& \left. ~~~~~+{\frac { 2337824\,\zeta_{{5}}}{9529569}}-{\frac {664\,\zeta_{{7}}}{16807}} \right) {\varepsilon}^{5}+\mathcal{O}(\varepsilon^6) \\
=& - 0.047619\,\varepsilon- 0.022244\,{\varepsilon}^{2}+ 0.031263\,{\varepsilon}^{ 3}- 0.061922\,{\varepsilon}^{4}+ 0.20454\,{\varepsilon}^{5}+\mathcal{O}(\varepsilon^6) \end{aligned} \end{align}
\begin{align} \begin{aligned} \eta_O^{(\mathrm{P})}(\varepsilon) =& -\frac{2}{7}\,\varepsilon-{\frac {355}{6174}}\,{\varepsilon}^{2}+ \left( {\frac {204 \,\zeta_{{3}}}{2401}}-{\frac{235495}{10890936}} \right) {\varepsilon}^{3} \\
& + \left( {\frac {153\,\zeta_4 }{2401}}-{\frac{157609181 }{19211611104}}+{\frac {99865\,\zeta_{{3}}}{3176523}}-{\frac {2000\, \zeta_{{5}}}{7203}} \right) {\varepsilon}^{4} \\
& +\left( {\frac {99865\, \zeta_4 }{4235364}}-{\frac {2500\,\zeta_6 }{7203}}-{\frac{97373066851}{4841325998208}}-{\frac {107248\, {\zeta_{{3}}}^{2}}{2470629}}+{\frac {202024997\,\zeta_{{3}}}{800483796 }} \right. \\
& \left. ~~~~~+{\frac {1860076\,\zeta_{{5}}}{3176523}}+{\frac {2288\,\zeta_{{7}}}{ 16807}} \right) {\varepsilon}^{5} +\mathcal{O}(\varepsilon^6) \\
=& - 0.28571\,\varepsilon- 0.057499\,{\varepsilon}^{2}+ 0.080517\,{\varepsilon}^{3 }- 0.18935\,{\varepsilon}^{4}+ 0.63749\,{\varepsilon}^{5}+\mathcal{O}(\varepsilon^6) \end{aligned} \end{align}
\begin{align} \begin{aligned} 1/\nu^{(\mathrm{P})}(\varepsilon) =&~ 2-{\frac {5\,\varepsilon}{21}}-{\frac {653\,{\varepsilon}^{2}}{18522}}+ \left( -{\frac{332009}{32672808}}+{\frac {356\,\zeta_{{3}}}{7203}} \right) {\varepsilon}^{3}+ \\
 &+\left( {\frac {89\,\zeta_4 }{ 2401}}-{\frac{59591131}{57634833312}}+{\frac {110219\,\zeta_{{3}}}{ 9529569}}-{\frac {3760\,\zeta_{{5}}}{21609}} \right) {\varepsilon}^{4}+ \\
&+ \left( -{\frac{119568216869}{14523977994624}}+{\frac {110219\,\zeta_4 }{12706092}}-{\frac {4700\,\zeta_6 }{ 21609}}+{\frac {298060003\,\zeta_{{3}}}{2401451388}}+\right. \\
&\left.~~~~~+{\frac {3242404\, \zeta_{{5}}}{9529569}}+{\frac {2952\,\zeta_{{7}}}{16807}}-{\frac { 134000\,{\zeta_{{3}}}^{2}}{7411887}} \right) {\varepsilon}^{5}+\mathcal{O}(\varepsilon^6) \\
=&~ 2 -0.238095\, \varepsilon -0.035255\, \varepsilon^2 + 0.049249\, \varepsilon^3 -0.12744 \,\varepsilon^4 +0.43287\, \varepsilon^5 +\mathcal{O}(\varepsilon^6) \end{aligned} \end{align}
\begin{align} \begin{aligned} \omega^{(\mathrm{P})}(\varepsilon) =&~ \varepsilon-{\frac {671}{882}}\,{\varepsilon}^{2}+ \left( {\frac{40639}{ 57624}}+{\frac {372\,\zeta_{{3}}}{343}} \right) {\varepsilon}^{3} \\
& + \left( {\frac {279\,\zeta_4 }{343}}-{\frac{317288185}{ 304946208}}-{\frac {348539\,\zeta_{{3}}}{151263}}-{\frac {1360\,\zeta_ {{5}}}{343}} \right) {\varepsilon}^{4} \\
& +\left( {\frac{601352852897}{ 691617999744}}-{\frac {348539\,\zeta_4 }{201684}}-{ \frac {1700\,\zeta_6 }{343}}+{\frac {207440\,{\zeta_{{3 }}}^{2}}{117649}}+{\frac {11664257531\,\zeta_{{3}}}{800483796}} \right. \\
& \left. ~~~~~+{ \frac{17305178\,\zeta_{{5}}}{453789}}-{\frac {55596\,\zeta_{{7}}}{ 2401}} \right) {\varepsilon}^{5} +\mathcal{O}(\varepsilon^6) \\
=&~ \varepsilon- 0.76077\,{\varepsilon}^{2}+ 2.0089\,{\varepsilon}^{3}- 7.0413\,{ \varepsilon}^{4}+ 30.216\,{\varepsilon}^{5}+\mathcal{O}(\varepsilon^6) ~. \end{aligned} \end{align}
Examining the numerical values one can easily see that the series for the
critical exponents has growing alternating coefficients, which is typical for
an asymptotic series. Therefore in order to obtain reliable estimates for
the exponents one needs to apply resummation methods.

\section{Resummation strategy}
\label{sec:resummation}

While our focus so far has been in relation to renormalizing $\phi^3$ theory
in six dimensions, the physical problems of interest are in lower
dimensions.
As noted we require a strategy to resum the asymptotic series in
$\varepsilon$ for the critical exponents in order to extract meaningful estimates
in for example three dimensions. In this case $\varepsilon$ itself would take the
value $3$ which is clearly not small as
$d = 6 - \varepsilon$. Therefore we devote this section to
discussing the various resummation techniques that we will apply to both the
Lee--Yang and percolation problems.

First we recall aspects of the resummation formalism that is well-established
in this area of quantum field theory. In general, an $\varepsilon$-expansion series, $f(\varepsilon)$, for a critical
exponent is asymptotic with factorially growing coefficients,
\cite{Lipatov:DivergencePT,BrezinGuillouZinnJustin:phi2N,mckane2019perturbation},
\begin{equation}
    f(\varepsilon) = \sum_{k=0}^{\infty} f_k \varepsilon^k, \qquad f_k \propto k!(-a)^k k^b \text{ for } k \rightarrow \infty ~,
    \label{hoa}
\end{equation}
where the constant $a > 0$ is related to 
the position of the closest singularity $-1/a$ in the 
Borel plane. This number is
the {\em same} for all exponents of the underlying model. By contrast the
parameter $b$ is related to the type of singularity and may take a different
value for different exponents. In the two cases 
considered here we will only
use the parameter $a$. 

For the Lee--Yang problem we take
$a=5/18$~\cite{Mckane:1978me,Kirkham_1979,kalagov2014higher}.
For percolation theory we use $a$~$=$~$15/28$~\cite{houghton1978high}.
In \cite{mckane1986asymptotic} the slightly lower value of $10/21$ was obtained,
but later on in \cite{caraccioloasymptotic} it was shown that there was a
discrepancy caused by an incorrect integration contour in
\cite{mckane1986asymptotic,brezin1987note}. If the contour is corrected then
the approach suggested in \cite{mckane1986asymptotic} leads to the same results
as \cite{houghton1978high}. 
Given that the precise value of $a$ is still not fully resolved yet,
in our subsequent resummation for percolation we have carried out the analysis
for both cases. It transpires that using either value of $a$ produces estimates that are almost the same. The resulting error bars are consistently of similar size and the difference of the two central values is completely negligible within these error bars.
 For this reason, we only provide the results based on the choice $a$~$=$~$15/28$.
Finally we note that even though the value of the implicit proportionality constant in eq.~\eqref{hoa} remains controversial \cite{mckane2019perturbation,mckane2021},
its precise value is not required for the
analysis we have performed for either problem.

In order to obtain reliable estimates for critical exponents we have applied
a number of different resummation techniques. These are Pad\'{e},
Pad\'{e}--Borel--Leroy (PBL), Borel resummation with conformal mapping (KP17)
\cite{KompanietsPanzer2017}, double-sided Pad\'{e} and constrained versions of
Pad\'{e}, PBL and KP17. The main purpose of implementing double-sided Pad\'{e}
and constrained resummations is to produce more accurate estimates especially
in lower dimensions such as $d$~$=$~$3$. For both problems the constraint
arises from the known exact values for exponents in two dimensions. Those for percolation are
derived from a minimal conformal field theory with $m$~$=$~$2$ and central charge $c$~$=$~$0$.
These conformal field theories have been
classified \cite{Friedan:1983xq}. 
It turns
out that taking into account the known exact value at $d$~$=$~$2$ significantly
improves 
estimates. For the Lee--Yang problem exact values for some exponents
are also are known in $d$~$=$~$1$. We will 
now discuss technical aspects of each of the
resummation methods we used separately.

\paragraph{Pad\'{e}, double-sided Pad\'{e}.}
Applying the method of Pad\'{e} approximants is regarded as one of the 
most simple
resummation methods. It does not require any knowledge about the
properties of the series considered. Specifically the approximant is
constructed as a rational function
\begin{equation}
    P_{[L/M]}(\varepsilon):=\frac{P_L(\varepsilon)}{P_M(\varepsilon)}
    \label{Pad\'{e}lm}
\end{equation}
where $P_L(\varepsilon)$ and $P_M(\varepsilon)$ are polynomials of order $L$ and $M$ respectively. %
The polynomials are chosen in such a way that the expansion of the approximant
up to $N=L+M+1$ coincides precisely with the initial series. For a double-sided
Pad\'{e} expansion information from both sides of the expansion interval is
used. For the lower end, which is $d$~$=$~$2$ in our case, we only know the first term of the
expansion from conformal field theory. So to find an estimate for the series up
to $\varepsilon^N$ one needs to consider approximants with $L+M+1=N+1$. Given that
$d$~$=$~$1$ data is available for the Lee--Yang problem too we need to consider
approximants with $L+M+1 = N+2$ in that case.

One of the problems with Pad\'{e} approximants is that different $L/M$
approximants can produce significantly different estimates and one has to
choose a ``proper'' approximant. Such subjective choices might be ill-conceived
and hence provide an incorrect estimate. In this article we follow the strategy
suggested in \cite{adzhemyan2019six} which is as follows. In order to obtain an
estimate and error bar (apparent accuracy) of order $N$ we will consider all
non-marginal approximants of order $N$ and $N-1$ and view them as ``independent
measurements'':
\begin{equation}
x_k := P_{[L_k/M_k]}(\varepsilon_{\text{phys}}) \quad \text{ for } \quad k =1,\ldots, n 
\end{equation}
where $\varepsilon_{\text{phys}}$ is the value of the expansion parameter where the estimate is computed. %
This gives the estimate and error bar as
\begin{equation}
\langle x\rangle =\frac{x_1+\ldots+x_n}{n}, \qquad \Delta x = t_{0.95,n} \sqrt{\frac{ (\langle x\rangle -x_1)^2+\ldots+(\langle x\rangle -x_n)^2}{n (n-1)}}\;.
\label{student}
\end{equation}
where $t_{0.95,n}$ is the $t$-distribution with $p=0.95$ confidence level. In
the situation where fewer than 3 approximants survive, we do not provide an
error bar since usually this error bar is unreliable. 
We consider Pad\'e approximants as marginal if the denominator in eq.~\eqref{Pad\'{e}lm} 
has a root in the interval $[0,2 \varepsilon_{\text{phys}}]$.

The standard Pad\'{e} technique also suffers from problems when the argument becomes
large because power law asymptotics $\varepsilon^{L-M}$ become dominant. In this case
the estimates have little predictive value. Thus we
limit ourselves to approximants with $|L-M|<3$ for which the asymptotics are
not so strong and become dominant at much larger values. The
double-sided Pad\'{e} is free of this problem because the asymptotic growth at
physical values of the expansion parameter is limited by the constraint imposed
at $d=2$ or $d=1$. %

\paragraph{Pad\'{e}--Borel--Leroy.}
By contrast the Pad\'{e}--Borel--Leroy method is one of the simplest methods of
Borel resummation. The series under consideration is assumed to be Borel
summable with factorially growing and sign-alternating coefficients as in eq.~\eqref{hoa}. In this case details of the large order behaviour such
as the parameters $a$ and $b$ are not relevant. To obtain an estimate and error
bar we also follow the technique suggested in \cite{adzhemyan2019six} which is
almost similar to the above Pad\'{e} strategy.

\paragraph{Borel resummation with conformal mapping.}
An extension of the previous approach is Borel resummation with conformal
mapping which has proved itself as one of the most reliable and precise
resummation methods for the $\varepsilon$ expansion
\cite{VladimirovKazakovTarasov:Calculation,KazakovTarasovShirkov:AnalyticContinuation,GuillouZinnJustin:Accurate,GuidaZinnJustin:CriticalON,KompanietsPanzer2017}.
It allows one to utilize information about the large order asymptotics and
other properties of the series under consideration. Meanwhile there are plenty
of realizations of this approach. For instance, in this paper we use the KP17
procedure which was developed in \cite{KompanietsPanzer2017} for studying the
$\varepsilon$ expansion of $\phi^4$ theory critical exponents at six loops. It is very reliable and
well-documented with the technical details provided in Section~V of
\cite{KompanietsPanzer2017}.

\paragraph{Constrained resummations.}
The idea of the constrained Borel resummation was proposed by R.~Guida and J.~Zinn-Justin 
in \cite{GuidaZinnJustin:CriticalON} where it was called
resummation with boundary condition. We have also applied this idea not only to the 
KP17 procedure, which is very similar to the method used in
\cite{GuidaZinnJustin:CriticalON}, but also to the above Pad\'{e} and PBL
methods. It should be noted that Pad\'{e} approximants are not Borel
resummable. It is more natural to apply constraints in the way described for
double-sided Pad\'{e}, but we retain Pad\'{e} approximants here as it provides
an additional view on constrained resummation. Also results from constrained
and double-sided Pad\'{e} are in fact different because Pad\'{e} approximants
apply to different series. Constrained resummation is based on the following
series transformation
\begin{equation}
 f(\varepsilon) =  \sum_{n=0}^\infty f_n \varepsilon^n = f(\varepsilon_{\text{bc}})+(\varepsilon_{\text{bc}}-\varepsilon) \sum_{n=0}^\infty h_n\varepsilon^n,
\label{bctransform}
\end{equation}
where $f(\varepsilon_{\text{bc}})$ is the known value of the expanded exponent at $\varepsilon_{\text{bc}}$ with $\text{bc}$ standing for \emph{boundary condition}. 
The resummation is then applied to
the series $h(\varepsilon)=\sum_{n=0}^\infty h_n \varepsilon^n$
before being substituted into
\eqref{bctransform} to obtain an estimate for $f(\varepsilon)$. Uncertainties are
computed for $h(\varepsilon)$ as described earlier for each method before being
transformed to an uncertainty for $f(\varepsilon)$ by standard algebraic rules. %

It should be noted that this approach allows one to apply only one constraint.
So for the Lee--Yang problem we will provide two constrained resummations for
each method. There will be one using the value of the exponent in $d=1$ and
another for $d=2$. We will distinguish these two constrained procedures as
<<c\{Method\}-\{$d_{bc}$\}>>. So for instance cPad\'{e}-1 is the constrained
Pad\'{e} with the $d$~$=$~$1$ constraint.

\paragraph{Overall estimate.}
Given that we will have a large number of estimates for each exponent 
from different methods we must present an overall final estimate to
summarize all our results. One of the problems is that there is no method which
provides the smallest uncertainty for every exponent. Here we introduce an
automatic algorithm that gives higher weight to results from methods which have
smaller uncertainty. So we will compute final estimates as a weighted average
of all estimates with weights $w_i$ proportional to the inverse uncertainties.
Specifically we define
\begin{equation}
    A_{\text{final}} = \frac{\sum_i A_i w_i}{\sum_i w_i}
\end{equation}
which allows
us to discard almost all estimates with very large errors.
If instead
we were to perform a simple averaging
that would significantly shift the final estimate to an incorrect value.

The error estimate is computed from two parts which are the weighted standard
deviation and weighted average of the uncertainties given by
\begin{equation}
  E_{\text{final}} = \sqrt{E_1^2+E_2^2}, \quad \textrm{where} \quad  E_1 = \sqrt{\frac{\sum_i (A_{\text{final}}-A_i)^2 w_i}{\sum_i w_i}}, \qquad E_2 = \frac{1}{\sum_i w_i} ~.
 \end{equation}
This strategy results in the following:
\begin{enumerate}
    \item methods with very large error bars almost always do not contribute to
an estimate and only slightly increase the error bar;
    \item if methods provide significantly different results with comparable
error bars, then the overall error bar increases;
    \item if all methods provide almost the same value then the error
decreases.
\end{enumerate}
In parallel to analysing the estimates from the five loop exponents we have
also repeated this exercise for Monte Carlo and series data in each integer dimension.
In this way we can compare data for exponents from perturbation theory and
simulations on the same level.

\section{Resummation analysis}
\label{sec:analysis}

Having outlined the technical description of
each of the resummation methods we have used to
extract critical exponents as well as how we
arrive at our error estimates, we devote this
section to recording the actual values for a 
wide range of exponents. For both Lee--Yang edge
singularity and percolation theory a table is
provided for each exponent. In addition we
illustrate that data with an associated figure. 
Each figure shows the exponent estimates with
errors for dimensions 2, 3, 4 and 5 together with
our overall Monte Carlo summary values where
these are available. The horizontal axis in each
figure corresponds to dimension $d$.

\subsection{Lee--Yang edge singularity}

For our Lee--Yang singularity analysis we focus on a small set of exponents.
As the main work of others has centred primarily around the exponent $\sigma$
we have compiled as comprehensive an amount of independent results as possible
in order to draw comparisons. For $\eta$ and $\nu$, aside from the three loop
work of \cite{Bonfim_1981}, we were only able to find one study of these
exponents which was \cite{An:2016lni}. That used
the functional renormalization
group approach and determined $\eta$ and $\sigma$
directly. An additional exponent termed $\nu_c$
was also computed. From the hyperscaling
relations
given in \cite{An:2016lni}, it can be related to our $\eta$ and $\nu$ via the
expression $\nu$~$=$~$\nu_c/\sigma$. We have used the values for $\eta$ and
$\nu_c$ of \cite{An:2016lni} to produce an independent estimate for $\nu$ to
benchmark our results against. 
Each of the tables follows a common theme.
The top part of each is a compilation of data
from numerical methods with sources. This is
followed by our resummation analysis for
both four
and five loops. The former is provided to gauge
convergence. In each case three lines 
summarize the unconstrained and constrained
estimates as well as the combination of both.
This can be compared to a similar estimate
from Monte Carlo (MC) and series (srs) data.

{\begin{table}
\begin{center}
\begin{tabular}{c|c|lllll}
&Reference & $d$~$=$~$1$ & $d$~$=$~$2$ & $d$~$=$~$3$ & $d$~$=$~$4$ & $d$~$=$~$5$ \\
\cline{2-7}
\noalign{\vskip\doublerulesep
         \vskip-\arrayrulewidth}
\cline{2-7}
&exact   & -1.0         & -0.8         & & & \\
\hline
\hline
\multirow{2}{*}{\rotatebox[origin=c]{90}{\parbox{.7cm}{\centering MC/\\srs}} }
& \multicolumn{1}{r|}{\cite{An:2016lni} (2016)} & & &                   -0.586(29)&    -0.316(16)&   -0.126(6)\\
\cline{2-7}
& overall & & &                   -0.586(29)&    -0.316(16)&   -0.126(6)\\

\hline
\hline
\multirow{12}{*}{\rotatebox[origin=c]{90}{4 loops} }& Pad\'{e}        & -1.177       & -0.8655      & -0.5893      & -0.38(11)    & -0.153(12)   \\
& KP17        & -1.2(9)      & -0.9(5)      & -0.6(3)      & -0.45(10)    & -0.161(8)    \\
& cPad\'{e}-1     &             & -0.77(4)     & -0.55(4)     & -0.34(2)     & -0.149(4)    \\
& cPad\'{e}-2     &             &             & -0.56(2)     & -0.344(15)   & -0.149(4)    \\
& cPBL-1      &             & -0.78(2)     & -0.56(3)     & -0.345(14)   & -0.150(3)    \\
& cPBL-2      &             &             & -0.570(13)   & -0.349(10)   & -0.151(2)    \\
& cKP17-1     &             & -0.763(8)    & -0.539(11)   & -0.332(9)    & -0.147(3)    \\
& cKP17-2     &             &             & -0.556(9)    & -0.338(8)    & -0.148(3)    \\
& Constr.~Pad\'{e} &            &            & -0.581(11)  & -0.356(10)  & -0.152(3)   \\
\cline{2-7}
& non constr. & -1.2(9)      & -0.9(5)      & -0.6(3)      & -0.42(11)    & -0.158(10)   \\
&constr. &             & -0.77(2)      & -0.56(2)      & -0.343(14)    & -0.149(3)    \\
&all &  -1.2(9)     & -0.77(3)      & -0.56(2)      & -0.35(2)      & -0.150(5)    \\

\hline
\hline
\multirow{13}{*}{\rotatebox[origin=c]{90}{5 loops} }& Pad\'{e}        & -1.249       & -0.9148      & -0.6185      & -0.39(9)     & -0.154(7)    \\
& PBL         & -0.9(5)      & -0.8(3)      & -0.55(12)    & -0.34(4)     & -0.150(3)    \\
& KP17        & -1.2(2)      & -0.91(11)    & -0.61(4)     & -0.36(7)     & -0.152(11)   \\
& cPad\'{e}-1     &             & -0.78(3)     & -0.56(2)     & -0.346(11)   & -0.151(2)    \\
& cPad\'{e}-2     &             &             & -0.570(13)   & -0.349(8)    & -0.1509(14)  \\
& cPBL-1      &             & -0.785(11)   & -0.564(11)   & -0.348(6)    & -0.1508(11)  \\
& cPBL-2      &             &             & -0.571(6)    & -0.350(4)    & -0.1511(8)   \\
& cKP17-1     &             & -0.765(7)    & -0.542(9)    & -0.334(8)    & -0.148(4)    \\
& cKP17-2     &             &             & -0.557(9)    & -0.338(8)    & -0.148(3)    \\
& Constr.~Pad\'{e} &            &            & -0.580(7)   & -0.356(6)   & -0.1521(13) \\

\cline{2-7}
& non constr. & -1.2(3)      & -0.9(2)      & -0.60(7)     & -0.36(6)     & -0.152(6)    \\
&constr. &             & -0.774(15)    & -0.565(15)    & -0.347(10)    & -0.151(2)    \\
&all &  -1.2(3)   & -0.78(3)      & -0.57(2)      & -0.347(12)    & -0.151(2)    \\

\end{tabular}
\end{center}
\caption{Estimates for Lee--Yang $\eta$.}
\label{lyetaests}
\end{table}}

\begin{figure}
	\centering
  \includegraphics[width=.9\textwidth]{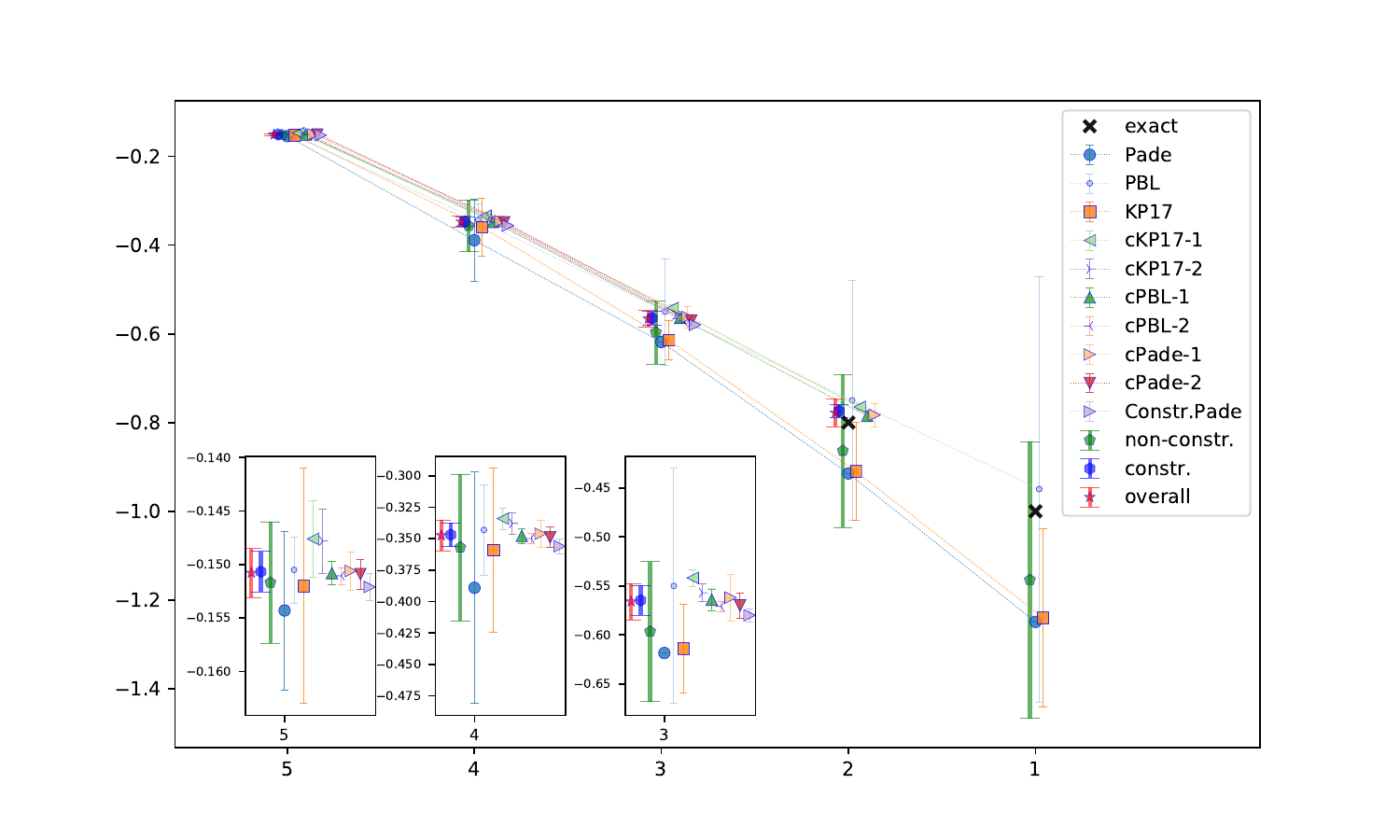}
	\caption{Plot of estimates for Lee--Yang $\eta$.}
	\label{perclyetafig}
\end{figure}

For the exponent $\eta$ our results are in Table~\ref{lyetaests}. We see that 
the estimates for the three key dimensions are stable from four to five loops.
However for the only available study that we could locate \cite{An:2016lni}
there is some overlap agreement for the 3 dimensional estimate in contrast to
those for higher dimensions. Although this exponent is ordinarily regarded as
one of the more difficult ones to reconcile between different methods since
it is very close to zero.
This is not the case here.

{\begin{table}
\begin{center}
\begin{tabular}{c|c|lllll}
&Reference & $d$~$=$~$1$ & $d$~$=$~$2$ & $d$~$=$~$3$ & $d$~$=$~$4$ & $d$~$=$~$5$ \\
\cline{2-7}
\noalign{\vskip\doublerulesep
         \vskip-\arrayrulewidth}
\cline{2-7}

\multirow{2}{*}{\rotatebox[origin=c]{90}{\parbox{.7cm}{\centering MC/\\srs}} }
& \multicolumn{1}{r|}{\cite{An:2016lni} (2016)} & & &                   (4.826146)& (1.187477) &  (0.696008)\\
\cline{2-7}
& overall & & &                   &  &   \\

\hline
\hline

\multirow{4}{*}{\rotatebox[origin=c]{90}{4 loops} }& Pad\'{e}        &             &             &             & 1.088        & 0.701(4)     \\
&KP17        & 1.6(1.2)     & 1.3(9)       & 1.1(6)       & 0.9(3)       & 0.69(3)      \\

\cline{2-7}
&non constr. & 1.6(1.2)     & 1.3(9)       & 1.1(6)       & 0.9(3)       & 0.699(8)     \\
&all         & 1.6(1.2)     & 1.3(9)       & 1.1(6)       & 0.9(3)       & 0.699(8)     \\
\hline
\hline

\multirow{4}{*}{\rotatebox[origin=c]{90}{5 loops} }& Pad\'{e}        &             &             &             &             & 0.7022(12)   \\
&PBL         & 2(2)  & 1.6(1.2)     & 1.3(5)       & 0.98(10)     & 0.696(3)     \\
&KP17        & 1.6(1.3)     & 1.4(1.0)     & 1.1(7)       & 0.9(2)       & 0.70(2)      \\
\cline{2-7}
&non constr. & 2(2)         & 1.5(1.1)     & 1.2(6)       & 0.95(14)     & 0.700(4)     \\
&all         & 2(2)         & 1.5(1.1)     & 1.2(6)       & 0.95(14)     & 0.700(4)     \\

\end{tabular}
\end{center}
\caption{Estimates for Lee--Yang $\nu$.}
\label{lynuests}
\end{table}}

{\begin{table}
\begin{center}
\begin{tabular}{c|c|lllll}
&Reference & $d$~$=$~$1$ & $d$~$=$~$2$ & $d$~$=$~$3$ & $d$~$=$~$4$ & $d$~$=$~$5$ \\
\cline{2-7}
\noalign{\vskip\doublerulesep
         \vskip-\arrayrulewidth}
\cline{2-7}
&exact   & -1.0         & -2.5         & & & \\
\hline
\hline

\multirow{2}{*}{\rotatebox[origin=c]{90}{\parbox{.7cm}{\centering MC/\\srs}} }
& \multicolumn{1}{r|}{\cite{An:2016lni} (2016)}& & &                   (4.826146)& (1.187477) &  (0.696008)\\
\cline{2-7}
&  overall & & &                   &    &   \\

\hline
\hline

\multirow{13}{*}{\rotatebox[origin=c]{90}{4 loops} }& Pad\'{e}        & -1.1(1.1)    & -3(5)  & 4(3)         & 1.18(10)     & 0.701(4)     \\
&PBL         & -1(2)  & -4(11)       & 3(3)         & 1.17(14)     & 0.700(6)     \\
&KP17        & -0.8(3)      & -2.1(1.1)    & 6(4)         & 1.23(5)      & 0.702(3)     \\
&cPad\'{e}-1     &             & -2.58(11)    & 4.5(3)       & 1.205(14)    & 0.7015(10)   \\
&cPad\'{e}-2     &             &             & 4.6(2)       & 1.208(10)    & 0.7016(8)    \\
&cPBL-1      &             & -2.56(8)     & 4.5(3)       & 1.208(10)    & 0.7017(6)    \\
&cPBL-2      &             &             & 4.64(14)     & 1.211(7)     & 0.7019(5)    \\
&cKP17-1     &             & -2.61(7)     & 4.4(3)       & 1.20(2)      & 0.701(2)     \\
&cKP17-2     &             &             & 4.58(12)     & 1.206(8)     & 0.7017(11)   \\
&Constr.~Pad\'{e}  &            &            & 4.74(2)     & 1.215(2)    & 0.7022(2)   \\
\cline{2-7}
&non constr. & -1.0(7)      & -2(3)  & 4(3)         & 1.20(9)      & 0.701(4)     \\
&constr. &             & -2.58(9)      & 4.67(13)      & 1.211(8)      & 0.7019(6)    \\
&all &  -1.0(7)       & -2.6(2)       & 4.7(2)        & 1.211(11)     & 0.7019(9)    \\
\hline
\hline
\multirow{13}{*}{\rotatebox[origin=c]{90}{5 loops} }&Pad\'{e}        & -1.0(4)      & -2.4(1.3)    & 5(2)         & 1.22(4)      & 0.7022(12)   \\
&PBL         & -2(2)        & -5(11)       & 3(2)         & 1.17(8)      & 0.701(2)     \\
&KP17        & -0.92(15)    & -2.3(5)      & 5.0(1.2)     & 1.22(2)      & 0.7021(11)   \\
&cPad\'{e}-1     &             & -2.55(8)     & 4.6(2)       & 1.209(8)     & 0.7019(4)    \\
&cPad\'{e}-2     &             &             & 4.65(14)     & 1.211(6)     & 0.7020(3)    \\
&cPBL-1      &             & -2.54(4)     & 4.60(11)     & 1.210(4)     & 0.7019(2)    \\
&cPBL-2      &             &             & 4.66(6)      & 1.212(2)     & 0.70199(14)  \\
&cKP17-1     &             & -2.61(9)     & 4.4(3)       & 1.211(15)    & 0.7020(5)    \\
&cKP17-2     &             &             & 4.58(14)     & 1.212(10)    & 0.7020(7)    \\
&Constr.~Pad\'{e}  &            &            & 4.71(7)     & 1.214(4)    & 0.7021(2)   \\
\cline{2-7}
&non constr. & -1.0(3)      & -2.4(1.2)    & 4(2)         & 1.21(4)      & 0.7019(14)   \\
&constr.     &              & -2.56(6)     & 4.61(14)     & 1.211(5)     & 0.7020(3)    \\
&all         & -1.0(3)      & -2.6(2)      & 4.6(2)       & 1.211(8)     & 0.7020(4)    \\

\end{tabular}
\end{center}
\caption{Estimates for Lee--Yang $\nu$ using the series for $1/\nu$.}
\label{lynuinvests}
\end{table}}

\begin{figure}
	\centering
  \includegraphics[width=.9\textwidth]{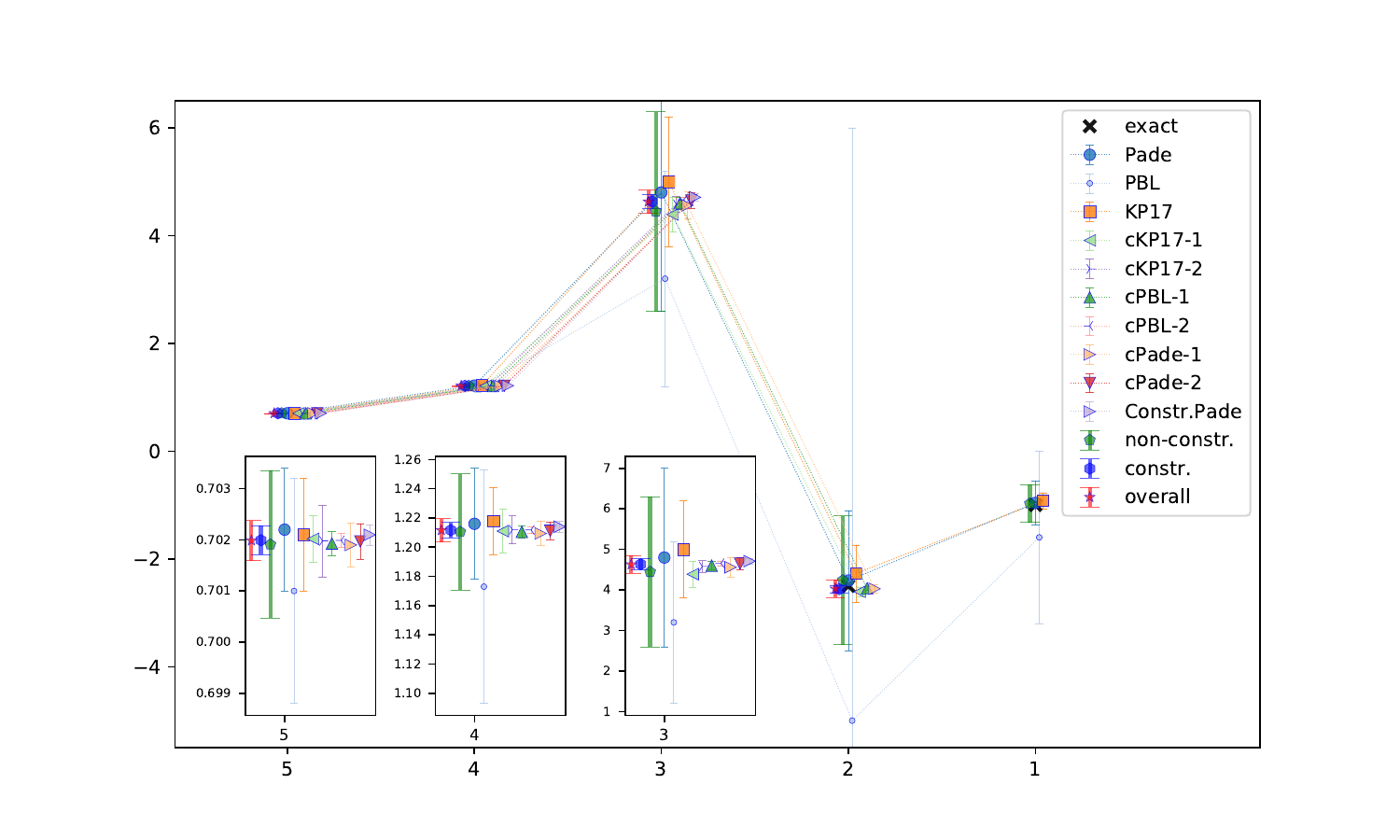}
	\caption{Plot of estimates (from $1/\nu$) for Lee--Yang $\nu$.}
	\label{perclynufig}
\end{figure}

Our direct estimate of $\nu$ suffered from singularity issues. To illustrate
this situation we have included Table~\ref{lynuests} where data from those few
techniques that did render a reliable estimate are presented. However the
resultant large error bars mean 
that these values cannot be regarded as
reliable. Instead we took a different tack 
and applied our resummation to the
series for $1/\nu$ before inverting the final numerical value. This gave
problem-free data for all our approaches which is recorded in Table~\ref{lynuinvests}. 
The behaviour of $\nu$ over the dimensions indicates that
there
is a maximum. Clearly $\nu$ is positive and increases as $d$ decreases to a
large value in $3$ dimensions before becoming negative in two dimensions 
where the exact values are known. The functional renormalization group results
have a similar behaviour and we are in qualitative accord at the very least.

{\begin{table}
\begin{center}
\begin{tabular}{c|c|lllll}
&Reference & $d$~$=$~$1$ & $d$~$=$~$2$ & $d$~$=$~$3$ & $d$~$=$~$4$ & $d$~$=$~$5$ \\
\cline{2-7}
\noalign{\vskip\doublerulesep
         \vskip-\arrayrulewidth}
\cline{2-7}

&exact   & -0.5         & -0.1667      & & & \\
\hline
\hline
\multirow{9}{*}{\rotatebox[origin=c]{90}{MC/srs} }
& \multicolumn{1}{r|}{\cite{Bonfim_1981} (1981)}& &          -0.146 -- -0.166&  0.079--0.091&   0.262--0.266&  0.399--0.400\\
& \multicolumn{1}{r|}{\cite{LaiFisher} (1995) }& &         -0.165(6)&       0.080(6)&      0.261(12)&    0.40(2) \\
&\multicolumn{1}{r|}{(hdsc) \cite{Butera:2012tq} (2012)}& &  -0.1662(5)& 0.077(2)&  0.258(5)&  0.401(9)\\
&\multicolumn{1}{r|}{ (hdbcc) \cite{Butera:2012tq}  (2012)}&  &-0.1662(5)&  0.076(2)&  0.261(4)&     0.402(2)\\
& \multicolumn{1}{r|}{\cite{ziff2} (2014)}& &                -0.161(8)&       0.0877(25)&    0.2648(16)&   0.402(5) \\
&\multicolumn{1}{r|}{ \cite{Gliozzi:2014jsa} (2014)} & &        -0.1664(5)&      0.085(1)&      0.2685(1)&    0.4105(5)\\
&\multicolumn{1}{r|}{ \cite{An:2016lni} (2016)} & &         &        0.0742(56)&    0.2667(32)&   0.4033(12)\\
&\multicolumn{1}{r|}{(FRG) \cite{Zambelli_2017}  (2017)} & & &                               &0.2648(6)    &0.40166(2)\\
&\multicolumn{1}{r|}{(LPA) \cite{Zambelli_2017}   (2017)}& &  -0.193&          0.0588&        0.245&        0.394\\
\cline{2-7}
& overall & & -0.1661(11) & 0.081(5)    & 0.267(2)    & 0.402(2)/0.407(4)   \\

\hline
\hline

\multirow{11}{*}{\rotatebox[origin=c]{90}{4 loops} } & Pad\'{e}        & -0.1(4)  & 0.0(2)  & 0.16(12)     & 0.27(4)      & 0.400(4)     \\
& PBL         & -0.1(4)  & 0.0(2)  & 0.16(12)     & 0.28(4)      & 0.401(4)     \\
& KP17        & 0.1(6)  & -0.3(4)  & 0.0(2)  & 0.24(2)      & 0.397(3)     \\
& cPad\'{e}-1     &             & -0.18(2)     & 0.07(2)      & 0.258(7)     & 0.3983(12)   \\
& cPad\'{e}-2     &             &             & 0.073(10)    & 0.259(6)     & 0.3983(10)   \\
& cKP17-1     &             & -0.17(3)     & 0.07(3)      & 0.258(11)    & 0.398(2)     \\
& cKP17-2     &             &             & 0.074(12)    & 0.258(8)     & 0.398(2)     \\
& Constr.~Pad\'{e} &            &            & 0.077(2)    & 0.260(2)    & 0.3985(5)   \\
\cline{2-7}
& non constr. & -0.1(5)  & -0.1(3)  & 0.1(2)  & 0.26(4)      & 0.399(4)     \\
&constr. &             & -0.18(3)      & 0.075(7)      & 0.259(5)      & 0.3984(7)    \\
&all &  -0.1(5)       & -0.16(9)      & 0.08(2)       & 0.259(9)      & 0.3985(11)   \\
\hline
\hline
\multirow{13}{*}{\rotatebox[origin=c]{90}{5 loops} } & Pad\'{e}        & -0.3113      & -0.0923      & 0.1002       & 0.26(3)      & 0.398(2)     \\
& PBL         & -0.2(2)  & -0.02(12)    & 0.13(5)      & 0.272(13)    & 0.3991(8)    \\
& KP17        &             & -0.21(10)    & 0.06(6)      & 0.257(14)    & 0.3982(12)   \\
& cPad\'{e}-1     &             & -0.18(2)     & 0.068(12)    & 0.257(4)     & 0.3983(8)    \\
& cPad\'{e}-2     &             &             & 0.072(6)     & 0.258(3)     & 0.3981(4)    \\
& cPBL-1      &             & -0.18(3)     & 0.07(2)      & 0.257(7)     & 0.3981(7)    \\
& cPBL-2      &             &             & 0.072(11)    & 0.258(5)     & 0.3981(6)    \\
& cKP17-1     &             & -0.17(2)     & 0.072(12)    & 0.258(4)     & 0.3983(5)    \\
& cKP17-2     &             &             & 0.075(6)     & 0.259(2)     & 0.3984(4)    \\
& Constr.~Pad\'{e} &            &            & 0.078(2)    & 0.2602(14)  & 0.3984(2)   \\

\cline{2-7}
& non constr. & -0.2(2)  & -0.12(14)    & 0.10(6)      & 0.26(2)      & 0.3986(13)   \\
&constr. &             & -0.18(2)      & 0.075(7)      & 0.259(3)      & 0.3983(5)    \\
&all &  -0.2(2)      & -0.17(5)      & 0.075(11)     & 0.259(5)      & 0.3983(6)    \\

\end{tabular}
\end{center}
\caption{Estimates for Lee--Yang $\sigma$.}
\label{lysigmaests}
\end{table}}

\begin{figure}
	\centering
  \includegraphics[width=.9\textwidth]{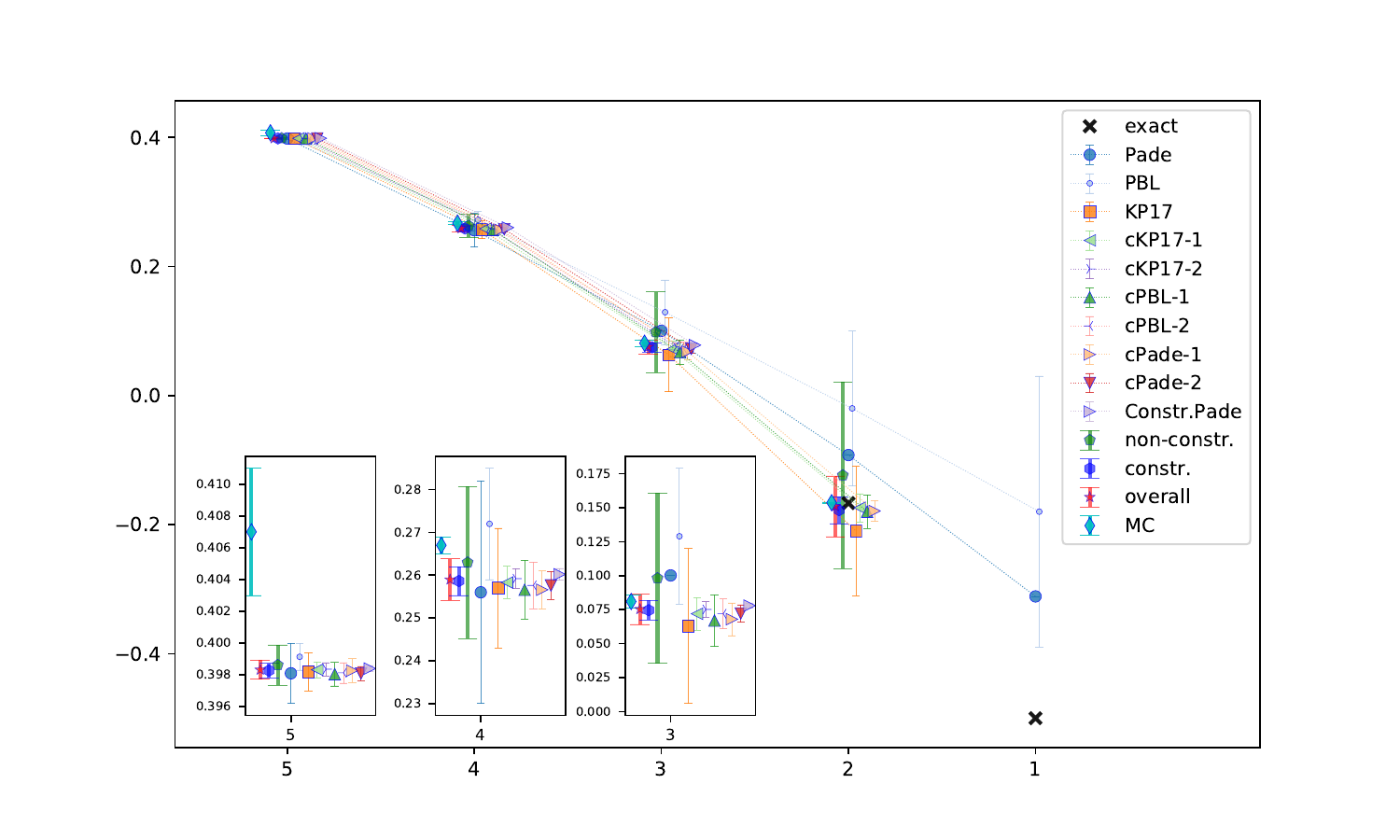}
	\caption{Plot of estimates for Lee--Yang $\sigma$.}
	\label{perclysigmafig}
\end{figure}

For $\sigma$ a similar picture emerges in Table~\ref{lysigmaests} where $hdsc$ and $hdbcc$
indicate results in $d$ dimensions from 
separate simulations. Also FRG denotes the 
functional renormalization group and LPA
indicates the use of the local potential
approximation. In compiling the overall
estimate and error for the first bank of the
table we have excluded the results of 
\cite{Bonfim_1981} and the LPA data of
\cite{Zambelli_2017} due to the absence of 
errors. We provide two estimates in 5 dimensions.
The first includes the FRG result of \cite{Zambelli_2017} while the second omits it.
Clearly with the increase in the order of $\varepsilon$ the individual constrained estimates are in good
agreement for $4$ and $5$ dimensions but less so for $3$ dimensions. This is
primarily because the monotonic decrease in the value of $\sigma$ from $6$ down
to $2$ dimensions where it is positive in the critical dimension but negative
in $1$ and $2$ means that at some non-integer dimension it will vanish. This
appears to be in the neighbourhood of $3$ dimensions as indicated by the
relatively small value in the hundredths. Indeed this is similar to the
situation alluded to for $\eta$ in certain problems. Moreover the same feature
is present in $3$ dimensional estimates from the other methods we have provided
in Table~\ref{lysigmaests}. This is ultimately reflected in our final five loop
estimate and in particular in 
the error. This should not overlook however the very
good agreement for $4$ and $5$ dimensions.

One way of gauging the accuracy of our Lee--Yang exponents is 
to use them to provide exponent estimates in a related model. 
One such example is the lattice animal problem, reviewed in 
\cite{Stauffer:1978kr}, which at criticality is related to 
the Lee--Yang edge singularity problem, \cite{Parisi:1980ia}. 
In particular it was shown there that the Lee--Yang critical 
exponents in $d$ dimensions are the same as those of the 
lattice animal problem in $(d+2)$ dimensions. In 
\cite{Adler1988} estimates were given for the exponents
denoted by $\gamma_H$ and $\nu_H$ by calculating general 
dimension series up to the $15$th order. The relation these 
two exponents have to the Lee--Yang $\sigma$ exponent studied 
here is \cite{Parisi:1980ia},
\begin{equation}
\gamma_H ~=~ 1 ~-~ \sigma ~~~,~~~
\nu_H ~=~ \frac{(1 + \sigma)}{d} ~.
\end{equation}
With these we have compiled Table~\ref{tableanimal} which 
records our estimates for both exponents using our 5 loop
cKP17-2 Lee--Yang values for $\sigma$. The table also records
the summary results of Table~VII of \cite{Adler1988} for
comparison. It is reassuring to note that our estimates are 
consistent with \cite{Adler1988}.

{\begin{table}[h]
\begin{center}
\begin{tabular}{c||c|c||c|c}
$d$ & $\gamma_H$ this work & $\gamma_H$ of \cite{Adler1988} &
$\nu_H$ this work & $\nu_H$ of \cite{Adler1988} \\
\hline
$3$ & $0.925(6)$ & $0.90(3)$ & $0.358(2)$ & $0.367(11)$ \\
$4$ & $0.741(2)$ & $0.70(4)$ & $0.314(1)$ & $0.325(10)$ \\
$5$ & $0.6016(4)$ & $0.59(3)$ & $0.2796(1)$ & $0.282(6)$ \\
\end{tabular}
\end{center}
\begin{center}
\caption{Comparison of exponents $\gamma_H$ and $\nu_H$ for the lattice animal
problem derived from our Lee--Yang cKP17-2 values of $\sigma$.}
\label{tableanimal}
\end{center}
\end{table}}

\begin{figure}
	\centering
  \includegraphics[width=.9\textwidth]{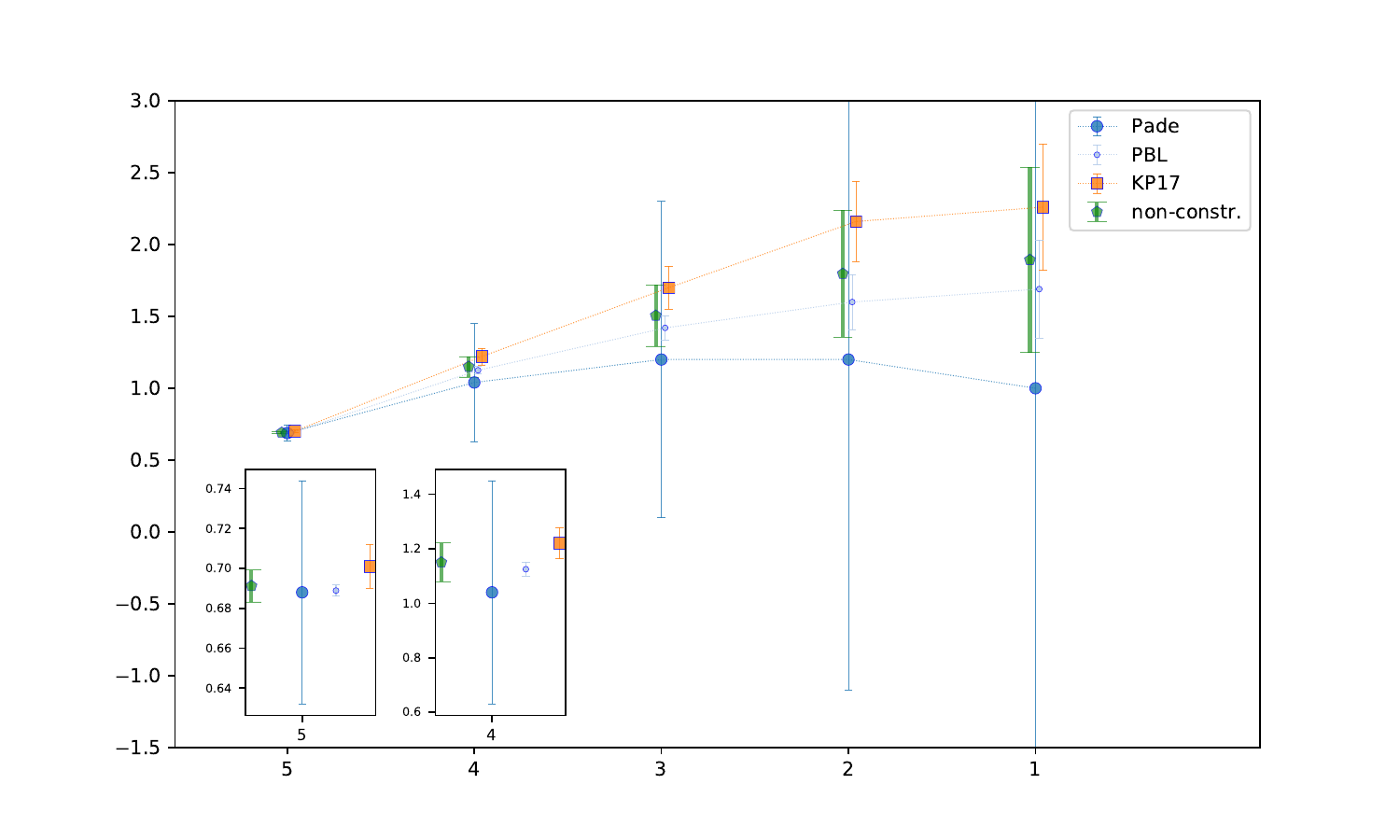}
	\caption{Plot of estimates for Lee--Yang $\omega$.}
	\label{perclyomegafig}
\end{figure}

{\begin{table}
\begin{center}
\begin{tabular}{c|c|lllll}
&Reference & $d$~$=$~$1$ & $d$~$=$~$2$ & $d$~$=$~$3$ & $d$~$=$~$4$ & $d$~$=$~$5$ \\
\cline{2-7}
\noalign{\vskip\doublerulesep
         \vskip-\arrayrulewidth}
\cline{2-7}

\multirow{5}{*}{\rotatebox[origin=c]{90}{4 loops} }& Pad\'{e}        & 1(4)  & 1(3)  & 1.3(1.4)     & 1.1(5)       & 0.69(10)     \\
&PBL         & 2.4(5)       & 2.1(3)       & 1.7(2)       & 1.22(6)      & 0.704(11)    \\
&KP17        & 2.3(6)       & 1.9(4)       & 1.6(3)       & 1.18(14)     & 0.69(3)      \\

\cline{2-7}
&non constr. & 2.3(8)       & 2.0(5)       & 1.6(3)       & 1.20(13)     & 0.70(2)      \\
&all         & 2.3(8)       & 2.0(5)       & 1.6(3)       & 1.20(13)     & 0.70(2)      \\

\hline
\hline

\multirow{5}{*}{\rotatebox[origin=c]{90}{5 loops} }& Pad\'{e}        & 1(4)  & 1(2)  & 1.2(1.1)     & 1.0(4)       & 0.69(6)      \\
& PBL         & 1.7(3)       & 1.6(2)       & 1.42(8)      & 1.12(3)      & 0.689(3)     \\
&KP17        & 2.3(4)       & 2.2(3)       & 1.70(15)     & 1.22(6)      & 0.701(11)    \\

\cline{2-7}
&non constr. & 1.9(6)       & 1.8(4)       & 1.5(2)       & 1.15(7)      & 0.691(8)     \\
&all         & 1.9(6)       & 1.8(4)       & 1.5(2)       & 1.15(7)      & 0.691(8)     \\

\end{tabular}
\end{center}
\caption{Estimates for Lee--Yang $\omega$.}
\label{lyomegaests}
\end{table}}

\clearpage

Finally the situation with the exponent $\omega$ is less clear. This is
primarily due to the lack of an exact value for this exponent in two
dimensions. Therefore we are only able to record the results of unconstrained
resummations which are given in Table~\ref{lyomegaests}. Equally we were
unable to compare the estimates with Monte Carlo studies. So no definite
conclusions should be drawn for $\omega$ in the Lee--Yang study.

\subsection{Percolation theory}

For percolation theory we recall the exact two dimensional values of the various exponents in
Table~\ref{tableexponentsd2} that were used when constraints were
implemented as indicated in Section~\ref{sec:resummation}.
We will use the 2 dimensional estimate
for the unconstrained resummations as a benchmark
to gauge whether our extrapolation is consistent
with the exact value. In several instances there
was either a significant undershoot or overshoot
for both the unconstrained and constrained case. 
This gave a strong indication that the higher
dimensional estimates could be unreliable. 
Consequently we examined the $\varepsilon$
expansion of various functions of the exponent,
such as its reciprocal. Some of these gave
improved estimates and a projection to 2
dimensions that was more in keeping with the
exact values. This is illustrated in
Table~\ref{tableexponentsd2} which also
summarizes our five loop two dimensional
estimates from the various resummation methods. 
The table also records the summary of \cite{percolationtable}; see also
\cite[Table~2]{stauffer2018introduction}. While
the exact values derive from
two dimensional conformal field theory we
note that a Monte Carlo study
\cite{Reynolds:1980zz} which predated
\cite{Friedan:1983xq} gave estimates that
were in remarkably good agreement. For instance,
the values of
$\alpha$~$=$~$-$~$0.708$~$\pm$~$0.030$,
$\beta$~$=$~$0.138(+0.006,-0.005)$,
$\gamma$~$=$~$2.432$~$\pm$~$0.035$,
$\delta$~$=$~$18.6$~$\pm$~$0.6$,
$\nu$~$=$~$1.354$~$\pm$~$0.015$ and
$\eta$~$=$~$0.204$~$\pm$~$0.006$ were
determined in \cite{Reynolds:1980zz}.

We note that here we use the value of $3/2$ for $\omega$
and $72/91$ for $\Omega$. This is in contrast with
the value of $2$ for $\omega$ used in \cite{Gracey:2015tta}
based on \cite{Nienhuis:1982} which also gave
an argument that $\Omega$ was $96/91$.
The discrepancy
between the exact 2 dimensional values for
$\omega$ and $\Omega$ was discussed at length in
\cite{ziff1}. In both cases the ratio $\omega/\Omega$
is the same and corresponds to the fractal dimension
of the system. Further a summary is given in
\cite{ziff1} of independent calculations of both
exponents. These appear to be more consistent with
the respective exact values of $3/2$ and $72/91$.
One test is the value of the exponent
$\Delta_1$~$=$~$\Omega/\sigma$ which is $2$
for $\Omega$~$=$~$72/91$ but $8/3$ for
\cite{Nienhuis:1982}. We note that naively
taking our central values gives $2.04545$ for
$\Delta_1$.

{ \renewcommand{\arraystretch}{1.4}
\begin{table}[h]
\begin{center}
\begin{tabular}{l||c|c||c||c|c|c|c}
         \multirow{2}{*}{exponent}& \multirow{2}{*}{exact}&\multirow{2}{*}{numerical}& {4 loop} & \multicolumn{4}{|c}{5 loop}\\
         \cline{4-8}
 & &  &  overall &Pad\'{e} & PBL & KP17& overall \\
 \hline
 $\alpha$ & $\nicefrac{-2}{3}$&  -0.667 &   -0.56(9) & -0.57(6)  &  -0.49(13) & -0.41(12) & -0.51(11) \\
 $\beta$ & $\nicefrac{5}{36}$&  0.139 &   0.2(3) & 0.3(3)  &  0.3(3) & 0.04(40) & 0.2(3) \\
 $\gamma$ & $\nicefrac{43}{18}$&  2.39 &   1.9(4) & 2.0(4)  &  1.8(3) & 2.2(3) & 2.0(4) \\
 $\delta$ & \multirow{2}{*}{$\nicefrac{91}{5}$}&  \multirow{2}{*}{18.2} &   4(2) &   &  3(5) & 4(2) & 4(3) \\
 $\delta$ $(1/\delta)$ & &   &   7(15) & 16.78  &  10(6) & -25(75) & 7(14) \\
 $\eta$ & $\nicefrac{5}{24}$&  0.208 &   -0.32(11) & -0.32(12)  &  -0.1(2) & -0.31(10) & -0.28(15) \\
 $\nu$ & \multirow{2}{*}{$\nicefrac{4}{3}$}&  \multirow{2}{*}{1.33} &   0.87(15) & 0.9315  &  0.86(13) & 0.8(3) & 0.9(2) \\
 $\nu$ $(1/\nu)$ & &   &   1.1(4) & 1.1(2)  &  0.99(13) & 1.17(12) & 1.1(2) \\
 $\sigma$ & $\nicefrac{36}{91}$&  0.396 &   0.46(4) & 0.44(4)  &  0.446(13) & 0.44(5) & 0.44(2) \\
 $\tau$ & $\nicefrac{197}{91}$&  2.06 &   2.08(10) & 2.07(4)  &  2.08(2) & 2.00(13) & 2.07(4) \\
  $\Omega$ & $\nicefrac{72}{91}$&  0.791 &   0.76(2) & 0.6(3)  &  0.96(5) &0.7(2) & 0.9(2) \\
  $\omega$ & $\nicefrac{3}{2}$&  1.5 &   2.6(2) & 2(2)  &  2.37(5) & 2.8(7) & 2.4(2) \\
\end{tabular}
\end{center}
\begin{center}
\caption{Exact values of exponents in $d$~$=$~$2$ together with our estimates.}
\label{tableexponentsd2}
\end{center}
\end{table}
}

We now discuss the results for each exponent in
alphabetical order. In Tables~\ref{percalphaests}--\ref{percomegaests} 
we have summarized the estimates for the various
exponents in the literature. One excellent
source we benefited from in this respect is the
table of \cite{percolationtable} which is regularly updated.
Included in this are our results for
the various resummations. In several of our
tables we also have a line designated as
$\varepsilon^4$. This records the results of the
constrained Pad\'{e} method at four loops given
in \cite{Gracey:2015tta}.

To open with, our estimates for $\alpha$ are presented in Table~\ref{percalphaests}. 
One aspect that is evident is the close agreement of the
estimates from both the constrained and unconstrained approaches 
for both 4 and
5 dimensions and to a lesser extent for $3$ dimensions. The latter is in essence
due to the overshoot of the two dimensional exact value for the unconstrained
methods. 
Indeed the five loop value has a more pronounced
discrepancy than at four loops.
Given the slower decrease in the constrained estimate as $d$ decreases, we would take the
position that those estimates are probably more reliable. 
Independent Monte Carlo data for each dimension
would be useful to compare with.

{\begin{table}[th]
\begin{center}
\begin{tabular}{c|c|llll}
&Reference & $d$~$=$~$2$ & $d$~$=$~$3$ & $d$~$=$~$4$ & $d$~$=$~$5$ \\
\cline{2-6}
\noalign{\vskip\doublerulesep
         \vskip-\arrayrulewidth}
\cline{2-6}         

&exact   & -0.6667      & & & \\ \hline
\hline
\multirow{10}{*}{\rotatebox[origin=c]{90}{4 loops} }
&Pad\'{e}        & -0.58(6)     & -0.67(4)     & -0.76(2)     & -0.873(3)    \\
&PBL         & -0.57(6)     & -0.66(4)     & -0.76(2)     & -0.872(3)    \\
&KP17        & -0.46(15)    & -0.60(7)     & -0.74(3)     & -0.869(4)    \\
&cPad\'{e}-2     &             & -0.708(11)   & -0.778(9)    & -0.875(3)    \\
&cPBL-2      &             & -0.710(13)   & -0.780(11)   & -0.876(4)    \\
&cKP17-2     &             & -0.72(2)     & -0.79(2)     & -0.874(6)    \\
&Constr.~Pad\'{e} &            & -0.7212(4)  & -0.782(14)  & -0.877(5)   \\

\cline{2-6}
&non constr. & -0.56(9)     & -0.65(5)     & -0.76(2)     & -0.871(4)    \\
&constr. &              & -0.720(3)    &-0.781(13)  &-0.876(4)    \\
&all         & -0.56(9)     & -0.719(12)   &-0.77(2)    &-0.874(5)    \\

\hline
\hline
\multirow{9}{*}{\rotatebox[origin=c]{90}{5 loops}}&Pad\'{e}        & -0.57(6)     & -0.66(3)     & -0.759(12)   & -0.872(2)    \\
&PBL         & -0.49(13)    & -0.62(6)     & -0.75(2)     & -0.871(2)    \\
&KP17        & -0.41(12)    & -0.59(4)     & -0.736(7)    & -0.8690(6)   \\
&cPad\'{e}-2     &             & -0.703(14)   & -0.773(9)    & -0.8739(13)  \\
&cPBL-2      &             & -0.69(4)     & -0.77(2)     & -0.872(4)    \\
&cKP17-2     &             & -0.72(2)     & -0.78(2)     & -0.870(6)    \\
&Constr.~Pad\'{e} &            & -0.723(2)   & -0.78(2)    & -0.880(10)  \\

\cline{2-6}
&non constr. & -0.51(11)    & -0.62(5)     & -0.745(15)   & -0.870(2)    \\
&constr. &              & -0.720(11)   &-0.77(3)    &-0.873(6)    \\
&all         & -0.51(11)    & -0.71(3)     &-0.76(2)    &-0.870(3)    \\

\end{tabular}
\end{center}
\caption{Estimates for $\alpha$ in percolation.}
\label{percalphaests}
\end{table}}

\begin{figure}[h]
	\centering
  \includegraphics[width=.9\textwidth]{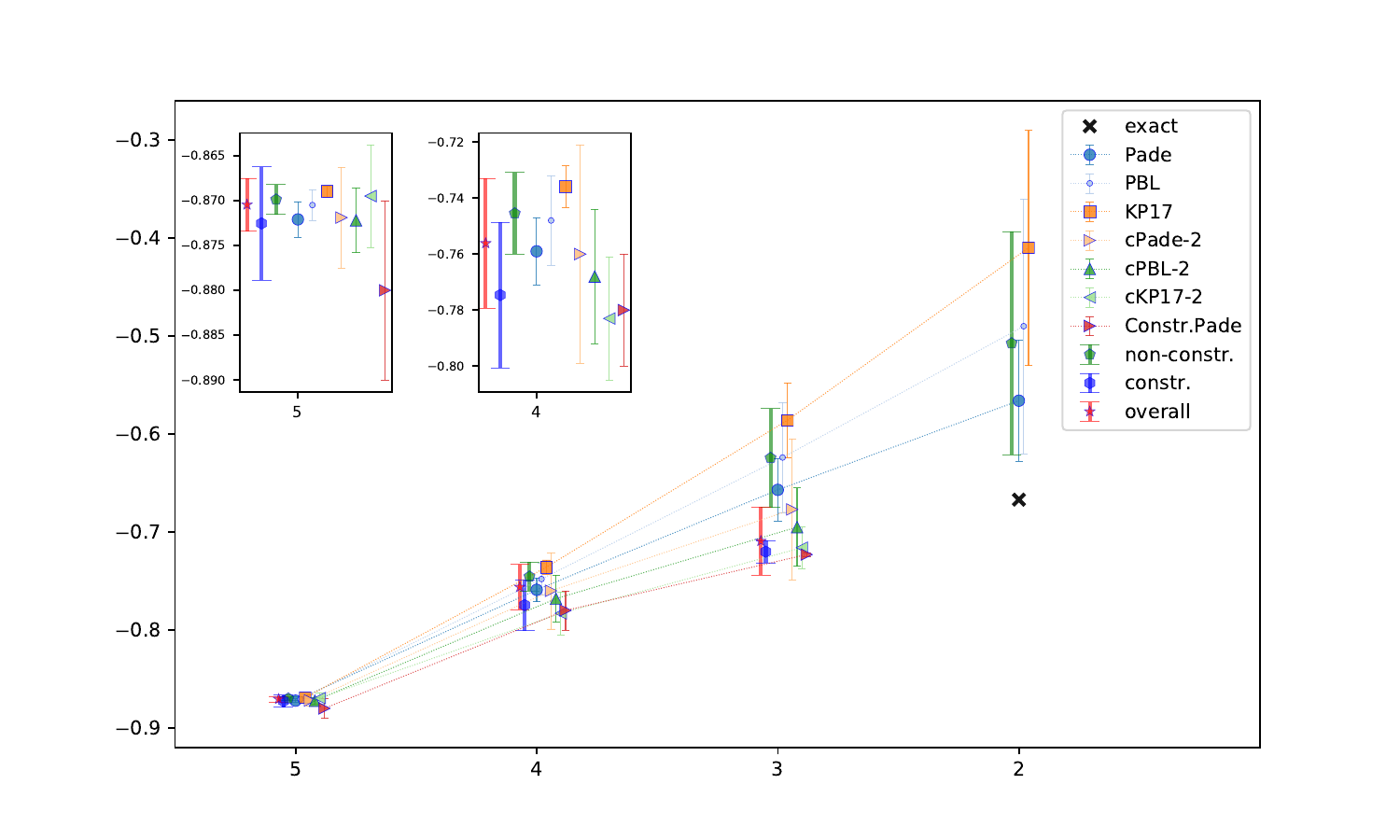}
	\caption{Plot of estimates for $\alpha$ in percolation.}
	\label{percalphafig}
\end{figure}

For $\beta$ the situation is much clearer in that we have several independent
results provided in Table~\ref{percbetaests}. Again there is a large degree of
stability within each of the comparable methods and loop order although the
unconstrained Pad\'{e} clearly overshoots the two dimensional exact result.
From four to five loops the final estimates are remarkably stable. Comparing,
however, with \cite{Adler:1990zz,Sur1976} the loop estimates are all slightly
larger even for 5 dimensions. 

{\begin{table}[h]
\begin{center}
\begin{tabular}{c|c|llll}
&Reference & $d$~$=$~$2$ & $d$~$=$~$3$ & $d$~$=$~$4$ & $d$~$=$~$5$ \\
\cline{2-6}
\noalign{\vskip\doublerulesep
         \vskip-\arrayrulewidth}
\cline{2-6}         
& exact   & 0.1389       & & & \\ 
\hline
\hline
$\varepsilon^4$& \multicolumn{1}{r|}{\cite{Gracey:2015tta} (2015)} & & 0.4273 & 0.6590 & 0.8457 \\
\hline
\hline
\multirow{3}{*}{\rotatebox[origin=c]{90}{MC/srs} }
& \multicolumn{1}{r|}{\cite{Sur1976} (1976)}&  & 0.41(1) & & \\
& \multicolumn{1}{r|}{\cite{Adler:1990zz} (1990)} & & 0.405(25) & 0.639(20) & 0.835(5) \\
\cline{2-6}
& overall &            & 0.409(14)   & 0.64(2)     & 0.835(5)    \\

\hline
\hline

\multirow{9}{*}{\rotatebox[origin=c]{90}{4 loops} }& Pad\'{e}        & 0.3(3)  & 0.49(13)     & 0.68(4)      & 0.847(4)     \\
& KP17        & 0.0(4)  & 0.40(9)      & 0.65(2)      & 0.845(4)     \\
& cPad\'{e}-2     &             & 0.426(8)     & 0.657(5)     & 0.8453(12)   \\
& cPBL-2      &             & 0.419(7)     & 0.653(5)     & 0.8444(12)   \\
& cKP17-2     &             & 0.424(11)    & 0.656(7)     & 0.845(2)     \\
& Constr.~Pad\'{e} &            & 0.4303 & 0.6599(2)   & 0.8458      \\

\cline{2-6}
& non constr. & 0.2(3)  & 0.43(11)     & 0.66(3)      & 0.846(4)     \\
& constr.     &              & 0.423(9)     & 0.656(6)     & 0.8449(14)   \\
& all         & 0.2(3)  & 0.42(2)      & 0.656(10)    & 0.845(2)     \\

\hline
\hline
\multirow{9}{*}{\rotatebox[origin=c]{90}{5 loops} }& Pad\'{e}        & 0.3(3)  & 0.49(13)     & 0.68(4)      & 0.845(4)     \\
& PBL         & 0.3(3)       & 0.48(13)     & 0.67(4)      & 0.847(4)     \\
& KP17        & 0.0(4)  & 0.39(6)      & 0.650(11)    & 0.8447(10)   \\
& cPad\'{e}-2     &             & 0.427(6)     & 0.658(4)     & 0.8455(6)    \\
& cPBL-2      &             & 0.417(5)     & 0.653(2)     & 0.8448(2)    \\
& cKP17-2     &             & 0.422(5)     & 0.655(3)     & 0.8450(5)    \\
& Constr.~Pad\'{e} &            & 0.42(2)     & 0.656(12)   & 0.8453(15)  \\

\cline{2-6}
& non constr. & 0.2(3)  & 0.44(10)     & 0.66(2)      & 0.845(2)     \\
&constr. &              & 0.420(8)     &0.655(4)    &0.8449(5)    \\
&all &      0.2(3)    & 0.42(2)      &0.655(7)    &0.8449(7)    \\

\end{tabular}
\end{center}
\caption{Estimates for $\beta$ in percolation.}
\label{percbetaests}
\end{table}}

\begin{figure}[h]
	\centering
  \includegraphics[width=.9\textwidth]{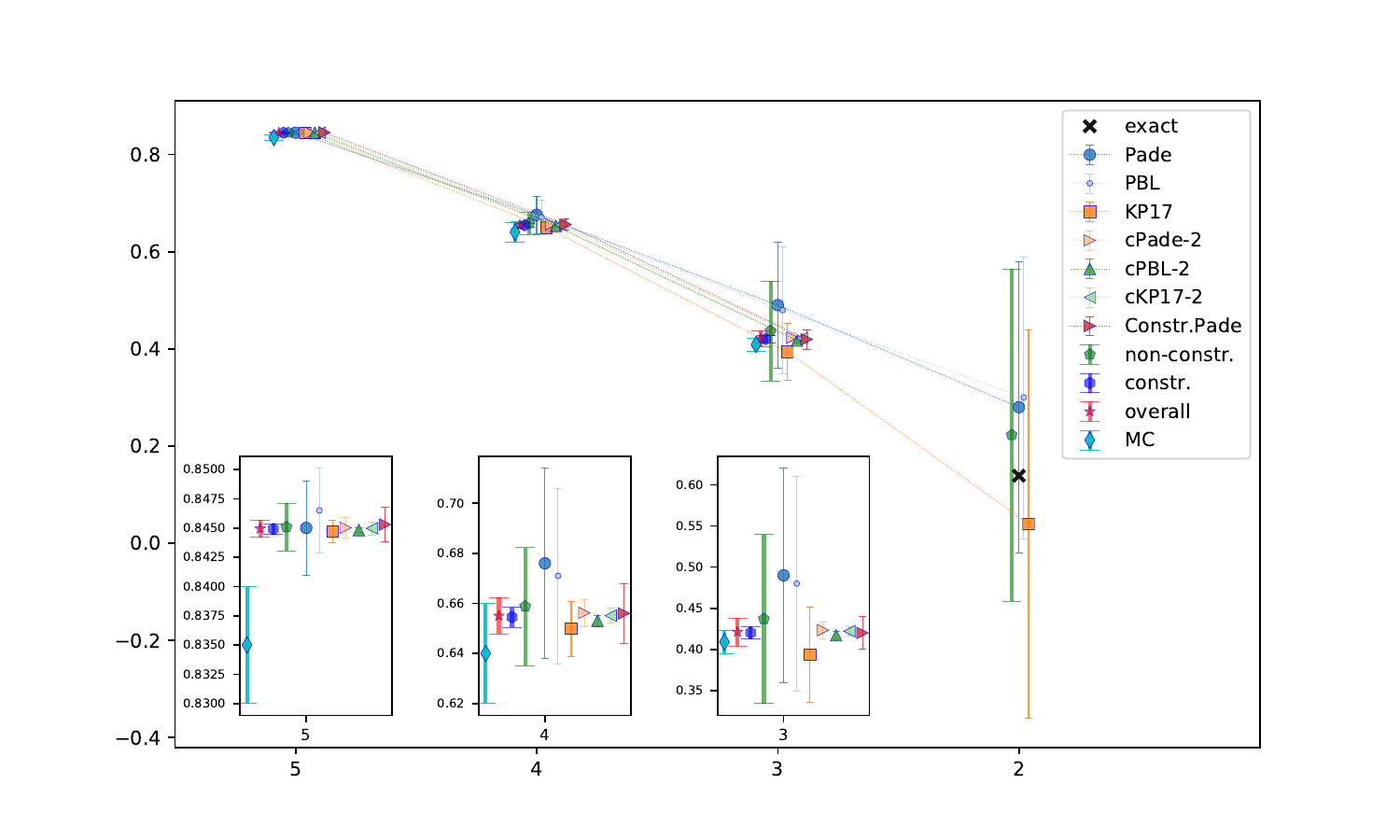}
	\caption{Plot of estimates for $\beta$ in percolation.}
	\label{percbetafig}
\end{figure}

{\begin{table}[h]
\begin{center}
\begin{tabular}{c|c|llll}
&Reference & $d$~$=$~$2$& $d$~$=$~$3$ & $d$~$=$~$4$ & $d$~$=$~$5$ \\
\cline{2-6}
\noalign{\vskip\doublerulesep
         \vskip-\arrayrulewidth}
\cline{2-6} 
&exact   & 2.3889       & & & \\ 
\hline
\hline
$\varepsilon^4$& \multicolumn{1}{r|}{\cite{Gracey:2015tta} (2015)}&  & 1.8357 & 1.4500 & 1.1817 \\
\hline
\hline
\multirow{3}{*}{\rotatebox[origin=c]{90}{MC/srs} }
&\multicolumn{1}{r|}{\cite{Sur1976} (1976)} &  & 1.6 & & \\
& \multicolumn{1}{r|}{\cite{Adler:1990zz} (1990)}&  & 1.805(20) & 1.435(15) & 1.185(5) \\
\cline{2-6} 
& overall &            & 1.805(20)     & 1.435(15)   & 1.185(5)    \\
\hline
\hline

\multirow{9}{*}{\rotatebox[origin=c]{90}{4 loops} }& Pad\'{e}        & 1.946        & 1.7(3)       & 1.44(6)      & 1.179(7)     \\
&PBL         & 1.9(3)       & 1.65(14)     & 1.40(5)      & 1.175(7)     \\
&KP17        & 1.9(8)       & 1.8(2)       & 1.44(6)      & 1.181(4)     \\
&cPad\'{e}-2     &             & 1.83(3)      & 1.45(2)      & 1.181(3)     \\
&cKP17-2     &             & 1.82(6)      & 1.44(4)      & 1.180(6)     \\
&Constr.~Pad\'{e} &            & 1.82(4)     & 1.44(2)     & 1.180(3)    \\

\cline{2-6} 
&non constr. & 1.9(4)       & 1.7(2)       & 1.42(6)      & 1.179(6)     \\
&constr. &              & 1.82(4)      &1.44(2)     &1.180(4)     \\
&all &  1.9(4)          & 1.81(8)      &1.44(3)     &1.180(5)     \\

\hline
\hline
\multirow{10}{*}{\rotatebox[origin=c]{90}{5 loops} } &Pad\'{e}        & 2.0(4)        & 1.8(2)       & 1.45(4)      & 1.181(4)     \\
&PBL         & 1.8(3)       & 1.6(2)       & 1.38(4)      & 1.176(3)     \\
&KP17        & 2.2(3)       & 1.77(13)     & 1.43(2)      & 1.179(2)     \\
&cPad\'{e}-2     &             & 1.83(3)      & 1.448(13)    & 1.181(2)     \\
&cPBL-2      &             & 1.83(4)      & 1.45(2)      & 1.181(3)     \\
&cKP17-2     &             & 1.81(5)      & 1.437(13)    & 1.1797(14)   \\
&Constr.~Pad\'{e} &            & 1.83(3)     & 1.444(14)   & 1.181(2)    \\

\cline{2-6} 
&non constr. & 2.0(4)       & 1.7(2)       & 1.42(4)      & 1.179(3)     \\
&constr. &              & 1.82(4)      &1.44(2)     &1.180(2)     \\
&all &    2.0(4)    & 1.81(8)      &1.44(3)     &1.180(3)     \\

\end{tabular}
\end{center}
\caption{Estimates for $\gamma$ in percolation.}
\label{percgammaests}
\end{table}}

\begin{figure}[h]
	\centering
  \includegraphics[width=.9\textwidth]{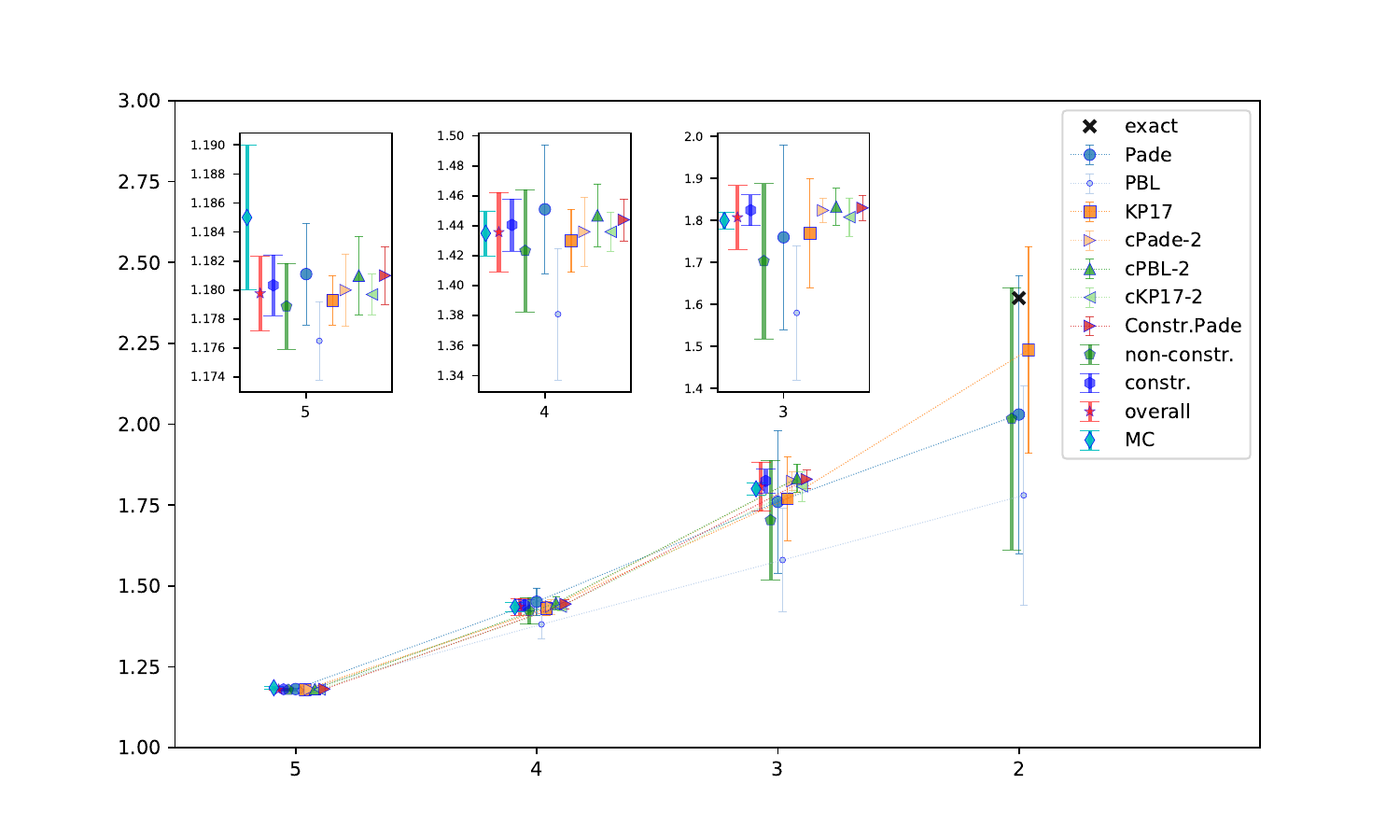}
	\caption{Plot of estimates for $\gamma$ in percolation.}
	\label{percgammafig}
\end{figure}

In the case of $\gamma$ the independent Monte Carlo data was only available
from the same articles as our comparison for $\beta$. There was a similar
picture for our estimates in that there was remarkable stability between four
and five loops for the three dimensions of interest as is clear from Table~\ref{percgammaests}. 
However the check estimate for two dimensions was much
closer to the exact value than that for $\beta$.
This may be one of the reasons why our final
values for five loops are in remarkably good agreement with the Monte Carlo
results of \cite{Adler:1990zz}. There was no error bar for the 3 dimensional
result of \cite{Sur1976}. So it is not clear how to interpret the central value
that is 12\% lower than both \cite{Adler:1990zz} and our estimate at five
loops.

{\begin{table}[h]
\begin{center}
\begin{tabular}{c|c|llll}
&Reference & $d$~$=$~$2$  & $d$~$=$~$3$ & $d$~$=$~$4$ & $d$~$=$~$5$ \\
\cline{2-6}
\noalign{\vskip\doublerulesep
         \vskip-\arrayrulewidth}
\cline{2-6} 
& exact   & 18.2         & & & \\ 
\hline
\hline
\multirow{4}{*}{\rotatebox[origin=c]{90}{MC/srs} }
&\multicolumn{1}{r|}{\cite{Sur1976} (1976)}&  & & 3.198(6) & \\
& \multicolumn{1}{r|}{\cite{Nakanishi:1980zz} (1980)} &  & 5.3 & 3.9 & 3.0 \\
&\multicolumn{1}{r|}{\cite{Lorenz:1998zz} (1998)}&  & 5.29(6) & & \\
\cline{2-6}

& overall &            & 5.29(6)     & 3.198(6)    &            \\
\hline
\hline

\hline
\multirow{9}{*}{\rotatebox[origin=c]{90}{4 loops} }& Pad\'{e}        & 4.756        & 3.742        & 3.1(2)       & 2.391(14)    \\
&KP17        & 4(2)         & 3.3(1.3)     & 3.1(4)       & 2.39(3)      \\
&cPad\'{e}-2     &             & 7.831        & 3.2(3)       & 2.40(2)      \\
&cPBL-2      &             & 11(4)        & 6(4)         & 2.6(4)       \\
&cKP17-2     &             & 10(4)        & 3(2)         & 2.40(11)     \\
&Constr.~Pad\'{e} &            & 5.1(2)      & 3.15(5)     & 2.393(5)    \\

\cline{2-6}
&non constr. & 4(2)         & 3.3(1.3)     & 3.1(3)       & 2.39(2)      \\
&constr. &              & 6(2)         &3.2(3)      &2.40(2)      \\
&all &  4(2)     & 5(2)         &3.2(3)      &2.40(2)      \\

\hline
\hline
\multirow{9}{*}{\rotatebox[origin=c]{90}{5 loops} }&Pad\'{e}        &             &             & 3.15(8)      & 2.396(5)     \\
&PBL         & 3(5)  & 3(3)  & 2.5(1.5)     & 2.3(2)       \\
&KP17        & 4(2)         & 3.3(1.5)     & 3.2(2)       & 2.396(6)     \\
&cPad\'{e}-2     &             & 7.831        & 3.3(3)       & 2.40(2)      \\
&cPBL-2      &             & 8.92(5)      & 3.84(2)      & 2.4061(12)   \\
&cKP17-2     &             & 10(4)        & 3.3(1.3)     & 2.40(5)      \\
&Constr.~Pad\'{e} &            & 5.16(7)     & 3.175(14)   & 2.3954(11)  \\

\cline{2-6}
&non constr. & 4(3)         & 3(2)         & 3.1(2)       & 2.395(12)    \\
&constr. &              & 7(2)         &3.4(3)      &2.401(6)     \\
&overall &  4(3)    & 7(2)         &3.4(3)      &2.400(7)     \\

\end{tabular}
\end{center}
\caption{Estimates for $\delta$ in percolation.}
\label{percdeltaests1}
\end{table}}

The situation with $\delta$ was much less
straightforward and we include two tables with
data.
In Table~\ref{percdeltaests1} we have applied our various resummation techniques directly to the
$\varepsilon$ expansion of $\delta$.
This illustrates one immediate difficulty in
obtaining reliable values for low $d$. 
The value for $\delta$ in exactly 6 dimensions is
$2$ whereas it is exactly $18.2$ in two dimensions. The latter places a big
restriction on the ability of the $\varepsilon$ expansion to manufacture large
corrections as 
$d$ decreases. This is clear in both our four and five loop
estimates from both our unconstrained and constrained resummations. Moreover
the skew that is necessary to meet the two dimensional constraint clearly
makes the convergence between loops in 3 dimensions impossible. Equally there
is no agreement with the few independent estimates which both have more
precise values. Although the direct $\varepsilon$ expansion approach is
problematic we have provided it here partly
to be transparent in the
application of our methods but also to highlight the hidden pitfalls in
naively applying various resummation methods. Equally to address this problem
we have instead tried a second approach which was to sum the $\varepsilon$
expansion of $1/\delta$, determine a numerical value and then invert this.
The results of this exercise are given in Table~\ref{percdeltaests2}. Many of
the failings with the direct evaluation are absent now. There is a degree of
stability between the various four and five loop results especially in 3
dimensions. Equally and more importantly we find 
that this and the 4 dimensional
results are certainly not out of line with the results of
\cite{Lorenz:1998zz,Sur1976} although our 3 dimensional result has large
errors. While the extrapolation to 2 dimensions has larger errors it does
accommodate the exact value. In some sense this much more consistent agreement
across Table~\ref{percdeltaests2} is an a posteriori justification of following
this second strategy.

{\begin{table}[h]
\begin{center}
\begin{tabular}{c|c|llll}
&Reference & $d$~$=$~$2$  & $d$~$=$~$3$ & $d$~$=$~$4$ & $d$~$=$~$5$ \\
\cline{2-6}
\noalign{\vskip\doublerulesep
         \vskip-\arrayrulewidth}
\cline{2-6} 
& exact   & 18.2         & & & \\ 
\hline
\hline
\multirow{4}{*}{\rotatebox[origin=c]{90}{MC/srs} }
&\multicolumn{1}{r|}{\cite{Sur1976} (1976)}&  & & 3.198(6) & \\
& \multicolumn{1}{r|}{\cite{Nakanishi:1980zz} (1980)} &  & 5.3 & 3.9 & 3.0 \\
&\multicolumn{1}{r|}{\cite{Lorenz:1998zz} (1998)}&  & 5.29(6) & & \\

\cline{2-6}

&overall &            & 5.29(6)     & 3.198(6)    &            \\
\hline
\hline

\multirow{9}{*}{\rotatebox[origin=c]{90}{4 loops} }& Pad\'{e}        & 8(10)        & 5(2)         & 3.1(2)       & 2.391(14)    \\
&PBL         & 8(10)        & 4.4(1.4)     & 3.1(2)       & 2.39(2)      \\
&KP17        & -20(120) & 6.3(1.5)     & 3.25(9)      & 2.399(4)     \\
&cPad\'{e}-2     &             & 5.13(11)     & 3.17(3)      & 2.395(3)     \\
&cPBL-2      &             & 5.19(8)      & 3.18(2)      & 2.396(2)     \\
&cKP17-2     &             & 5.15(12)     & 3.17(3)      & 2.395(5)     \\
&Constr.~Pad\'{e}  &            & 5.2(2)      & 3.17(6)     & 2.394(6)    \\

\cline{2-6}
&non constr. & 7(15)        & 5(2)         & 3.2(2)       & 2.395(9)     \\
&constr.     &              & 5.17(12)     & 3.18(3)      & 2.395(3)     \\
&all         & 7(15)        & 5.2(3)       & 3.18(5)      & 2.395(5)     \\

\hline
\hline
\multirow{9}{*}{\rotatebox[origin=c]{90}{5 loops} }&Pad\'{e}        & 16.78        & 6(2)         & 3.22(15)     & 2.397(7)     \\
&PBL         & 10(6)        & 4.7(6)       & 3.13(7)      & 2.393(3)     \\
&KP17        &  {-25(75)} & 6.1(1.5)     & 3.25(9)      & 2.398(4)     \\
&cPad\'{e}-2     &             & 5.16(5)      & 3.179(14)    & 2.3956(12)   \\
&cPBL-2      &             & 5.18(7)      & 3.179(14)    & 2.3956(11)   \\
&cKP17-2     &             & 5.16(3)      & 3.175(8)     & 2.3952(10)   \\
&Constr.~Pad\'{e}  &            & 5.1(2)      & 3.17(3)     & 2.395(2)    \\

\cline{2-6}
&non constr. & 7(14)        & 5.2(1.2)     & 3.19(11)     & 2.396(5)     \\
&constr.     &              & 5.16(6)      & 3.176(14)    & 2.3954(12)   \\
&all         & 7(14)        & 5.16(15)     & 3.18(3)      & 2.396(2)     \\

\end{tabular}
\end{center}
\caption{Estimates for $\delta$ in percolation using the series for $1/\delta$.}
\label{percdeltaests2}
\end{table}}

\begin{figure}[h]
	\centering
  \includegraphics[width=.9\textwidth]{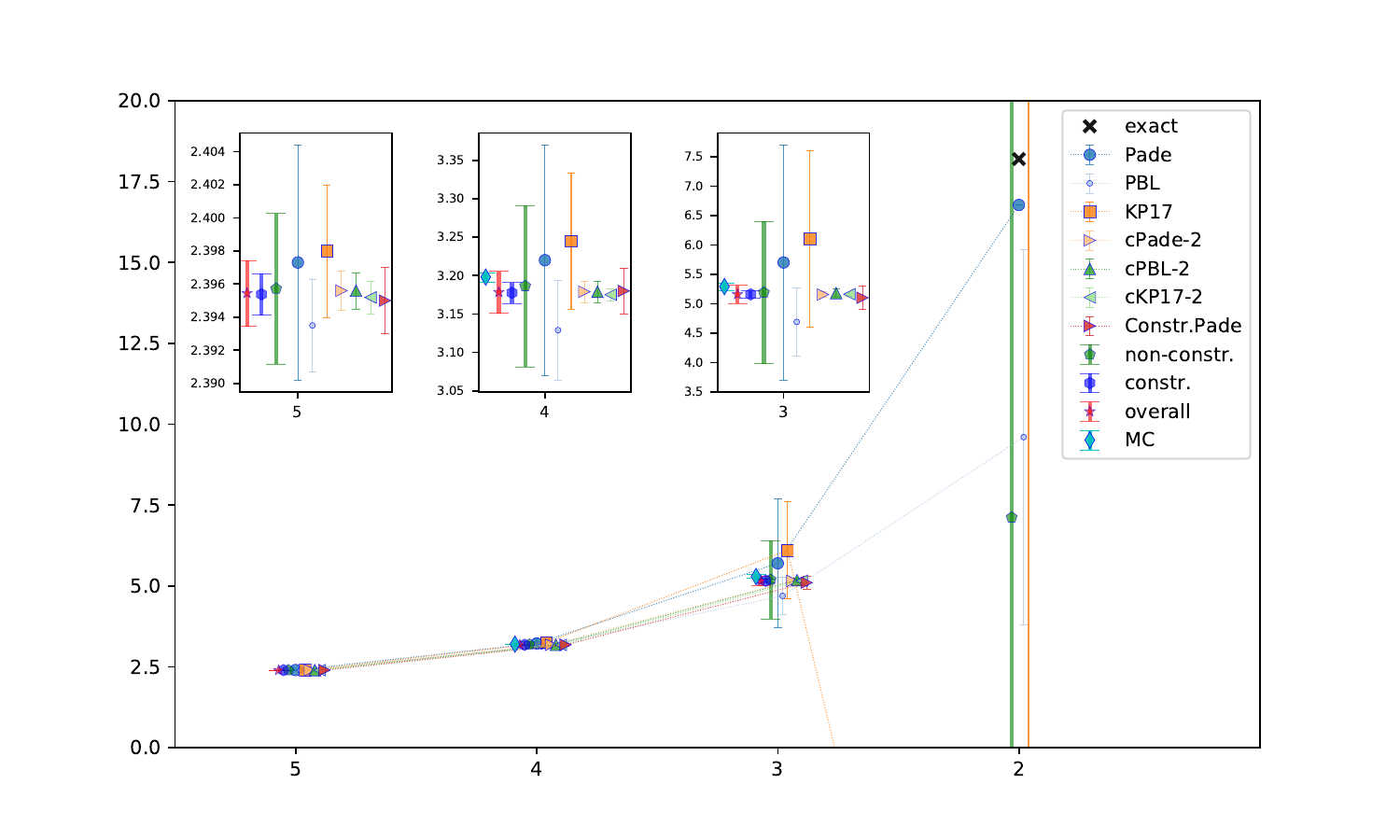}
	\caption{Plot of estimates (from $1/\delta$) for $\delta$ in percolation.}
	\label{percdeltafig}
\end{figure}

{\begin{table}[h]
\begin{center}
\begin{tabular}{c|c|llll}
&Reference & $d$~$=$~$3$ & $d$~$=$~$3$ & $d$~$=$~$4$ & $d$~$=$~$5$ \\
\cline{2-6}
\noalign{\vskip\doublerulesep
         \vskip-\arrayrulewidth}
\cline{2-6}
&exact   & 0.2083       & & & \\ 
\hline
\hline

$\varepsilon^4$& \multicolumn{1}{r|}{\cite{Gracey:2015tta} (2015)}  & & -0.0470 & -0.0954 & -0.0565 \\
\hline
\hline
\multirow{6}{*}{\rotatebox[origin=c]{90}{MC/srs} }
&\multicolumn{1}{r|}{\cite{Adler:1990zz} (1990)}&  & -0.07(5) & -0.12(4) & -0.075(20) \\
&\multicolumn{1}{r|}{\cite{Ballesteros:1996ct} (1996)} & & & -0.0944(28) & \\
&\multicolumn{1}{r|}{\cite{Lorenz:1998zz} (1998)} & & -0.046(8) & & \\
&\multicolumn{1}{r|}{\cite{Jan1998} (1998)} & & -0.059(9) & & \\
&\multicolumn{1}{r|}{\cite{Tiggemann2001} (2001)} & & & -0.0929(9) & \\
\cline{2-6}
& overall &            & -0.054(14)  & -0.094(4)   & -0.075(20)    \\
\hline
\hline

\multirow{8}{*}{\rotatebox[origin=c]{90}{4 loops} }&Pad\'{e}        & -0.32(13)    & -0.21(6)     & -0.13(2)     & -0.059(3)    \\
&KP17        & -0.33(10)    & -0.23(5)     & -0.14(2)     & -0.061(3)    \\
&cPad\'{e}-2     &             & 0.01143      & -0.06287     & -0.055(7)    \\
&cPBL-2      &             & 0.04(9)  & -0.06(3)     & -0.051(6)    \\
&cKP17-2     &             & -0.09(12)    & -0.10(3)     & -0.057(3)    \\

\cline{2-6}
&non constr. & -0.32(11)    & -0.22(5)     & -0.13(2)     & -0.060(3)    \\
&constr.     &              & -0.02(12)    & -0.08(4)     & -0.055(5)    \\
&all         & -0.32(11)    & -0.15(13)    & -0.12(4)     & -0.057(5)    \\

\hline
\hline

\multirow{11}{*}{\rotatebox[origin=c]{90}{5 loops} }&Pad\'{e}        & -0.32(12)    & -0.22(5)     & -0.13(2)     & -0.059(2)    \\
&PBL         & -0.1(2)  & -0.14(10)    & -0.11(3)     & -0.057(3)    \\
&KP17        & -0.31(10)    & -0.23(5)     & -0.14(2)     & -0.061(3)    \\
&cPad\'{e}-2  &             & -0.003(13)    & -0.074(7)     & -0.056(4)    \\
&cPBL-2      &             & 0.06(7)  & -0.03(5)  & -0.049(5)    \\
&cKP17-2     &             & -0.04(2)     & -0.091(8)    & -0.0553(9)   \\
&Scaling Pad\'{e}  & & (-0.0566) & (-0.0833) & (-0.0545) \\
&Hyperscaling Pad\'{e} &  & (-0.0176) & (-0.0790) & (-0.0540) \\
\cline{2-6}
&non constr. & -0.28(15)    & -0.20(7)     & -0.13(2)     & -0.059(3)    \\
&constr. &              & -0.01(4)  &-0.08(2)    &-0.055(3)    \\
&all &  -0.28(15)    & -0.06(10)    &-0.10(3)    &-0.056(3)    \\

\end{tabular}
\end{center}
\caption{Estimates for $\eta$ in percolation.}
\label{percetaests1}
\end{table}}

\begin{figure}[h]
	\centering
  \includegraphics[width=.9\textwidth]{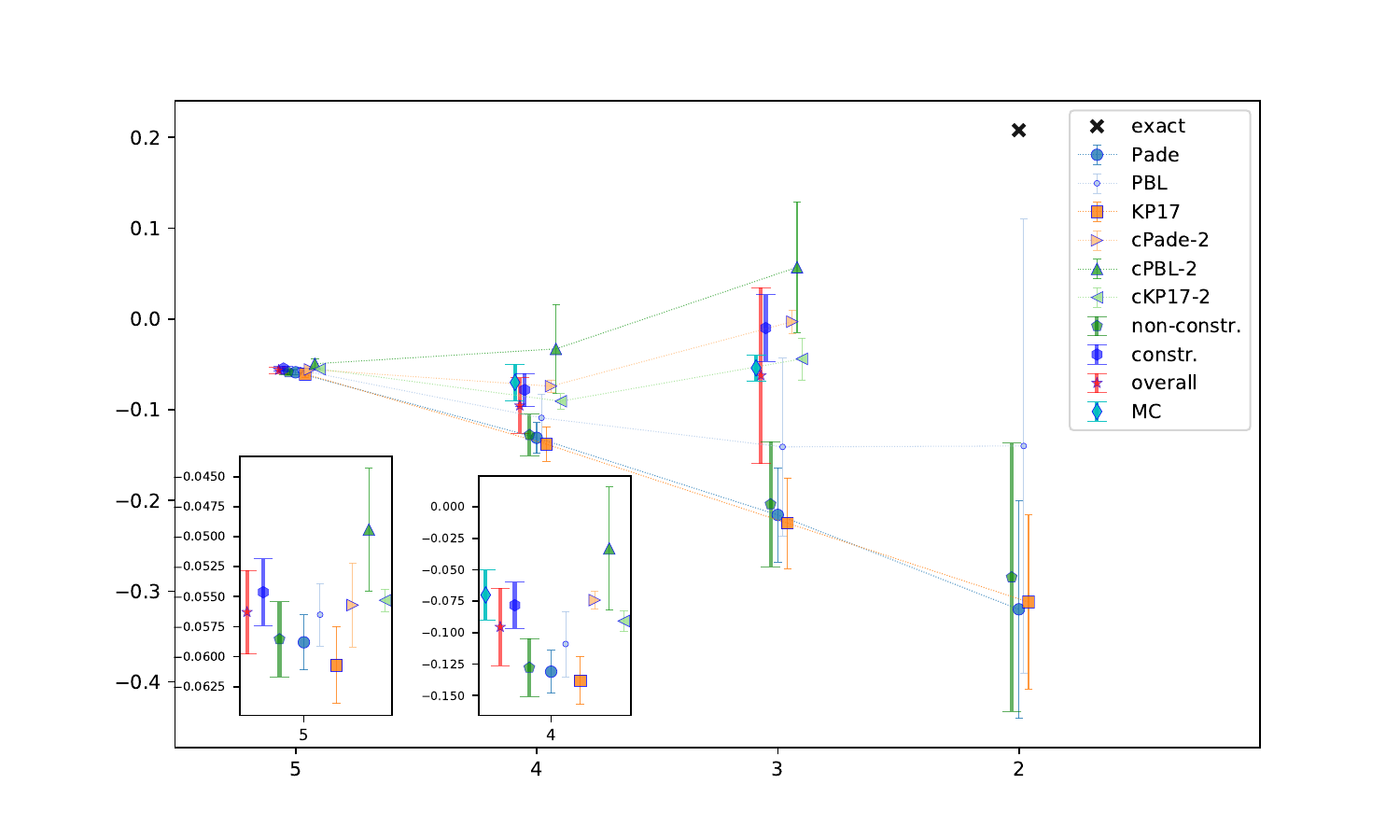}
	\caption{Plot of estimates for $\eta$ in percolation.}
	\label{percetafig}
\end{figure}

We now turn to one of the more widely measured 
exponents which is $\eta$ with data
recorded in Table~\ref{percetaests1}.
As is the case with $\eta$ in other problems its
value in percolation is relatively small and
close to zero being associated with the wave
function anomalous dimension. Consequently it is difficult to obtain an accurate value
for it that is in consistent agreement with other
estimates. This is evident in the numerical
values of
\cite{Adler:1990zz,Lorenz:1998zz,Jan1998,Tiggemann2001,Ballesteros:1996ct}
in 3 and 4 dimensions although those of \cite{Tiggemann2001,Ballesteros:1996ct}
are the most accurate and do overlap. On the $\varepsilon$ expansion side the
Pad\'{e} resummations suffered from the presence
of at least one singularity in 
$2$~$<$~$d$~$<$~$6$ or were not monotonic. For
this case 
we indirectly estimated $\eta$ using 
the respective scaling and hyperscaling
relations
\begin{equation}
\eta ~=~ 2 ~-~ \frac{\gamma}{\nu} ~,\quad
\eta ~=~ \frac{ d + 2 - (d-2) \delta }{1+\delta} ~.
\end{equation}
The numerical values for each of these are
bracketed to indicate that they were not derived directly. A more difficult issue to circumvent
for $\eta$ in the constrained approach is the
relatively large and positive exact value in 2
dimensions. This means that $\eta$ would have to
be identically zero somewhere between $3$ and $2$
dimensions as $d$ decreases. So while our final 3
dimensional estimate of $\eta$ has shown an
improvement from four to
five loops this is to be balanced against a necessarily large error.

{\begin{table}[h]
\begin{center}
\begin{tabular}{c|c|llll}
&Reference & $d$~$=$~$2$ & $d$~$=$~$3$ & $d$~$=$~$4$ & $d$~$=$~$5$ \\
\cline{2-6}
\noalign{\vskip\doublerulesep
         \vskip-\arrayrulewidth}
\cline{2-6} 
&exact   & 1.3333       & & & \\ 
\hline
\hline
$\varepsilon^4$&\multicolumn{1}{r|}{\cite{Gracey:2015tta} (2015)}&  & 0.8960 & 0.6920 & 0.5746 \\
\hline
\hline
\multirow{13}{*}{\rotatebox[origin=c]{90}{MC/srs} }& \multicolumn{1}{r|}{\cite{Levenshtein1975} (1975)} &  & 0.80(5) & & \\
& \multicolumn{1}{r|}{\cite{Sur1976} (1976)}&  & 0.8(1) & & \\
& \multicolumn{1}{r|}{\cite{Adler:1990zz} (1990) }&  & 0.872(70) & 0.678(50) & 0.571(3) \\
&\multicolumn{1}{r|}{\cite{Ballesteros:1996ct} (1997) }&  & & 0.689(10) & \\
&\multicolumn{1}{r|}{\cite{Lorenz:1998zz} (1998) }&  & 0.875(1) & & \\
&\multicolumn{1}{r|}{\cite{Ballesteros:1999fx} (2000)}&  & 0.8765(18) & & \\
&\multicolumn{1}{r|}{\cite{Wang2013} (2013)} &  & 0.8764(12) & & \\
&\multicolumn{1}{r|}{\cite{Hu2014} (2014)}&  & 0.8751(11) & & \\
&\multicolumn{1}{r|}{\cite{Xu2014} (2014)}&  & 0.87619(11) & & \\
&\multicolumn{1}{r|}{\cite{Koza2016} (2016)} &  & 0.8774(13) & 0.6852(28) & 0.5723(18) \\
&\multicolumn{1}{r|}{\cite{LeClair:2018edq} (2018)}&  & & 0.693 & \\
&\multicolumn{1}{r|}{\cite{Tan:2020htb} (2020)}&  & & 0.6845(6) & 0.5757(7) \\
\cline{2-6}
& overall &            & 0.876(4)    & 0.685(2)    & 0.574(2)    \\
\hline
\hline
\multirow{9}{*}{\rotatebox[origin=c]{90}{4 loops} }& Pad\'{e}        & 0.8812       & 0.78(8)      & 0.68(2)      & 0.573(2)     \\
&PBL         & 0.88(10)     & 0.77(5)      & 0.66(2)      & 0.572(3)     \\
&KP17        & 0.8(3)       & 0.9(2)       & 0.68(3)      & 0.574(3)     \\
&cPad\'{e}-2     &             & 0.9496       & 0.697(15)    & 0.575(2)     \\
&cKP17-2     &             & 0.95(12)     & 0.69(4)      & 0.574(3)     \\
&Constr.~Pad\'{e}  &            & 0.896(3)    & 0.6916(12)  & 0.5746(2)   \\

\cline{2-6}
&non constr. & 0.87(15)     & 0.79(9)      & 0.67(2)      & 0.573(3)     \\
&constr. &              & 0.897(10)    &0.692(4)    &0.5746(5)    \\
&all &  0.87(15)    & 0.89(4)      &0.690(9)    &0.5744(11)   \\

\hline
\hline
\multirow{9}{*}{\rotatebox[origin=c]{90}{5 loops} }& Pad\'{e}        & 0.9315       & 0.81(5)      & 0.681(13)    & 0.5737(10)   \\
&PBL         & 0.86(13)     & 0.78(4)      & 0.669(6)     & 0.5730(2)    \\
&KP17        & 0.8(3)       & 0.85(8)      & 0.682(11)    & 0.5737(7)    \\
&cPad\'{e}-2     &             & 0.9496       & 0.697(15)    & 0.575(2)     \\
&cPBL-2      &             & 0.93(3)      & 0.703(14)    & 0.575(2)     \\
&cKP17-2     &             & 0.88(7)      & 0.69(2)      & 0.5738(12)   \\
&Constr.~Pad\'{e}  &            & 0.89(2)     & 0.689(6)    & 0.5743(7)   \\

\cline{2-6}
&non constr. & 0.9(2)       & 0.80(6)      & 0.675(11)    & 0.5732(5)    \\
&constr. &              & 0.90(4)      &0.692(13)   &0.563(13)    \\
&all &  0.9(2)    & 0.87(7)      &0.685(15)   &0.570(9)     \\
\end{tabular}
\end{center}
\caption{Estimates for $\nu$ in percolation.}
\label{percnuests1}
\end{table}}

The situation for $\nu$ is similar as can be seen in Tables~\ref{percnuests1}
and \ref{percnuests2} where we note that there has been a significant number
of independent measurements 
of $\nu$ especially in $3$ dimensions. This has
provided us with a very reliable benchmark to compare with. Again we had to
apply our resummation methods to both $\nu$ and $1/\nu$ since the estimate of the former 
undershot
the exact 2 dimensional value. This was not
the case for the latter which we would regard as our results for $\nu$.
From Table~\ref{percnuests2} it is clear that our 4 and 5 dimensional five
loop estimates are in very good accord with the Monte Carlo values and are an
improvement on the four loop ones. The situation for 3 dimensions is good and
also closer to numerical values.

{\begin{table}[h]
\begin{center}
\begin{tabular}{c|c|llll}
&Reference & $d$~$=$~$2$ & $d$~$=$~$3$ & $d$~$=$~$4$ & $d$~$=$~$5$ \\
\cline{2-6}
\noalign{\vskip\doublerulesep
         \vskip-\arrayrulewidth}
\cline{2-6} 
&exact   & 1.3333       & & & \\ 
\hline
\hline
$\varepsilon^4$&\multicolumn{1}{r|}{\cite{Gracey:2015tta} (2015)}  &  & 0.8960 & 0.6920 & 0.5746 \\
\hline
\hline
\multirow{13}{*}{\rotatebox[origin=c]{90}{MC/srs} }
& \multicolumn{1}{r|}{\cite{Levenshtein1975} (1975)} &  & 0.80(5) & & \\
& \multicolumn{1}{r|}{\cite{Sur1976} (1976)}&  & 0.8(1) & & \\
& \multicolumn{1}{r|}{\cite{Adler:1990zz} (1990) }&  & 0.872(70) & 0.678(50) & 0.571(3) \\
&\multicolumn{1}{r|}{\cite{Ballesteros:1996ct} (1997) }&  & & 0.689(10) & \\
&\multicolumn{1}{r|}{\cite{Lorenz:1998zz} (1998) }&  & 0.875(1) & & \\
&\multicolumn{1}{r|}{\cite{Ballesteros:1999fx} (2000)}&  & 0.8765(18) & & \\
&\multicolumn{1}{r|}{\cite{Wang2013} (2013)} &  & 0.8764(12) & & \\
&\multicolumn{1}{r|}{\cite{Hu2014} (2014)}&  & 0.8751(11) & & \\
&\multicolumn{1}{r|}{\cite{Xu2014} (2014)}&  & 0.87619(11) & & \\
&\multicolumn{1}{r|}{\cite{Koza2016} (2016)} &  & 0.8774(13) & 0.6852(28) & 0.5723(18) \\
&\multicolumn{1}{r|}{\cite{LeClair:2018edq} (2018)}&  & & 0.693 & \\
&\multicolumn{1}{r|}{\cite{Tan:2020htb} (2020)}&  & & 0.6845(6) & 0.5757(7) \\
\cline{2-6}
& overall &            & 0.876(4)    & 0.685(2)    & 0.574(2)    \\
\hline
\hline
\multirow{9}{*}{\rotatebox[origin=c]{90}{4 loops} }&Pad\'{e}        & 1.1(3)       & 0.82(9)      & 0.68(2)      & 0.573(3)     \\
&PBL         & 1.0(3)       & 0.80(9)      & 0.67(2)      & 0.572(3)     \\
&KP17        & 1.5(7)       & 0.91(10)     & 0.69(2)      & 0.574(2)     \\
&cPad\'{e}-2     &             & 0.896(3)     & 0.6916(12)   & 0.5746(2)    \\
&cPBL-2      &             & 0.896(2)     & 0.6918(11)   & 0.5746(2)    \\
&cKP17-2     &             & 0.90(3)      & 0.685(10)    & 0.5739(9)    \\
&Constr.~Pad\'{e}  &             & 0.895(5)      & 0.691(2)      & 0.5745(3)    \\
\cline{2-6}
&non constr. & 1.1(4)       & 0.84(10)     & 0.68(2)      & 0.573(2)     \\
&constr.     &              & 0.896(4)     & 0.691(2)     & 0.5745(3)    \\
&all         & 1.1(4)       & 0.894(14)    & 0.691(5)     & 0.5744(7)    \\

\hline
\hline
\multirow{9}{*}{\rotatebox[origin=c]{90}{5 loops} }&Pad\'{e}        & 1.1(2)       & 0.84(7)      & 0.681(13)    & 0.5737(10)   \\
&PBL         & 0.99(13)     & 0.81(4)      & 0.675(8)     & 0.5733(6)    \\
&KP17        & 1.17(12)     & 0.86(3)      & 0.683(6)     & 0.5738(5)    \\
&cPad\'{e}-2     &             & 0.895(3)     & 0.6914(13)   & 0.5746(2)    \\
&cPBL-2      &             & 0.88(3)      & 0.687(11)    & 0.5741(11)   \\
&cKP17-2     &             & 0.89(3)      & 0.685(5)     & 0.5738(3)    \\
&Constr.~Pad\'{e}  &             & 0.899(11)     & 0.6913(12)    & 0.5744(5)    \\

\cline{2-6}
&non constr. & 1.1(2)       & 0.84(5)      & 0.680(9)     & 0.5736(7)    \\
&constr.     &              & 0.894(9)     & 0.690(3)     & 0.5743(5)    \\
&all         & 1.1(2)       & 0.89(3)      & 0.689(5)     & 0.5741(6)    \\

\end{tabular}
\end{center}
\caption{Estimates for $\nu$ in percolation using
the series for $1/\nu$.}
\label{percnuests2}
\end{table}

}

\begin{figure}[h]
	\centering
  \includegraphics[width=.9\textwidth]{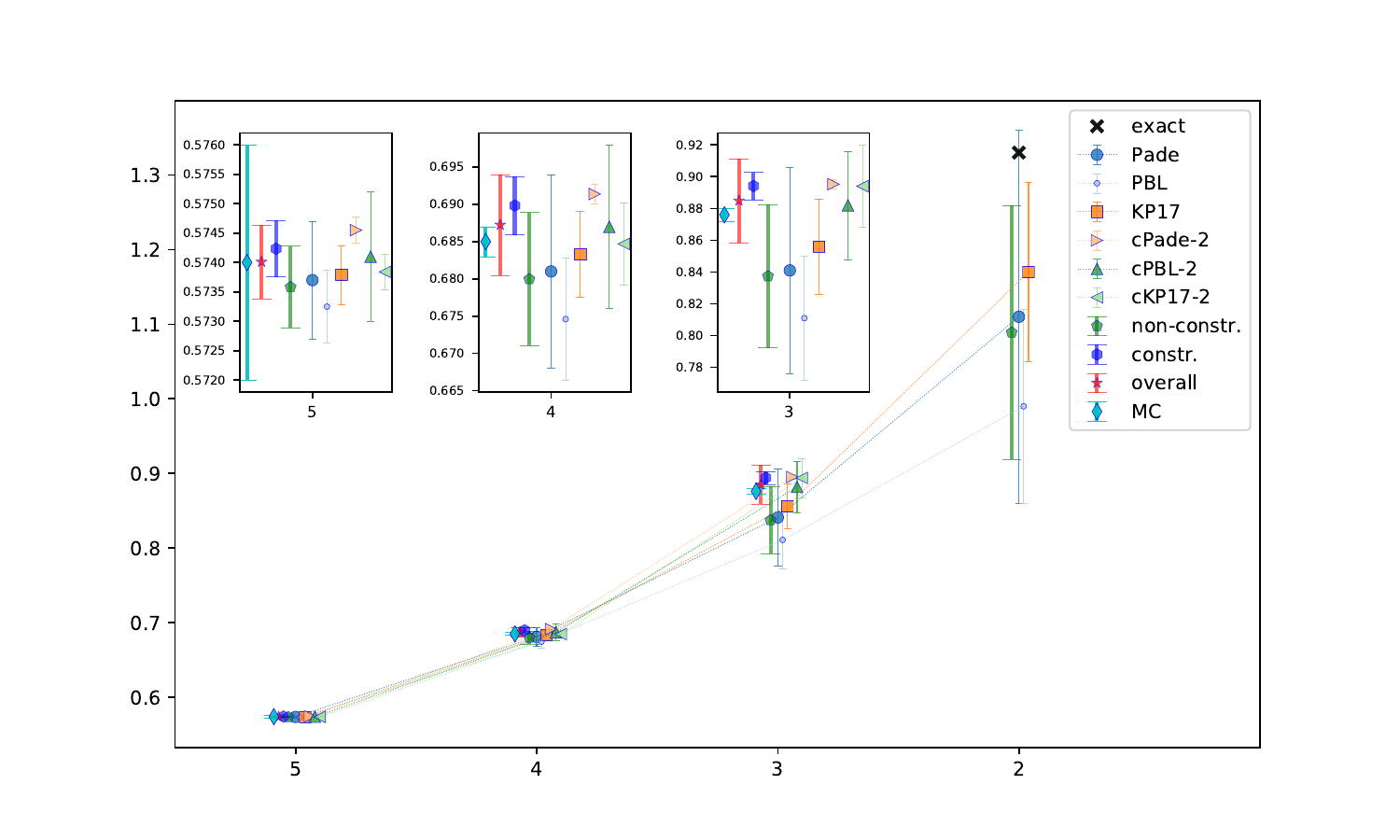}
	\caption{Plot of estimates (from $1/\nu$) for $\nu$ in percolation.}
	\label{percnufig}
\end{figure}

{\begin{table}[h]
\begin{center}
\begin{tabular}{c|c|llll}
&Reference & $d$~$=$~$2$ & $d$~$=$~$3$ & $d$~$=$~$4$ & $d$~$=$~$5$ \\
\cline{2-6}
\noalign{\vskip\doublerulesep
         \vskip-\arrayrulewidth}
\cline{2-6} 
&exact   & 0.3956       & & & \\ 
\hline
\hline

$\varepsilon^4$&\multicolumn{1}{r|}{\cite{Gracey:2015tta} (2015)}&  & 0.4419 & 0.4642 & 0.4933 \\
\hline
\hline
\multirow{5}{*}{\rotatebox[origin=c]{90}{MC/srs} }
&\multicolumn{1}{r|}{\cite{Sykes1976} (1976)} &  & 0.42(6) & & \\
&\multicolumn{1}{r|}{\cite{Ballesteros:1996ct} (1996)} &  & 0.4522(8) & & \\
&\multicolumn{1}{r|}{\cite{Lorenz:1998zz} (1998)}&  & 0.445(10) & & \\
&\multicolumn{1}{r|}{\cite{Xu2014} (2014)} &  & 0.45237(8) & & \\
\cline{2-6}
& overall &            & 0.4523(13)  &            &            \\
\hline
\hline

\multirow{9}{*}{\rotatebox[origin=c]{90}{4 loops} } & Pad\'{e}        & 0.45(5)      & 0.47(3)      & 0.483(10)    & 0.494(2)     \\
&PBL         & 0.45(3)      & 0.470(14)    & 0.484(5)     & 0.4946(10)   \\
&KP17        & 0.47(6)      & 0.48(4)      & 0.475(13)    & 0.4937(13)   \\
&cPad\'{e}-2     &             & 0.442(2)     & 0.476(6)     & 0.4934(10)   \\
&cPBL-2      &             & 0.440(6)     & 0.473(3)     & 0.4930(5)    \\
&cKP17-2     &             & 0.448(9)     & 0.478(5)     & 0.4939(7)    \\
&Constr.~Pad\'{e} &            & 0.444(11)   & 0.475(6)    & 0.4932(13)  \\

\cline{2-6}
&non constr. & 0.46(4)      & 0.47(2)      & 0.482(9)     & 0.4943(13)   \\
&constr. &              & 0.443(6)     &0.475(5)    &0.4934(8)    \\
&all &  0.46(4)    & 0.447(13)    &0.477(7)    &0.4937(11)   \\

\hline
\hline
\multirow{9}{*}{\rotatebox[origin=c]{90}{5 loops} }&Pad\'{e}        & 0.44(4)      & 0.46(2)      & 0.480(6)     & 0.4940(8)    \\
&PBL         & 0.446(13)    & 0.465(5)     & 0.4814(15)   & 0.49409(14)  \\
&KP17        & 0.44(5)      & 0.45(2)      & 0.478(4)     & 0.4939(3)    \\
&cPad\'{e}-2     &             & 0.4425       & 0.476(6)     & 0.4934(7)    \\
&cPBL-2      &             & 0.443(15)    & 0.475(9)     & 0.4935(11)   \\
&cKP17-2     &             & 0.452(7)     & 0.4789(14)   & 0.49396(13)  \\
&Constr.~Pad\'{e} &            & 0.447(9)    & 0.477(5)    & 0.4937(8)   \\

\cline{2-6}
&non constr. & 0.44(2)      & 0.462(12)    & 0.480(3)     & 0.4940(3)    \\
&constr. &              & 0.448(12)    &0.478(4)    &0.4939(4)    \\
&overall &  0.44(2)  & 0.454(14)    &0.479(4)    &0.4940(3)    \\
\end{tabular}
\end{center}
\caption{Estimates for $\sigma$ in percolation.}
\label{percsigmaests}
\end{table}}

\begin{figure}[h]
	\centering
  \includegraphics[width=.9\textwidth]{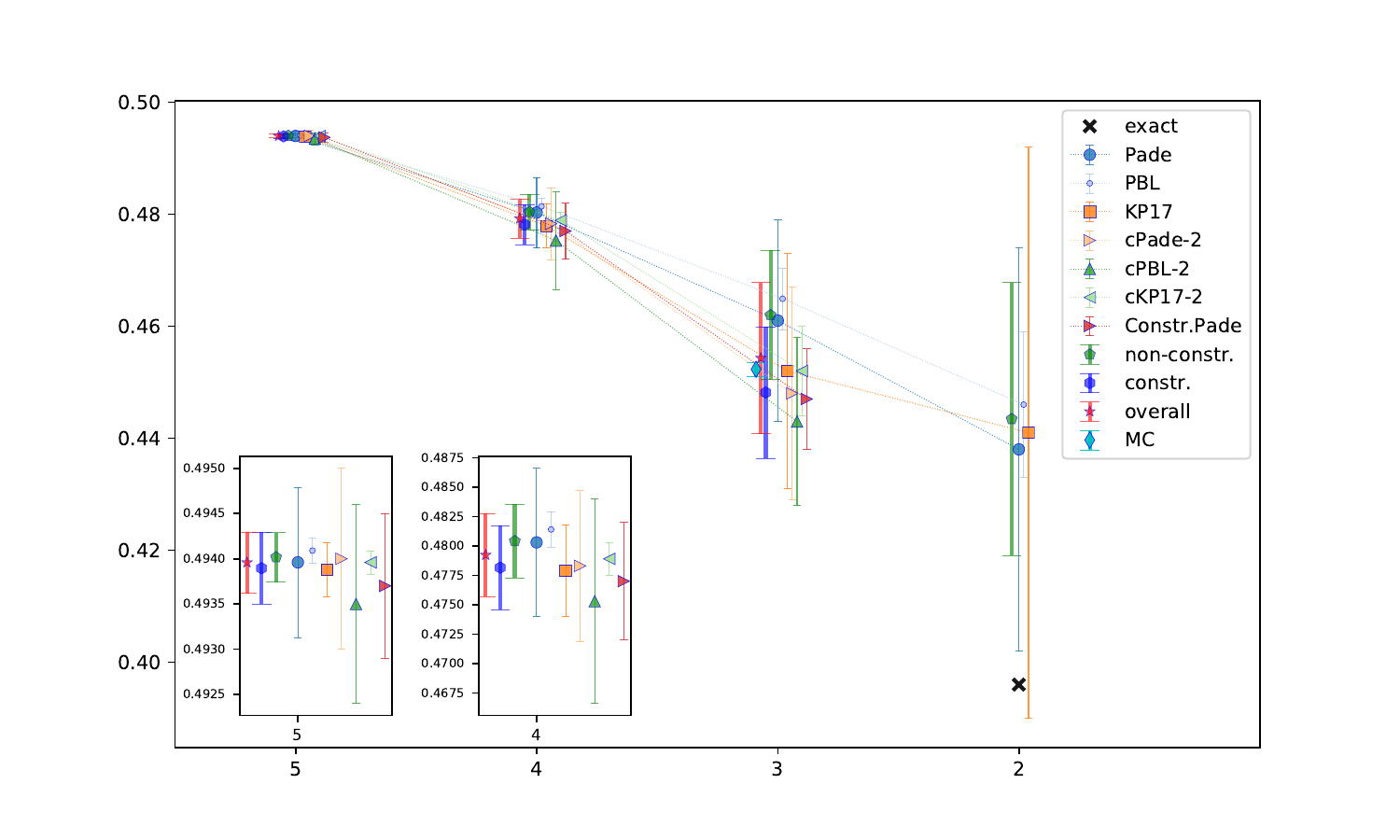}
	\caption{Plot of estimates for $\sigma$ in percolation.}
	\label{percsigmafig}
\end{figure}

Examining Table~\ref{percsigmaests} which summarizes our investigation for $\sigma$, we 
note that the only independent estimates are in $3$ dimensions. We have clearly
obtained excellent agreement with these which
exceed the situation with other
exponents.
This outcome has to be qualified by noting that in strictly 6
dimensions $\sigma$ is $0.5$ with the exact result in 2 dimensions being just
under $0.4$. In other words the behaviour of $\sigma$ in the intervening
dimensions is a shallow decrease as $d$ decreases which underpins the very
good overlap. 
Consequently even though there have been no measurements of
$\sigma$ in 4 and 5 dimensions that
we can compare with, we would
expect our five loop estimates to be in line with any future Monte Carlo
studies in those dimensions.

{\begin{table}[h]
\begin{center}
\begin{tabular}{c|c|llll}
&Reference & $d$~$=$~$2$ & $d$~$=$~$3$ & $d$~$=$~$4$ & $d$~$=$~$5$ \\
\cline{2-6}
\noalign{\vskip\doublerulesep
         \vskip-\arrayrulewidth}
\cline{2-6} 
&exact   & 2.0549       & & & \\ 
\hline
\hline

$\varepsilon^4$&\multicolumn{1}{r|}{\cite{Gracey:2015tta} (2015)} &  & 2.1888 & 2.3124 & 2.4171 \\
\hline
\hline
\multirow{11}{*}{\rotatebox[origin=c]{90}{MC/srs} }
&\multicolumn{1}{r|}{\cite{Nakanishi:1980zz} (1980)}&  & & 2.26 & 2.33 \\
&\multicolumn{1}{r|}{\cite{Ballesteros:1996ct} (1996)} &  & 2.18906(8) & 2.3127(6) & \\
&\multicolumn{1}{r|}{\cite{Jan1998} (1998)}&  & 2.186(2) & & \\
&\multicolumn{1}{r|}{\cite{Lorenz:1998zz} (1998)}&  & 2.189(2) & & \\
&\multicolumn{1}{r|}{\cite{Paul2001} (2001)}&  & & 2.313(3) & 2.412(4) \\
&\multicolumn{1}{r|}{\cite{Tiggemann2001} (2001)}&  & 2.190(2) & 2.313(2) & \\
&\multicolumn{1}{r|}{\cite{Tiggemann2006} (2006)}&  & 2.189(1) & & \\
&\multicolumn{1}{r|}{\cite{Xu2014} (2014)}&  & 2.18909(5) & & \\
& \multicolumn{1}{r|}{(site) \cite{Mertens2018} (2018)}&  & 2.1892(1) & 2.3142(5) & 2.419(1) \\
&\multicolumn{1}{r|}{(bond) \cite{Mertens2018} (2018)}&  & 2.1890(2) & 2.311(2) & 2.422(4) \\
&\multicolumn{1}{r|}{\cite{Xun2019} (2019)} &  & & 2.3135(5) & \\
\cline{2-6}
&overall &            & 2.1891(4)    & 2.3133(12)  & 2.418(4)   \\
\hline
\hline
\multirow{9}{*}{\rotatebox[origin=c]{90}{4 loops} }& Pad\'{e}        & 2.10(7)      & 2.20(5)      & 2.318(13)    & 2.4180(15)   \\
&PBL         & 2.11(7)      & 2.22(3)      & 2.323(10)    & 2.4186(14)   \\
&KP17        & 1.9(2)       & 2.16(11)     & 2.31(2)      & 2.417(2)     \\
&cPad\'{e}-2     &             & 2.195(4)     & 2.315(3)     & 2.4176(5)    \\
&cPBL-2      &             & 2.193(3)     & 2.314(2)     & 2.4174(3)    \\
&cKP17-2     &             & 2.194(4)     & 2.315(3)     & 2.4175(9)    \\
&Constr.~Pad\'{e} &            & 2.197(9)    & 2.317(6)    & 2.4179(10)  \\

\cline{2-6}
&non constr. & 2.08(10)     & 2.20(5)      & 2.318(14)    & 2.418(2)     \\
&constr.     &              & 2.194(5)     & 2.315(3)     & 2.4176(6)    \\
&all         & 2.08(10)     & 2.194(9)     & 2.315(5)     & 2.4176(8)    \\

\hline
\hline
\multirow{9}{*}{\rotatebox[origin=c]{90}{5 loops} }& Pad\'{e}        & 2.07(4)      & 2.19(3)      & 2.314(8)     & 2.4174(7)    \\
&PBL         & 2.08(2)      & 2.202(8)     & 2.317(2)     & 2.4176(2)    \\
&KP17        & 2.00(13)     & 2.17(4)      & 2.312(9)     & 2.4173(8)    \\
&cPad\'{e}-2     &             & 2.194(2)     & 2.3146(14)   & 2.4174(2)    \\
&cPBL-2      &             & 2.193(3)     & 2.3145(14)   & 2.4174(2)    \\
&cKP17-2     &             & 2.1938(12)   & 2.3150(8)    & 2.4175(2)    \\
&Constr.~Pad\'{e} &            & 2.194(2)    & 2.3148(14)  & 2.4174(2)   \\

\cline{2-6}
&non constr. & 2.07(4)      & 2.20(2)      & 2.316(5)     & 2.4176(5)    \\
&constr.     &              & 2.194(2)     & 2.3149(12)   & 2.4175(2)    \\
&all         & 2.07(4)      & 2.194(4)     & 2.315(2)     & 2.4175(3)    \\

 \end{tabular}
\end{center}
\caption{Estimates for $\tau$ in percolation.}
\label{perctauests}
\end{table}}

\begin{figure}[h]
	\centering
  \includegraphics[width=.9\textwidth]{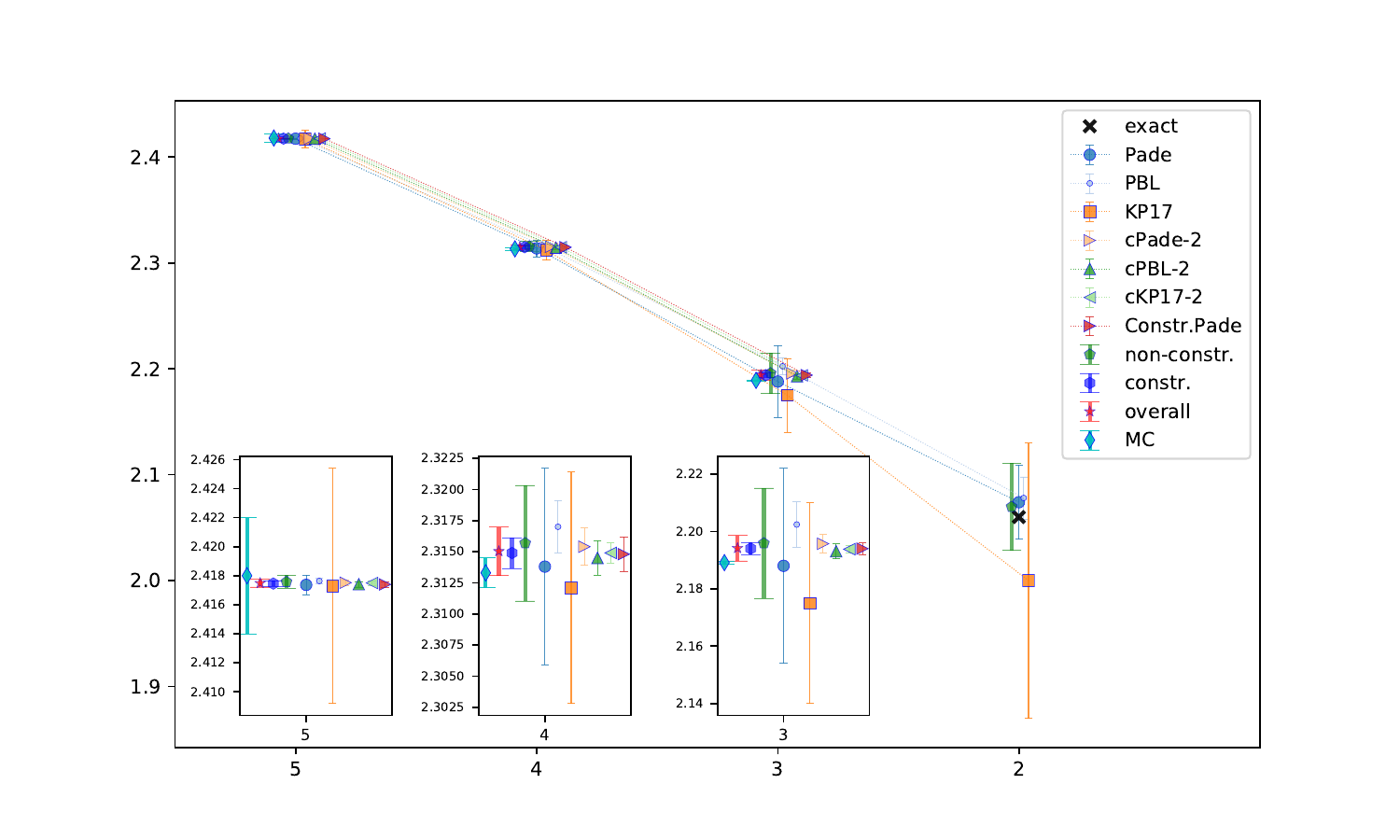}
	\caption{Plot of estimates for $\tau$ in percolation.}
	\label{perctaufig}
\end{figure}

There is a similar picture for $\tau$ which is apparent in Table~\ref{perctauests}. 
Again the behaviour of $\tau$ from 2 to 6 dimensions is
relatively flat due to the exact values in these dimensions
being of a similar order. In this case there are independent data in 4 and 5
dimensions. In the absence of these we could have repeated our final comment
for $\sigma$. So it is reassuring that our five loop 4 and 5 dimensional
estimates are indeed in excellent agreement with the independent studies. The
situation with our 3 dimensional result is in similar accord.

{\begin{table}[h]
\begin{center}
\begin{tabular}{c|c|llll}
&Reference & $d$~$=$~$2$ & $d$~$=$~$3$ & $d$~$=$~$4$ & $d$~$=$~$5$ \\
\cline{2-6}
\noalign{\vskip\doublerulesep
         \vskip-\arrayrulewidth}
\cline{2-6}        
&exact   & 0.7912       & & & \\ 
\hline
\hline
$\varepsilon^4$&\multicolumn{1}{r|}{\cite{Gracey:2015tta} (2015)} &  & & 0.4008 & 0.2304 \\
\hline
\hline
\multirow{10}{*}{\rotatebox[origin=c]{90}{MC/srs} }
&\multicolumn{1}{r|}{\cite{Adler:1990zz} (1990)} &  & 0.50(9) & 0.31(5) & 0.27(7) \\
&\multicolumn{1}{r|}{\cite{Ballesteros:1996ct} (1996)}&  & & 0.37(4) & \\
&\multicolumn{1}{r|}{\cite{Lorenz:1998zz} (1998)}&  & 0.64(2) & & \\
&\multicolumn{1}{r|}{\cite{Jan1998} (1998)}&  & 0.73(8) & & \\
&\multicolumn{1}{r|}{\cite{Ballesteros:1999fx} (1999)} &  & 0.64(5) & & \\
&\multicolumn{1}{r|}{\cite{Gimel2000} (2000)}&  & 0.65(2) & & \\
&\multicolumn{1}{r|}{\cite{Tiggemann2001} (2001)}&  & 0.60(8) & 0.5(1) & \\
&\multicolumn{1}{r|}{\cite{Mertens2018} (2018)}&  & 0.77(3) & & \\
&\multicolumn{1}{r|}{\cite{Xun2019} (2019)}&  & & 0.40(4) & \\
\cline{2-6}
& overall &            & 0.66(8)     & 0.38(8)     &  0.27(7)          \\
\hline
\hline
\multirow{9}{*}{\rotatebox[origin=c]{90}{4 loops} }& Pad\'{e}        & 0.6(3)       & 0.5(2)       & 0.37(7)      & 0.201(14)    \\
&PBL         & 0.768(4)     & 0.583(2)     & 0.3956(8)    & 0.20593(14)  \\
&KP17        & 0.7(2)       & 0.51(15)     & 0.36(7)      & 0.205(15)    \\
&cPad\'{e}-2     &             & 0.58(3)      & 0.39(2)      & 0.205(6)     \\
&cPBL-2      &             & 0.59425(2)   & 0.399952(15) & 0.206613(5)  \\
&cKP17-2     &             & 0.593(5)     & 0.399(5)     & 0.206(2)     \\
&Constr.~Pad\'{e} &            & 0.599(4)    & 0.405(5)    & 0.209(3)    \\

\cline{2-6}
&non constr. & 0.76(2)      & 0.581(13)    & 0.395(5)     & 0.2059(6)    \\
&constr.     &              & 0.5943(5)    & 0.4000(4)    & 0.20661(10)   \\
&all         & 0.76(2)      & 0.594(2)     & 0.3999(10)   & 0.2066(2)    \\

\hline
\hline
\multirow{9}{*}{\rotatebox[origin=c]{90}{5 loops} }& Pad\'{e}        & 0.6(3)       & 0.50(15)     & 0.36(6)      & 0.21(3)      \\
&PBL         & 0.96(5)      & 0.68(2)      & 0.432(5)     & 0.2112(3)    \\
&KP17        & 0.7(2)       & 0.5(2)       & 0.47(9)      & 0.215(14)    \\
&cPad\'{e}-2     &             & 0.58(3)      & 0.39(2)      & 0.205(5)     \\
&cPBL-2      &             & 0.608(6)     & 0.411(5)     & 0.2095(12)   \\
&cKP17-2     &             & 0.595(3)     & 0.400(3)     & 0.2067(12)   \\
&Constr.~Pad\'{e} &            & 0.597(2)    & 0.403(2)    & 0.2079(11)  \\

\cline{2-6}
&non constr. & 0.9(2)       & 0.65(8)      & 0.43(2)      & 0.2112(9)    \\
&constr.     &              & 0.598(7)     & 0.404(6)     & 0.208(2)     \\
&all         & 0.9(2)       & 0.60(2)      & 0.409(15)      & 0.210(2)     \\

\end{tabular}
\end{center}
\caption{Estimates for $\Omega$ in percolation.}
\label{percOmegaests}
\end{table}}

For the exponent $\Omega$ there are again a large
number of independent measurements 
that we have shown in Table~\ref{percOmegaests} though mostly in
3 dimensions. However these and those in 4 dimensions have a wide range of
central values as well as a range of errors that are 10\% or more in some cases. By
contrast to $\sigma$ and $\tau$ there is a much larger variation in $\Omega$
across the dimensions rising from zero in strictly 6 dimensions as $d$
decreases. Despite this both our constrained and unconstrained resummations as well
as from four to five loops appear to have a
degree of stability.
Although our check estimate against the exact two
dimensional value is not as good as for
other exponents, it is encouraging that our 3 dimensional
five loop estimate has overlap with independent data. This is
clearly more pronounced for 4 and 5 dimensions. For instance, our final five loop 4
dimensional estimate of $0.41(2)$ for $\Omega$ is consistent with the
recent 4 dimensional value of $0.40(4)$, \cite{Xun2019}.

Finally we arrive at the correction to scaling exponent $\omega$ with the situation summarized
in Table~\ref{percomegaests}. As
mentioned earlier for our constrained resummations we have taken the exact
value in 2 dimensions to be $3/2$. The independent data is mostly available
in 3 dimensions and like $\Omega$ covers a wide range from around $1.0$ to
$1.62$. Again similar to $\Omega$, $\omega$ rises from zero to a large 2
dimensional value. The overall picture is that
in 3 dimensions our five loop estimates are
close to one standard deviation away from our calculated global value from the
Monte Carlo data. Though the
constrained data in 3 dimensions is in better
accord. However the 2 dimensional projected
value for our unconstrained resummations is not close to the exact value.
What would be useful are independent numerical studies in 4 and 5 dimensions to compare our
results with.

{\begin{table}[h]
\begin{center}
\begin{tabular}{c|c|llll}
&Reference & $d$~$=$~$2$ & $d$~$=$~$3$ & $d$~$=$~$4$ & $d$~$=$~$5$ \\
\cline{2-6}
\noalign{\vskip\doublerulesep
         \vskip-\arrayrulewidth}
\cline{2-6}   
&exact   & 1.5          & & & \\ 
\hline
\hline
$\varepsilon^4$&\multicolumn{1}{r|}{\cite{Gracey:2015tta} (2015)}&  & 1.6334 & 1.2198 & 0.7178 \\
\hline
\hline
\multirow{6}{*}{\rotatebox[origin=c]{90}{MC/srs} }
&\multicolumn{1}{r|}{\cite{Adler:1990zz} (1990)}&  & 1.26(23) & 0.94(15) & 0.96(26) \\
&\multicolumn{1}{r|}{\cite{Ballesteros:1996ct} (1996)}&  & 1.13(10) & & \\
&\multicolumn{1}{r|}{\cite{Lorenz:1998zz} (1998)}&  & 1.61(5) & & \\
&\multicolumn{1}{r|}{\cite{Ballesteros:1999fx} (1999)}&  & 1.62(13) & & \\
&\multicolumn{1}{r|}{\cite{Kozlov2010} (2010)} &  & 1.0(2) & & \\
\cline{2-6}
& overall &         & 1.4(3)              &    0.94(15)        &  1.0(3)     \\
\hline
\hline

\multirow{9}{*}{\rotatebox[origin=c]{90}{4 loops} }& Pad\'{e}        & 2(2)  & 1.5(1.3)     & 1.2(5)       & 0.73(10)     \\
&PBL         & 2.58(4)      & 1.99(2)      & 1.388(7)     & 0.7544(14)   \\
&KP17        & 2.2(8)       & 1.8(4)       & 1.3(2)       & 0.77(3)      \\
&cPad\'{e}-2     &             & 1.36(7)      & 1.12(6)      & 0.70(2)      \\
&cPBL-2      &             & 1.38(6)      & 1.13(6)      & 0.70(2)      \\
&cKP17-2     &             & 1.38(9)      & 1.13(9)      & 0.70(3)      \\
&Constr.~Pad\'{e}  &            & 1.35(4)     & 1.10(5)     & 0.70(2)     \\

\cline{2-6}
&non constr. & 2.6(2)       & 1.98(9)      & 1.38(4)      & 0.755(6)     \\
&constr.     &              & 1.37(6)      & 1.12(6)      & 0.70(2)      \\
&all         & 2.6(2)       & 1.6(3)       & 1.30(13)     & 0.74(2)      \\

\hline
\hline
\multirow{9}{*}{\rotatebox[origin=c]{90}{5 loops} }&Pad\'{e}        & 2(2)  & 1.5(1.1)     & 1.2(4)       & 0.75(10)     \\
&PBL         & 2.37(5)      & 1.88(2)      & 1.349(6)     & 0.7483(6)    \\
&KP17        & 2.8(7)       & 2.1(4)       & 1.43(12)     & 0.76(2)      \\
&cPad\'{e}-2     &             & 1.37(8)      & 1.13(7)      & 0.71(2)      \\
&cPBL-2      &             & 1.44(9)      & 1.19(8)      & 0.73(2)      \\
&cKP17-2     &             & 1.39(9)      & 1.14(9)      & 0.71(3)      \\
&Constr.~Pad\'{e} &            & 1.35(5)     & 1.10(6)     & 0.69(3)     \\

\cline{2-6}
&non constr. & 2.4(2)       & 1.89(9)      & 1.35(3)      & 0.749(3)     \\
&constr.     &              & 1.39(11)      & 1.15(11)      & 0.71(4)      \\
&all         & 2.4(2)       & 1.7(3)       & 1.31(9)      & 0.746(11)    \\

\end{tabular}
\end{center}
\caption{Estimates for $\omega$ in percolation.}
\label{percomegaests}
\end{table}}

\begin{figure}
	\centering
  \includegraphics[width=.9\textwidth]{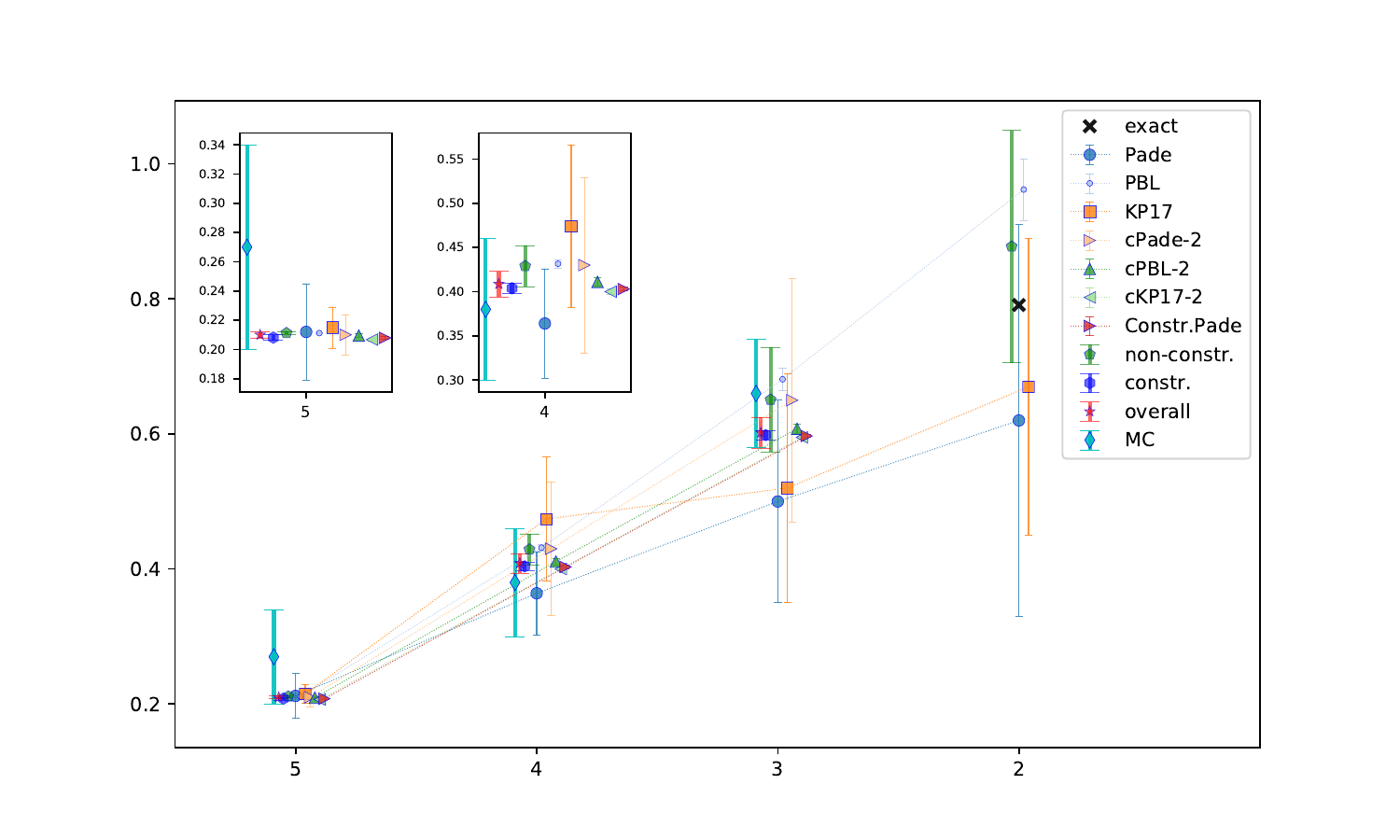}
	\caption{Plot of estimates for $\Omega$ in percolation.}
	\label{percOmegafig}
\end{figure}

\clearpage

\begin{figure}
	\centering
  \includegraphics[width=.9\textwidth]{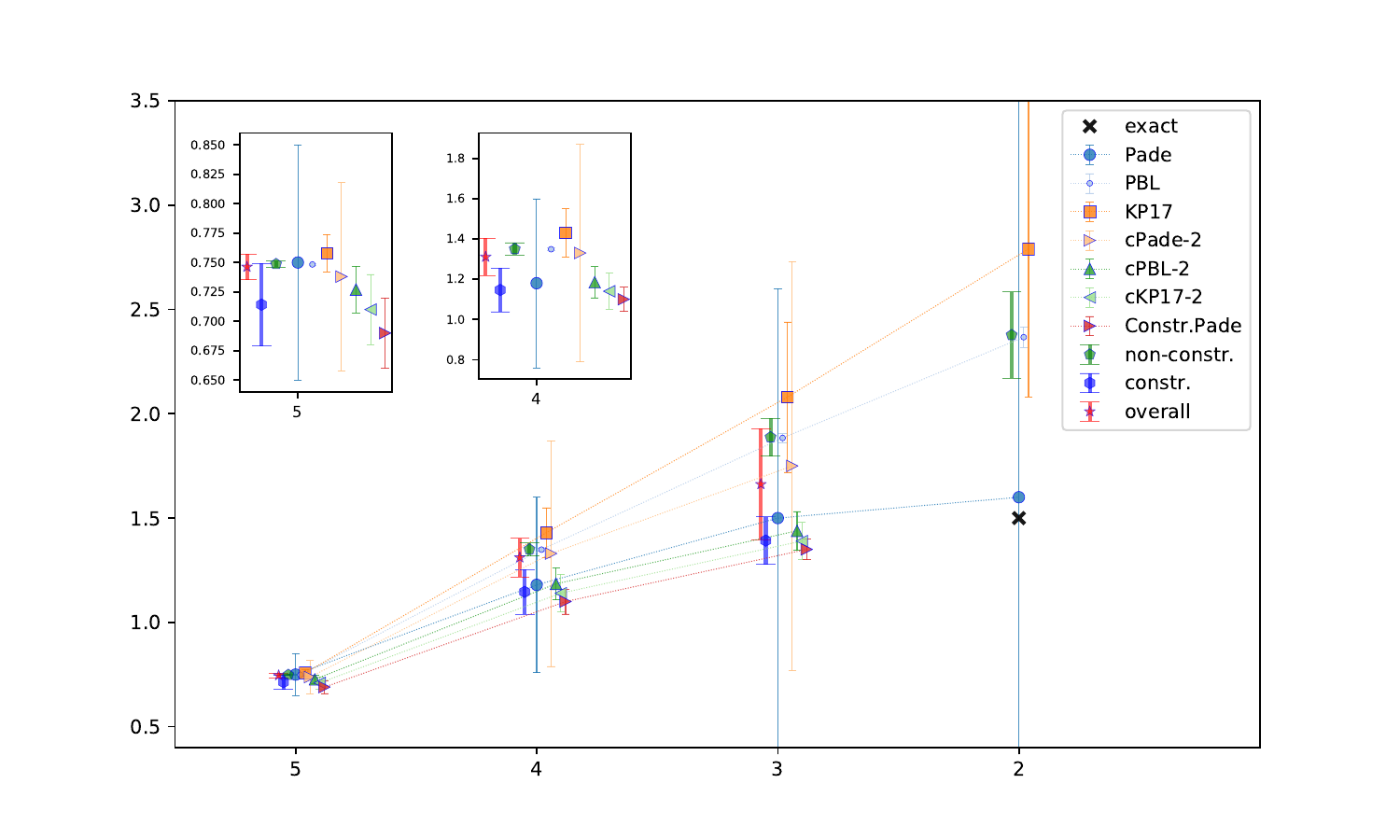}
	\caption{Plot of estimates for $\omega$ in percolation.}
	\label{percomegafig}
\end{figure}

{\begin{table}[h]
\begin{center}
\begin{tabular}{c||lll}
exponent & $d$~$=$~$3$ & $d$~$=$~$4$ & $d$~$=$~$5$ \\
\hline
   $\sigma$&   0.452(7) &  0.4789(14) &  0.49396(13) \\
   $\tau$  &   2.1938(12) &  2.3150(8) & 2.4175(2) \\
\hline
   $\alpha$ &    -0.64(4) &-0.75(2) & -0.870(1) \\
   $\beta$   &     0.429(4) & 0.658(1) &  0.8452(2) \\
   $\gamma$  &     1.78(3)  &1.430(6) & 1.1792(7) \\
   $\delta$  &     5.16(4)  &3.175(8) & 2.3952(12) \\
   $\eta$    &   -0.03(1)  &-0.084(4) & -0.0547(10) \\
   $\nu$    &      0.88(2)  &0.686(2) & 0.5739(1) \\
\end{tabular}
\end{center}
\caption{Estimates of percolation exponents
from hyperscaling relations using
$O(\varepsilon^5)$ cKP17-2 estimates of 
$\sigma$ and $\tau$ as input.}
\label{perctotalhyper}
\end{table}}

For the most part our focus in this section has
been on providing estimates for various
exponents. To do this we used the $\varepsilon$
expansion expressions for $\eta$ and $\nu$ and
then derived $O(\varepsilon^5)$ expansions for
the other exponents through the scaling and
hyperscaling relations of (\ref{scalerelns}).
We then applied various different summation
methods to each exponent to arrive at estimates
for the three dimensions of interest. One
question arises as a consequence of this and
concerns whether the individual estimates are
then consistent with the scaling and
hyperscaling relations. Equally another
question is which pair of exponents could
reasonably be regarded as the ones from which
all the other estimates can be accurately
described and thereby be our final exponent
estimates. To this end we have presented our
analysis in Table \ref{perctotalhyper}. In
particular we have chosen $\sigma$ and $\tau$ as
our two independent exponents and note that the
results using the cKP17-2 technique are the most
reasonable to use for final predictions. This is
because the two dimensional constraints have
been implemented and these two exponents are
accurate with a tight error bar in each
dimension. Using these cKP17-2 values we have
computed the exponents for $\alpha$, $\beta$,
$\gamma$, $\delta$, $\eta$ and $\nu$ using (\ref{scalerelns}). In the Table the independent
values for $\sigma$ and $\tau$ are given above
the rule with the values derived from scaling
relations recorded below. For the remaining two
exponents $\omega$ and $\Omega$, their cKP17-2 
values have not been included as they are not
accessible by scaling or hyperscaling relations
and those estimates are already in their
respective tables. 

Comparing these scaling relation values of Table
\ref{perctotalhyper} with those from the direct
summation ones given in the earlier tables, none
look far out of line. One exception might
possibly be the 3 dimensional value of $\alpha$
as it appears to be more consistent with the
unconstrained analysis. However looking at the
extrapolation to 2 dimensions suggests it is
probably on a better trajectory. The lack of
Monte Carlo or series computations for this
exponent means that we have no way of gauging
our $\alpha$ estimates against an independent
analysis. For the other exponents the view is
that the individual resummations are consistent
with the scaling relations and also with
independent data from other methods where that
is substantial. In the case of $\delta$ where
there are only a few such other cases the
estimates in Table \ref{perctotalhyper} are not
inconsistent. Our final 3 dimensional estimate
for $\eta$ appears to be on the low side but
this exponent is difficult to measure
accurately. For $\nu$ our 4 and 5 dimensional
values are remarkably close to the global
average we compiled. While that for 3
dimensions is not as accurate, it does lie 
within the errors of the global estimate. 

In presenting our final estimates in this Table
we need to be clear in stating that this shows
our results are consistent with scaling
relations both by constructing the $\varepsilon$
equations directly and resumming them before
comparing the estimates obtained from using two
independent summed values in the (hyper)scaling
relations. So in essence it is a self-consistency check. By contrast it is
usually the case that in the Monte Carlo and
series approaches the focus is on one or two
specific exponents. Those values are then used
to generate the remaining exponents via
(\ref{scalerelns}) rather than make extra
measurements. So until all techniques have
achieved a large degree of computational 
accuracy it is probably the case that in
comparing exponent estimates the overall picture
is still not perfect.

\section{Conclusions} 

\label{sec:conclusions}

The main result of this paper is clearly the provision of the five loop
renormalization group functions of $\phi^3$ theory in six dimensions in the
$\MSbar$ scheme. This level of precision could not have been achieved without
the use of the graphical function formalism developed in
\cite{Schnetz:2013hqa,Schnetz_2018}. That method was originally pioneered in
four dimensions to renormalize $\phi^4$ theory to {\em seven} loops in the same
scheme \cite{Schnetz_2018,Schnetz2020phi4loop7}. To achieve the level of five
loop accuracy here required the extension of graphical functions to six
dimensions that was provided in \cite{Borinsky2020prep,Borinsky2020prep2}. An
indication of the advantage of such new and powerful techniques can be gained
from the dates that the previous loop order results became available for the
cubic theory. The one and two loop renormalization group functions were
recorded in the early 70's \cite{Macfarlane:1974vp} with the three loop
extension appearing within seven years, \cite{Bonfim_1980,Bonfim_1981}. Similar
to the extension of $\phi^4$ theory from five to six loops, there was a
quarter of a century lull before $\phi^3$ theory was renormalized at four loops
\cite{Gracey:2015tta}. The relatively quick extension, in terms of time, to five
loops here is suggestive that with suitable investment in the underlying
mathematics of the graphical functions approach, higher loop orders are
potentially within reach in this and other theories.

Such higher order computations are not purely academic exercises since the
second part of our study was to extract improved estimates for critical
exponents in two important problems. Both the Lee--Yang edge singularity and
percolation problems at criticality lie in separate universality classes but
both have a $\phi^3$ continuum field theory at their heart. The main difference
is that the respective versions of the cubic theory are decorated with
different symmetry properties. Given that we have renormalized the pure
$\phi^3$ Lagrangian at five loops, it was a relatively simple exercise to
include the respective symmetry decorations and determine the two sets of
renormalization group functions. From these the $\varepsilon$ expansion of the
critical exponents were constructed to $\mathcal O(\varepsilon^5)$ before a variety of
resummation techniques were applied to extract numerical estimates. Moreover,
we devoted a significant part of determining estimates to a careful error
analysis using the same formalism provided in \cite{KompanietsPanzer2017} for
$\phi^4$ theory. What ought to be recognized is that on the whole not only do
the five loop results improve upon the four loop results, as well as showing
convergence, but also that there is good agreement with other techniques in
both problems for the dimensions of interest. These include Monte
Carlo and strong coupling methods to name but a few. In making this remark it
should not be overlooked that this includes 3 dimensions where we have summed
from $d$~$=$~$6$~$-$~$\varepsilon$ down to $d$~$=$~$3$ with a large value of
$\varepsilon$. In one sense this confirms the role of $\phi^3$ theory as being in
the same universality class. What was useful in
making this comparison between the exponents from
discrete systems and the continuum field theory
was collating the available data for the former
to produce a global average. The associated error
bars were produced with the same routine that we
used for the five loop results from the various
resummations. In this respect we were
endeavouring to compare the picture in the
discrete and continuum sides in the same way.

The determination of the five loop
renormalization group functions in the core
cubic theory opens up the possibility of 
studying related six dimensional cubic theories
to the same level of precision. For instance the
conformal bootstrap formalism represents a
powerful tool to calculate exponents. It was used in \cite{Pang2016} to determine
exponents for $\phi^3$ theory in a variety of
representations of the Lie group $F_4$. While
the three loop comparison in that article was
in reasonable agreement for various exponents
calculated with the bootstrap, the four loop
study of \cite{Gracey_2017_F4} gave an
improvement towards convergence. It would
therefore be interesting, given the accuracy of
results in \cite{Pang2016}, to extend to the
$F_4$ study of \cite{Gracey_2017_F4} to
five loops. Aside from this particular symmetry
of the cubic theory, an intriguing property of
$\phi^3$ with a bi-adjoint symmetry was observed
in \cite{Gracey:2020baa}. For any Lie group it 
transpires that this theory is asymptotically 
free due to the {\em two} loop coefficient of
the $\beta$ function; the one loop term is
identically zero. One feature of the four
loop result was the appearance of higher order
group Casimirs at three loops. This is one order
earlier than in four dimensional non-abelian
gauge theories. So the bi-adjoint theory offers
a window into the type of group Casimirs that 
could appear in six loop gauge theory $\beta$
functions such as QCD. Although such problems are worth
pursuing, the more interesting extension of our
current work would clearly be to six loops. This
would obviously provide a higher level of
precision with which to compare future numerical
studies of the discrete spin models in the same universality class. The graphical functions
method that was used here, and was successful in
extending $\phi^4$ theory to seven loops
\cite{Schnetz_2018,Schnetz2020phi4loop7} should
be regarded as a starting point to achieve the
six loop renormalization of $\phi^3$ theory. 

\paragraph{Acknowledgements.}
We warmly thank A.J.~McKane 
for discussions on the resummation of 
asymptotic series in quantum field theory. Also
JAG thanks R.M.~Ziff for encouragement and
discussions, and in particular that surrounding
the value of $\omega$ in two dimensional
percolation theory. 
We also thank J. Adler for comments and
pointing out \cite{Adler1988}.
The work of MB was supported by the NWO Vidi grant 680-47-551 ``Decoding Singularities of Feynman graphs''.
The work of JAG was supported
by a DFG Mercator Fellowship and in part through
the STFC Consolidated grant ST/T000988/1. 
The work of MVK was supported by Foundation for
the Advancement of Theoretical Physics ``BASIS'' 
through Grant 18-1-2-43-1. 
The work of OS was supported by the DFG grant SCHN~1240.

\providecommand{\href}[2]{#2}\begingroup\raggedright\endgroup
\end{document}